\documentclass[journal]{rmaa}

\usepackage{paralist}

\usepackage{psfrag,color}

\usepackage[latin1]{inputenc}

\usepackage{graphicx}
\usepackage{lscape}
\usepackage{txfonts}
\usepackage{natbib}
\usepackage{url}
\usepackage[T1]{fontenc}

\usepackage[bookmarks=false]{hyperref}

\newcommand{\fuden}{10$^{-17}$ erg~s$^{-1}$~cm$^{-2}$~arcsec$^{-1}$}
\newcommand{\fuDEN}{10$^{-16}$ erg~s$^{-1}$~cm$^{-2}$~arcsec$^{-1}$}
\newcommand{\funits}{10$^{-16}$ erg~s$^{-1}$~cm$^{-2}$~\AA$^{-1}$}
\newcommand{\Funits}{10$^{-16}$ erg~s$^{-1}$~cm$^{-2}$}

\newcommand{\FunitsA}{10$^{-16}$ erg~s$^{-1}$~cm$^{-2}$~arcsec$^{-2}$}
\newcommand{\degree}{\ensuremath{^\circ}}

\DeclareRobustCommand{\ion}[2]{%
\relax\ifmmode
\ifx\testbx\f@series
{\mathbf{#1\,\mathsc{#2}}}\else
{\mathrm{#1\,\mathsc{#2}}}\fi
\else\textup{#1\,{\mdseries\textsc{#2}}}%
\fi}

\newcommand{\HII}{\ion{H}{ii}~}

\title{Pipe3D, a pipeline to analyse Integral Field Spectroscopy Data: II. Analysis sequence and CALIFA dataproducts.}

\author{
S.\,F. S\'anchez\altaffilmark{1},
E.\,P\'erez\altaffilmark{2},
P.\,S\'anchez-Bl\'azquez\altaffilmark{3},
R.\,Garc\'\i a-Benito\altaffilmark{2},
H.\,J.\,Ibarra-Mede\altaffilmark{6},
J.J.\,Gonz\'alez\altaffilmark{1},
F.F.\,Rosales-Ortega\altaffilmark{4},
L.\,S\'anchez-Menguiano\altaffilmark{2},
Y. Ascasibar\altaffilmark{5},
T. Bitsakis\altaffilmark{1},
D. Law\altaffilmark{6},
M.\,Cano-D\'\i az\altaffilmark{1},
C.\,L\'opez-Cob\'a\altaffilmark{1},
R.\,A. Marino\altaffilmark{7},
A. Gil de Paz\altaffilmark{8},
A. R. L\'{o}pez-S\'{a}nchez\altaffilmark{9},
J. Barrera-Ballesteros\altaffilmark{6},
L. Galbany\altaffilmark{10,11},
D. Mast\altaffilmark{12,13},
V. Abril-Melgarejo\altaffilmark{1},
A. Roman-Lopes\altaffilmark{14}
}

\altaffiltext{1}{Instituto de Astronom\'\i a, Universidad Nacional Auton\'oma de M\'exico, A.P. 70-264, 04510 M\'exico, D.F.,  Mexico.}
\altaffiltext{2}{Instituto de Astrof\'{\i}sica de Andaluc\'{\i}a (CSIC), Glorieta de la Astronom\'\i a s/n, Aptdo. 3004, 18080 Granada, Spain.}
\altaffiltext{3}{Departamento de F\'isica Te\'orica, Universidad Aut\'onoma de Madrid, 28049 Madrid, Spain.}
\altaffiltext{4}{Instituto Nacional de Astrof{\'i}sica, {\'O}ptica y Electr{\'o}nica, Luis E. Erro 1, 72840 Tonantzintla, Puebla, Mexico. }
\altaffiltext{5}{Universidad Autónoma de Madrid, 28049 Madrid, Spain; Astro-UAM, UAM, Unidad Asociada CSIC}
\altaffiltext{6}{Space Telescope Science Institute, 3700 San Martin Drive, Baltimore, MD 21218, USA}
\altaffiltext{7}{CEI Campus Moncloa, UCM-UPM, Departamento de Astrof\'{i}sica y CC. de la Atm\'{o}sfera, Facultad de CC. F\'{i}sicas, Universidad Complutense de Madrid}
\altaffiltext{8}{Australian Astronomical Observatory, PO BOX 296, Epping, NSW 1710, Australia.}
\altaffiltext{9}{Instituto de Astrof\'\i sica de Canarias (IAC), 38205 La Laguna, Tenerife, Spain}
\altaffiltext{10}{Millennium Institute of Astrophysics, Chile}
\altaffiltext{11}{Departamento de Astronom\'\i a, Universidad de Chile, Camino El Observatorio 1515, Las Condes, Santiago, Chile}
\altaffiltext{12}{Observatorio Astron\'omico, Laprida 854, X5000BGR, C\'ordoba, Argentina.}
\altaffiltext{13}{Consejo de Investigaciones Cient\'{i}ficas y T\'ecnicas de la Rep\'ublica Argentina, Avda. Rivadavia 1917, C1033AAJ, CABA, Argentina.}
\altaffiltext{14}{Departamento de F\'\i sica y Astronom\'\i a, Universidad de La Serena, Cisternas 1200, La Serena, Chile.}

\shortauthor{S\'anchez}
\shorttitle{Pipe3D, analysis sequence}

\fulladdresses{

\item Instituto de Astronom\'\i a,Universidad Nacional Auton\'oma de Mexico, A.P. 70-264, 04510, M\'exico,D.F. (sfsanchez@astro.unam.mx)

}

\listofauthors{S. F. S\'anchez}
\indexauthor{S. F. S\'anchez}

\abstract{We present {\sc
    Pipe3D}\footnote{\url{ftp://ftp.caha.es/CALIFA/dataproducts/DR2/Pipe3D_soft/Pipe3D.tgz}},
  an analysis pipeline based on the {\sc FIT3D} fitting tool,
  developed to explore the properties of the stellar populations and
  ionized gas of Integral Field Spectroscopy data. {\sc Pipe3D} was
  created to provide with coherent, simple to distribute, and
  comparable dataproducts, independently of the origin of the data,
  focused on the data of the most recent IFU surveys (e.g., CALIFA,
  MaNGA, and SAMI), and the last generation IFS instruments (e.g.,
  MUSE).  Along this article we describe the different steps involved
  in the analysis of the data, illustrating them by showing the dataproducts
  derived for NGC~2916, observed by CALIFA and P-MaNGA.
As a practical use of the pipeline we present the complete set of dataproducts derived for the 200 datacubes that comprises the V500 setup of the CALIFA Data Release 2 (DR2),
making them freely available through the network. 
Finally, we explore the hypothesis that the properties of the stellar populations and ionized gas of
galaxies at the effective radius are representative of the overall average
ones, finding that this is indeed the case.}

\resumen{Presentamos {\sc Pipe3D}, un dataducto de analisis basado en el paquete de ajustes {\sc FIT3D}, desarrollado para explorar las propiedades de las poblaciones estelares y el gas ionizado en datos de espectroscop\'\i a de campo integral. {\sc Pipe3D} se desarroll\'o para obtener productos derivados de una forma coherente, f\'acil de distribuir y de comparar independientemente del origen de los datos, enfocado al an\'alisis de datos de muestreos recientes de espectroscop\'\i a 3D (por ej., CALIFA, MaNGA y SAMI), y la nueva generación de estos instrumentos (por ej., MUSE). A lo largo de este articulo describimos los diferentes pasos incluidos dentro del an\'alisis d elos datos, ilustrandolos mediante los productos derivados para NGC~2916, observada por CALIFA y P-MaNGA. Como un ejemplo pr\'actico del uso de este dataducto se presentan los datos completos obtenidos para 200 cubos que conforman la segunda distribuci\'on de datos CALIFA para la configuraci\'on de V500, distribuy\'endolos de forma libre a trav\'es de la red. Finalmente, exploramos la hip\'otesis seg\'un la cual las propiedades de las poblaciones estelares y el gas ionizado en las galaxias al radio efectivo son representativas del promedio a lo largo de toda la galaxia, encontrando que de hecho este es el caso.}

\addkeyword{methods: data analysis}
\addkeyword{techniques: spectroscopic}
\addkeyword{surveys}
\addkeyword{galaxies: structure}

\begin{document}
\maketitle

\section{Introduction}
\label{intro}

Integral field Spectroscopy (IFS) is steadily becoming a common user
technique after several years of being limited to a handful of
specialists across the world. In particular, IFS is nowadays widely
used in the study of the spectroscopic properties of galaxies
and their evolution along cosmological
times. This is evident in the observational pattern, that has evolved
from studies focused on limited samples or individual objects
\citep[e.g.][]{bego05,rosales11} to the study of large samples of
galaxies in the last decade \citep[e.g.][]{rosa15a}.

After the success of prototyping surveys, like SAURON \citep{bacon01},
a new set of observational programs has flourished, either at low
redshift: Atlas3D \citep{cappellari10}, Disk Mass Survey
\citep{bershady10}, CALIFA \citep{sanchez12a}, and the on-going MaNGA
\citep{manga} and SAMI \citep{sami} surveys, or at high redshifts:
e.g. SINS \citep[][]{sins}. Despite of their differences, like
the number of galaxies observed and/or the number of spaxels
sampling each galaxy, the total amount of spectra of each of these
surveys is similar, to an order of magnitude, to the total number of
spectra in the Sloan Digital Sky Survey \citep{york00}, as recently
highlighted by \citet{sanchez15X}. Moreoever, the advent of new instrumentation
able to produce even larger datasets for a single galaxy
\citep[e.g., MUSE][]{muse}, and their presumable use in survey mode,
will increase orders of magnitude beyond these current numbers and very fast
the number of IFS spectra to be analyzed. For this reason, it is necessary
to develop new tools capable of analyzing spectra of different surveys in a consistent and
automatic way. 

In order to address this problem we developed {\sc Pipe3D}. This
article is the second in a series focused on the description of this
pipeline, a spectroscopic analysis tool developed to characterize the
properties of the stellar populations and ionized gas emission lines
in the spatially resolved data of optical IFU surveys described
before. In the first article of this series, \citet{FIT3D}, hereafter
PaperI, we described in detail the basic fitting algorithms behind
{\sc Pipe3D}, included in a package named {\sc FIT3D}. In that article
we focused on the description of how the algorithms work on an
individual spectrum, on the definition of the different parameters
recovered, and on the estimation of the accuracy of the 
numerical values recovered, and the limitations of the methodology. In the
current article we focus on the description of how {\sc Pipe3D}
handles a complete datacube, indicating step-by-step the different
analysis performed to the data to generate the dataproducts. In order
to illustrate the process we use real data extracted from the
different on-going IFS surveys.

The sequence of the article is as follows: In Section \ref{data} we
describe the datasets that have been used to illustrate how the
pipeline works. In Section \ref{ana} we describe the full analysis
process, step by step, including (i) the description of the
pre-processing of the data, required to perform an homogeneous
analysis for different datasets (Section \ref{pre}); (ii) The analysis
of the central spectrum (Sec. \ref{cen}), with a detailed description of
the study of the stellar population (Sec. \ref{cen_ssp}); (iii) In
Section \ref{binning} we explain the spatial binning scheme adopted in
{\sc Pipe3D} in order to increase the S/N of the stellar continuum,
indicating the main differences with the most common used one; (iv)
The analysis of the stellar population in the different spatial-bins
and the corresponding analysis of the emission lines are described in
Sections \ref{ssp} and \ref{ssp_elines}; (v) The {\it dezonification}
procedure and how a emission line pure datacube is generated is described in
Sec. \ref{ssp_des}; (vi) The analysis of the stellar indices for the
spatially binned spectra is described in Section \ref{indices}; (vii)
In Sections \ref{strong} and \ref{weak} we describe the procedures
adopted to analyze the strong and weak emission lines spaxel-wise for
the  emission line datacube; (viii) Section \ref{pack} summarizes how the
dataproducts are packed in a set of datacubes in order to be
distributed in a simple way; (ix) A practical use of {\sc Pipe3D} is
described in Section \ref{test}, including the distribution of all the
 dataproducts derived for the V500 setup of the CALIFA DR2 galaxies
\citep{rgb15}; Finally, the summary and conclusions from this article
are included in Section \ref{sum}.

\begin{figure*}[!t]
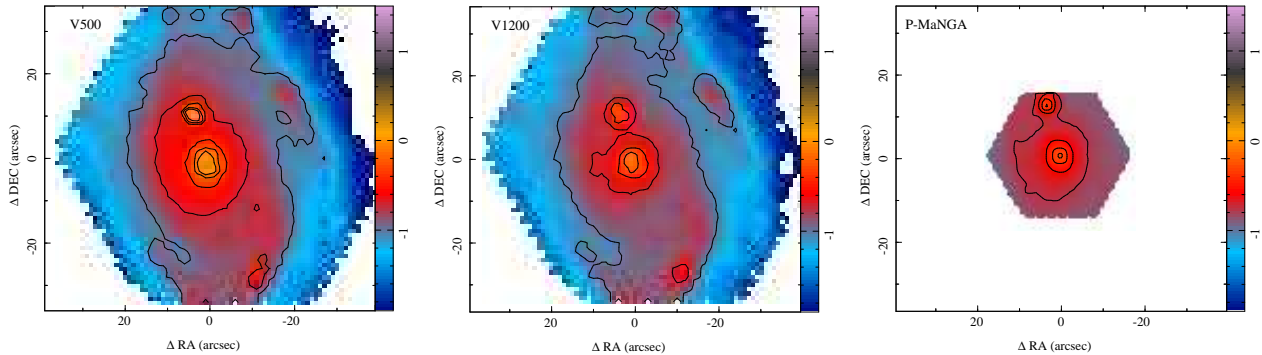

  \includegraphics[angle=270,width=0.33\linewidth]{figs/NGC2916_V500.V.ps}
  \includegraphics[angle=270,width=0.33\linewidth]{figs/NGC2916_V1200.B.ps}
  \includegraphics[angle=270,width=0.33\linewidth]{figs/NGC2916_MaNGA.V.ps}
  \caption{Broad-band image maps synthesized from the V500 ($V$-band), V1200 ($B$-band) and P-MaNGA ($V$-band) datacubes in logarithm scales of \fuDEN. The contours represent the intensity level starting at \fuden\ and with successive steps of \fuden.}\label{fig:V_map}
\end{figure*}

\section{Data}
\label{data}

Along this article we describe the different steps of the analysis
pipeline illustrating the intermediate results using the following IFU
data of the galaxy NGC~2916: (i) the datacubes provided by the CALIFA
survey \citep{sanchez12a}, in both the high and low spectral resolution modes,
and (ii) the datacubes provided by the P-MaNGA studies \citep{manga}.
This galaxy was selected since it was already used by \citet{cid-fernandes13}
and \citet{cid-fernandes14} to illustrate the use of their own analysis
pipeline, based on {\sc starlight}.

The details of the CALIFA survey, the sample, observational strategy,
and reduction are explained in \citet{sanchez12a}. All galaxies were
observed using PMAS \citep{roth05} in the PPAK configuration
\citep{kelz06}, covering an hexagonal field of view (FoV) of
74$\arcsec$$\times$64$\arcsec$, which is sufficient to map the full
optical extent of the galaxies up to two to three disk effective
radii. This is possible because of the diameter selection of the
sample \citep{walcher14}. The observing strategy guarantees complete
coverage of the FoV, with a final spatial resolution of
FWHM$\sim$2.5$\arcsec$, corresponding to $\sim$1 kpc at the average
redshift of the survey \citep{rgb15}. The sampled wavelength range and
spectroscopic resolution (3745-7500 \AA,
$\lambda/\Delta\lambda\sim$850, V500 setup) are more than sufficient
to explore the most prominent ionized gas emission lines from
[\ion{O}{ii}]$\lambda$3727 to [\ion{S}{ii}]$\lambda$6731 at the
redshift of our targets, on one hand, and to deblend and subtract the
underlying stellar population, on the other
\citep[e.g.,][]{kehrig12,cid-fernandes13,cid-fernandes14,sanchez13,sanchez14}.
In addition the objects are observed using a higher resolution setup,
covering only the blue end of the spectral range (3700-4800\AA,
$\lambda/\Delta\lambda\sim$1650, V1200 setup). The exposure time in
this second setup is three times larger than in the previous one to
ensure a similar depth of the corresponding data. The dataset was
reduced using version 1.5 of the CALIFA pipeline, whose modifications
with respect to the ones presented in \citet{sanchez12a} and
\citet{husemann13} are described in detail in \citet{rgb15}. In
summary, the data fulfill the predicted quality-control requirements
with a spectrophotometric accuracy better than a 6\% in the entire wavelength range.

The details of the MaNGA survey, its sample, observational strategy,
and reduction are explained in \citet{manga} and \citet{law15}. The MaNGA
instrument was developed under the framework of the SDSS-IV project. It
deploys 17 science integral field units (IFUs), each one
composed of an hexagonal array of fibers, across a field of view of 3
degree diameter attached to the 2.5m Sloan Telescope \citep{gunn06}. Individual science IFUs range in size from 19
fibers (12.5$\arcsec$ diameter) to 127 fibers (32.5$\arcsec$
diameter), with a diameter of 2$\arcsec$/fiber, and a 56\% effective
filling factor.  The fiber-end are coupled with the BOSS spectrographs
\citep{smee13}, that provides  a continuous wavelength coverage
from 3600 \AA\ to 10300 \AA\ at a spectral resolution R$\sim$2000
(R$\sim$1600 at 4000\AA , and R$\sim$2300 at 8500\AA), with a total
system throughput of $\sim$25\%. More details on the MaNGA setup are
given by \citet{drory15}.

The P-MaNGA, or MaNGA prototype, observations were obtained for three
galaxy fields in January 2013, as a testing phase of the instrument,
spectrograph, observing procedures, and data reduction. They comprise
a heterogeneous sample of galaxies, including four objects selected
from the CALIFA survey for photometric and astrometric calibration
purposes: IC 0944, NGC 2916, UGC 05124, UGC 06036 \citep[e.g.][]{manga,belf15,li15,wilk15}.  Like in the case of the
CALIFA survey, a three dithering scheme was adopted to obtain a
complete spatial coverage, filling the gaps between the adjacent
fibers. The raw data was reduced using a prototype of the MaNGA Data
Reduction Pipeline (DRP), which is described in detail by
Law et al. (in preparation). In essence, the data reduction comprises all the usual
steps required to extract the fiber-based spectra from the CCDs,
perform the wavelength calibration, correct for the fiber-to-fiber
transmission, subtract the sky spectrum, flux calibration and re-arrange spatially the
spectra \citep[e.g.][]{sanchez06a}. The MaNGA and P-MaNGA data have
the same spectral resolution, similar to the CALIFA-V1200 one, and
similar spatial resolution than the CALIFA data
\citep[$\sim$2.5$\arcsec$][]{manga}.

The final product of the data reduction from both surveys is a regular grid datacube,
with $x$ and $y$ coordinates that indicate the right ascension and
declination of the target, and the $z$ coordinate a common step in wavelength, 
case of CALIFA, or in logarithm of the wavelength, in the case of
P-MaNGA. For simplicity the P-MaNGA cubes were transformed to the same
format of the CALIFA ones. In both cases the pipelines also provide
the propagated error cube and a proper mask cube of bad pixels. In the
case of CALIFA they also include a prescription of how to handle the
errors when performing spatial binning (due to covariance between
adjacent pixels after image reconstruction).  Although we describe
here the analysis of this particular dataset that comprises galaxies
in common between these two surveys, {\sc Pipe3D} is capable of
analyzing data from any of the three major on-going IFU
surveys: MaNGA, CALIFA, and SAMI \citep{sami}. There are very few
galaxies in common between the three surveys, since the redshift
footprint overlap, but the sample selection are quite different. In
a companion article (S\'anchez et al., in preparation) we will provide with
the dataproducts for the early-data release of the SAMI
survey \citep{EDR_SAMI}.

\section{Analysis Sequence}
\label{ana}

{\sc Pipe3D} analyzes each individual datacube in a fully automatic
way, without using any additional external information on the object
to be analyzed (like redshift, astrometry, and so on).  Here we
describe the different individual steps taken and the 
dataproducts provided.

\subsection{Cube pre-processing}
\label{pre}

Prior to any analysis, a preprocessing of the datacubes is required in order to
(i) standardize the input format and (ii)
determine which areas within the FoV of the data are suitable for the
analysis.

\begin{figure*}[!t]
  \includegraphics[angle=270,width=1.0\linewidth]{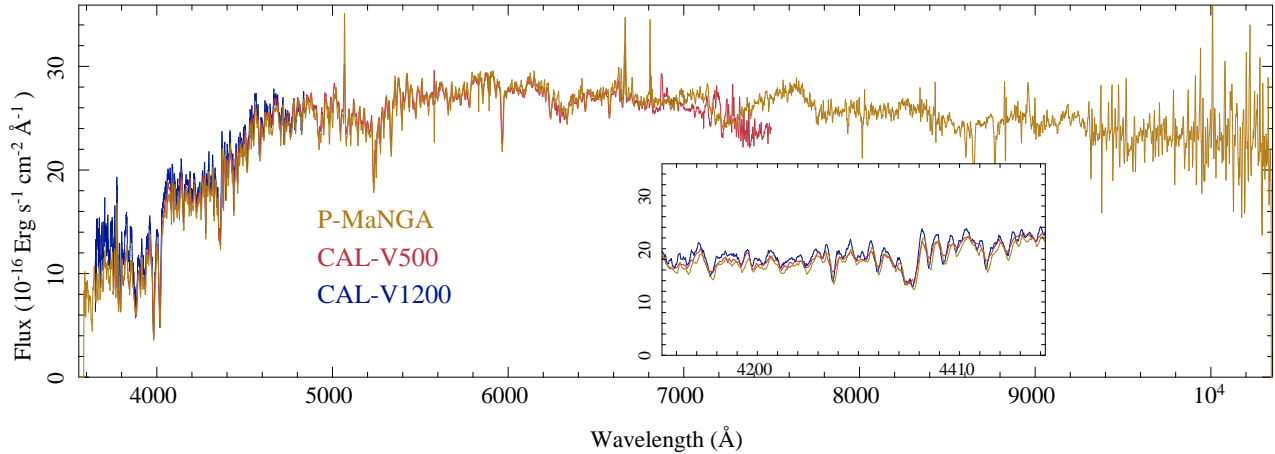}
  \caption{Central spectrum of NGC2916 for 5$\arcsec$ aperture centered at the peak emission of the galaxy extracted from the V500 (in red) and the V1200 (blue) CALIFA setups, and the P-MaNGA datacubes (orange). { The inset shows a zoom area centred in the H$\delta$ and H$\gamma$ spectral regions to highlight the similarities between the datasets.} }\label{fig:cen_spec}
\end{figure*}

Most  IFU surveys, and in particular CALIFA, MaNGA, and SAMI,
provide a FITS format file including a datacube as the final product of the
reduction.  In that cube, created using different
interpolation/image-reconstruction schemes, the X and Y coordinates
correspond to the spatial dimension (i.e., RA and DEC), and the third
coordinate corresponds to the wavelength. All of them include several
extensions in the FITS files that store, not only the physical flux
intensity at each location and wavelength, but also the propagated error
associated with those fluxes, a mask to indicate which pixels within
the cube should or should not be taken into account, and finally even
the weight of the covariance in the error propagation. However, the
actual format is different for each survey \citep[e.g., see ][for a
 few examples]{husemann13,rgb15}. {\sc Pipe3D} requires that all
the input cubes are in the same format, which corresponds to the
configuration adopted for the CALIFA datacubes, since it was originally
developed for this survey.

The input file FITS  format is described in \citet{husemann13}, and it
comprises a set of datacubes stored as extensions of the same
file. The first extension corresponds to the measured flux densities,
corrected for Galactic extinction in units of \funits, with the
wavelength solution following a linear step of a fixed spectral
sampling ($d_\lambda$). The second extension corresponds to the 1$\sigma$ noise 
level of each pixel as formally propagated by the pipeline. Those two
extensions are mandatory for {\sc Pipe3D}. In addition, if there is a third
extension it is identified as the bad-pixel mask, where the pixels not usable 
are indicated with a 1. Any further extensions will be ignored by the code.

Therefore, in the case of CALIFA data it is not required to perform
any modification of the original cubes. But in MaNGA and SAMI there
are different modifications that have to be taken into account. In the
case of MaNGA the spectral sampling should be transformed from the
logarithmic scale to a linear one (at least in the current format of
the MaNGA data). { Note that this transformation does not alter
the spectral resolution of the data, or fix it to a particular value.
We have just re-sampled the data}. In addition the order and meaning of the extensions
should be re-arranged to produce the required input file. Finally, in
the case of SAMI the blue and red datacubes correspond to two
different and discontinuous spectral ranges \citep{sami}, should be
glued in a single dataset to cover the maximum wavelength range
observed by the final setup of this survey (3720-7426 \AA). Take into
account that this final wavelength range is different than the largest
accessible one by the SAMI instrument, due to the selected setup for
the red spectra \citep{sami}. The spectra included in the red datacube
provided by the SAMI pipeline are degraded to the instrumental
resolution of the blue datacube before creating a COMBO datacube.
{ That procedure is mandatory if we want to analyze the blue
and red-arm spectra together as a single spectrum per spaxel}.
Then
the cubes are just combined by using the two datasets and
interpolating the spectra to a common linear spectral sampling
(adopting the one of the blue datacube). Obviously, the
COMBO datacubes have a blank wavelength range between $\sim$5800
and $\sim$6300 \AA. Finally, all the cubes are converted to the same
flux units, 10$^{-16}$ erg~s$^{-1}$~cm$^{-2}$~\AA$^{-1}$~spaxel$^{-1}$,
to facilitate comparison of the results. The cubes are corrected
for galactic extinction (when feasible) using the information in
the header, the Milky Way extinction law by \citet{cardelli89}, and a
Milky Way specific dust attenuation of $R_V=$3.1.

The next step selects the areas of interest within the FoV. 
Due to the nature of the IFU systems provided by the
three surveys, the useful FoV follows either a fixed hexagonal shape
(CALIFA), an hexagonal shape of different size (MaNGA), or a circular
shape (SAMI). Even more, in many cases the FoV covers foreground stars
that should be masked, either using a proper catalog of field stars or
a mask provided by the user. Finally due to the gradients in the
surface brightness of galaxies across the FoV there are areas with too
little S/N to perform any reliable analysis of the stellar continuum,
even in the case of a proper spatial binning. Those areas should be
masked for the analysis of the continuum, but not (in general) for
the analysis of the emission lines since the spatial pattern of both
components (and therefore of the S/N distribution) are in general
decoupled. In {\sc Pipe3D} we mask all the areas with a S/N$<$3 in the
wavelength range 5590-5680 \AA. This range was
selected to avoid possible contamination by strong night sky emission
lines, and at the same time not too strong contamination by emission
lines in the galaxy. This masking is needed since at low S/N the
noise is not dominated by the Poissonian errors of the intensity of
the astronomical target, but by other effects such as the sky
brightness and sky subtraction, or by the electron noise, which
requires to perform a binning of a huge area to increase the S/N to an
acceptable level. This is a problem since at large areas the co-added
spectra lack coherence in their properties, e.g., different
kinematics, different stellar populations, and different sources of gas
ionization.

\begin{figure*}[!t]
  \includegraphics[angle=270,width=1.0\linewidth]{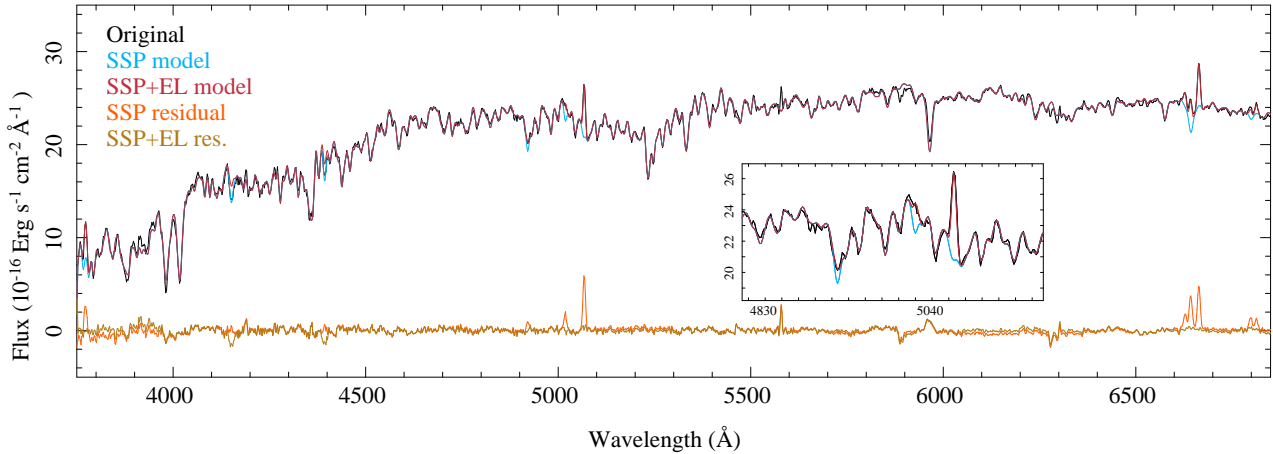}
  \caption{Results of the SSP { and emission line} fitting procedure using FIT3D for the central spectrum of NGC 2916 extracted from the V500 datacube of the CALIFA survey, shown in Fig. \ref{fig:cen_spec}. The  black line shows the original spectrum, along with the best fitted stellar population (light blue), and the best fitted combination of stellar population and emission lines (red). Finally the  emission line pure spectrum, after subtracting the best model for the stellar population, is shown as a solid orange line, and the residual of the subtraction of the best fitted model including both the stellar population and the emission line model is shown as a light green line. { The inset shows the same spectra for the wavelength range between H$\beta$ and \ion{O}{iii}, to highlight the quality of the fitting.}}\label{fig:fit_cen_spec}\end{figure*}

\subsection{Analysis of the central spectrum}
\label{cen}

Initially the pipeline extracts the central spectrum of each
datacube, defined as the 5$\arcsec$ diameter (2.5$\arcsec$ radius)
aperture spectrum in the case of CALIFA (P-MaNGA), centred at the peak
intensity in a broad-band image of the corresponding object.  The
broad-band image in the observed frame is synthesized by
convolving the filter response curve through the datacube. For the
CALIFA V500 and the P-MaNGA we use the $V$ band filter, while
for the CALIFA V1200 we use the $B$ band. Figure
\ref{fig:V_map} shows a comparison between the three broad-band
images, illustrating the similarities in terms of spatial resolution
between the three different datasets. The absolute flux intensities
differ within the expectations for the CALIFA and P-MaNGA datasets.
We must recall here that the current estimations of spectrophotometric
accuracies for CALIFA are of the order of $\sim 3-4$ \% \citep{rgb15}, 
while for P-MaNGA they are of the order of $\sim 15$\% \citep{belf15}.
The lower photometric accuracy of the P-MaNGA
observations arises because the prototype MaNGA hardware was designed
to explore a variety of alternative flux calibration methods in order
to determine the optimal approach for the main survey.  In contrast,
the full MaNGA survey-mode data reach spectrophotometric accuracies
$\sim 3$\% (Yan et al., submitted). The P-MaNGA dataset were originally 
reduced using a preliminary
version of the pipeline, and therefore there are some inaccuracies
associated with the reduction that are expected to be larger than those
of the current version of the MaNGA datacubes \citep{law15}.

An example of the central spectra extracted from each datacube is
shown in Figure \ref{fig:cen_spec}. For each one of them we applied
the stellar population and emission line fitting procedures described
in PaperI { (Sec. 2)}. First, each spectrum was fitted using a very simple
template including two SSPs plus a spectrum of an emission line
source, for the non-linear analysis, with main aim of estimating the
systemic velocity of the galaxy, its central velocity dispersion and
the dust attenuation. In this first analysis a wide range of
non-linear parameters is explored.  A priory, the explored range of
systemic velocities covers the full redshift range of the
survey considered. The velocity dispersion covers the range between 0
and 400 km/s, including most of the known central velocity dispersion values
for galaxies. Finally, the dust attenuation for the stellar population
covers a range between $A_{\rm V}$=0 to 1.6 mag.  This latter parameter
is derived from the range of dust attenuation values observed in most 
galaxies \citep[e.g.][]{char00,calz01}. The number of SSPs in this template is
limited for the shake of the speed, due to the
strong dependence of the computational time with the number of SSPs in
the template and the range of  parameters explored.

If there is a hint of the expected non-linear parameters, like a
published systemic velocity and velocity dispersion from previous
analysis (e.g., from SDSS spectroscopy), or the expected dust
attenuation, the pipeline can restrict the range of 
parameters explored and speed up the process.

After the non-linear parameters are derived, each spectrum is fitted,
in the linear phase, with a limited stellar population library that
includes 12 SSPs, as described below. This provides with a simple but
robust estimation of the properties of the stellar populations and the
shape of the underlying continuum \citep[e.g.][]{sanchez13}.

All the SSP templates used so far, were extracted from the MILES
project \citep{miles, vazdekis10, falc11}. We selected this template
on the basis of the results of PaperI, were we demonstrate that is is
optimal for the analysis of the stellar population { based on
  simulations (Sec. 3.1 and 3.2 of that paper)}. The main reason is
that it is based on one of the best spectrophotometrically calibrated
library of stellar spectra. { The template library adopted for the 
estimation of the non-linear parameters of the central spectrum of the galaxies comprises two
extreme stellar populations}: (i) a young ($\sim$90 Myr) and low metallicity
($Z/Z_{\odot}=0.2$) stellar population, and (ii) an old ($\sim$17.8
Gyr) and high metallicity ($Z/Z_{\odot}=1.5$) one. In addition it
includes an empirical spectrum characteristic of an emission line
nebula, corresponding to the integrated spectrum across a FoV of
$5\arcmin\times6\arcmin$ of the Orion Nebula \citep{sanchez07c}. The
choice of the spectra included in this template was the result of
different experiments, guesses and errors, along the past five years
of analyzing the CALIFA data, and nearly two years of analyzing MaNGA
and SAMI data, in order to recover the non-linear parameters in a
consistent way with the values reported for the central SDSS spectra
\citep[e.g.][]{marmol-queralto11,sanchez12a}.

{ The template library adopted for the estimation of the properties
  of the stellar populations of the central spectrum} comprises a grid
of SSPs including four stellar ages (0.09, 0.45, 1.00, and 17.78 Gyr),
and three metallicities (0.0004, 0.019, and 0.03), subsolar, solar,
and supersolar. This template library, {\tt miles12} hereafter, is the
same used in many previous CALIFA studies, for instance,
\citet{sanchez12a,sanchez13,sanchez14} and \citet{jkbb15b}.  { Note
  that we use a very simple template library in this case since the
  main goal of the analysis of the central spectrum is to derive the
  systemic velocity and the central velocity dispersion
  properties. The results of the analysis of the stellar population is not
  used anymore by the pipeline, and the template is adopted just to
  speed-up the computing process.}


\subsubsection{Detailed analysis of the stellar population}
\label{cen_ssp}

After a first guess has been derived of the systemic velocity 
and the central velocity dispersion, on the basis of the
analysis described above, the procedure is repeated restricting the
exploration of the kinematic parameters within a range of $\pm$300 km/s
around the estimated systemic velocity, and $\pm$50\% around the
estimated velocity dispersion. The dust attenuation is explored in the
same range of values. For this second iteration we select a template
with 3 SSPs for the non-linear exploration (i.e., the derivation of
the velocity, velocity dispersion, and dust attenuation, as described
in PaperI, { Sec. 2.1}), including the two extreme ones described above and an
intermediate population with an age of $\sim$1 Gyr and metallicity
$Z/Z_{\odot}=0.4$.  For the linear exploration (i.e., the detailed
analysis of the stellar population by a multi-SSP decomposition), 
a more complex stellar library was considered, defined as {\tt gsd156}
in PaperI  { ( Sec. 3.1, of that paper)}. This library is described in detail in
\citet{cid-fernandes13}. It comprises 156 templates that cover 39
stellar ages (1 Myr to 13 Gyr), and 4 metallicities ($Z/Z_{\odot}=$
0.2, 0.4, 1, and 1.5). These templates were extracted from a
combination of the synthetic stellar spectra from the GRANADA
\citep{martins05} and the SSP libraries provided by the MILES
project. This SSP template has been extensively used within the CALIFA
collaboration in different studies \citep[e.g.][]{eperez13,
  cid-fernandes13, rosa14}. The only difference with respect to these
studies is that the spectral resolution of the library was not fixed
to the spectral resolution of the CALIFA V500 setup data (FWHM$\sim$6
\AA), to allow its use for datasets with different resolution (like
the ones provided by MaNGA and the CALIFA V1200 setup). This SSP-library
uses the \citet{Salpeter:1955p3438} Initial Mass Function (IMF). Although 
the current implementation of the pipeline uses this particular SSP library,
{\sc Pipe3D} is not restricted to this particular one, since it
can be exchanged by modifying a configuration parameter in the main script.

As described in PaperI, { Sec. 2}, {\sc FIT3D} allows to fit the stellar continuum
and the emission lines adopting an iterative procedure. In the case of
{\sc Pipe3D} we fit the strongest emission lines in the 
optical wavelength range, fitting together the following emission
lines: (i) [\ion{O}{ii}]$\lambda$3727; (ii) H$\delta$; (iii)
H$\gamma$; (iv) H$\beta$, [\ion{O}{iii}]$\lambda$4959, and
[\ion{O}{iii}]$\lambda$5007; (v) [\ion{N}{ii}]$\lambda$6548, H$\alpha$,
[\ion{N}{ii}]$\lambda$6583, [\ion{S}{ii}]$\lambda$6717, and
[\ion{S}{ii}]$\lambda$6731. In this way, we define a set of wavelength ranges
including the indicated set of emission lines, and they are all fitted
together, assuming that they have similar kinematic properties. In
addition we fix certain line intensity ratios, such as the relative
strength of the [\ion{O}{iii}] and [\ion{N}{iii}] doublets.

The result of this analysis is illustrated by Figure
\ref{fig:fit_cen_spec}, where the best model including the stellar
population and the  emission lines is shown, along  with
the residuals from the different analysis, for the central spectrum of
NGC 2916 extracted from the V500 datacube of the CALIFA survey.
{ In this figure it is possible to appreciate the quality of the
fitting of both the stellar populations and the emission lines, that
has been extensively quantified in Paper I, Sec. 3 and 4. In particular,
it is possible to appreciate that we can recover the emission line
fluxes even in the case of severe absorptions (e.g., in the case of H$\beta$). }


\begin{figure*}[!t]
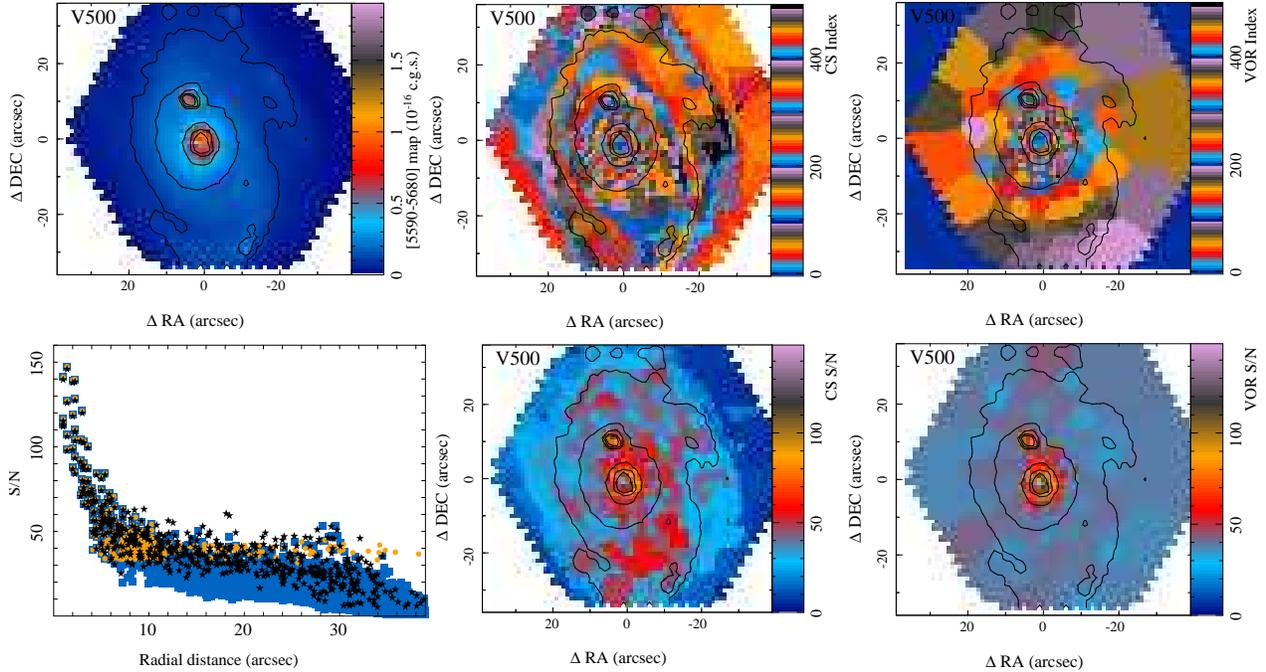

\includegraphics[angle=270,clip,trim=25 10 10 10,width=0.33\linewidth]{figs/signal_V500.ps}\includegraphics[angle=270,clip,trim=5 10 10 10,width=0.33\linewidth]{figs/cont_seg_SN.ps}\includegraphics[angle=270,clip,trim=5 10 10 10,width=0.33\linewidth]{figs/cont_seg_VOR.ps}
\includegraphics[angle=270,clip,trim=5 10 10 0,width=0.33\linewidth]{figs/rad_SN.ps}\includegraphics[angle=270,clip,trim=5 0 10 10,width=0.33\linewidth]{figs/SN_cont_seg_SN.ps}\includegraphics[angle=270,clip,trim=5 10 10 10,width=0.33\linewidth]{figs/SN_cont_seg_VOR.ps}
  \caption{{\it top-left panel:} Narrow-band intensity map derived by summing the fluxes within the wavelength range  5590--5680 \AA\ for the
CALIFA V500-datacube of NGC~2916; {\it top-central panel:} Segmentation map derived for the same datacube using a continuum plus S/N binning scheme, as outlined in the text; {\it top-right panel:} Segmentation map derived for the same datacube using the most frequently used S/N voronoi binning scheme; {\it bottom-left panel:} Radial distribution of the signal-to-noise for the original datacube (blue squares), the segmented cube based on Voronoi binning (orange stars), and the continuum plus S/N segmented cube (black circles), for the same datacube; {\it bottom-central panel:} S/N map for each of the spatial bins created using a continuum plus S/N binning scheme (the one on the top-central panel), for the same datacube; {\it bottom-right panel:} S/N map for each of the spatial-bins created using a S/N voronoi binning scheme (the one on the top-right panel), for the same datacube. In all the maps the contours are the same as the ones presented in Fig. \ref{fig:V_map}, left-panel.}\label{fig:bin}
\end{figure*}

\subsection{Spatial binning}
\label{binning}

The central spectra described in the previous section have, in
general, a S/N well above 50 for most of the galaxies included in the
IFU surveys of our interest
\citep[e.g.][]{sanchez12a,manga}. Therefore, they are above the S/N
threshold for which the simulations from PaperI  { (Sec. 3 and Table 1)} suggest that
the properties of the stellar populations are well recovered (i.e.,
within an error of $\sim$0.1 dex). However, as the surface-brightness of
the galaxies declines as a function of the galactocentric distance, the
S/N decreases rapidly in the outer regions \citep[e.g., Fig. 13,][]{sanchez12a}, 
and therefore the results from any analysis
of the stellar continuum become unreliable, as already noticed by
several authors \citep[e.g.][]{capp03,cid-fernandes13,cid-fernandes14}.


In order to overcome this problem  a binning scheme is frequently adopted
to aggregate spaxels in the outer regions in order to
increase the signal to noise. This is a mathematical problem that goes
beyond the field of integral field spectroscopy, although it is
broadly addressed in this field. A set of solutions have been proposed
on the basis of different assumptions and goals, in
addition to the main basic one, i.e., to increase the S/N preserving
as much as possible the spectroscopic properties of the data. 

One of the most simple methods was proposed by \citet{sam84}, the
so called Quadtree algorithm.  This method consists of a recursive
partition of the FoV into axis-aligned squares. The initial square
corresponds to the entire FoV. Then the FoV is divided in four areas
of equal size. Subsequently each of the sub-squares are equally
divided. If a certain goal S/N, required as  input to the algorithm,
is not achieved in the next iteration, then the procedure stops for a
particular square. On the other hand, if it is achieved, the procedure
continues until the original pixel (spaxel) size is reached. This
algorithm is extensively explored in \citet{capp03}. The two main
problems with this procedure is that (1) it depends on the actual
orientation of the FoV with respect to the original geometry of the
galaxies, (2) for the dataset discussed here, with an intrinsic
non-square (or rectangular shape), the method should be adapted, and
(3) it does not preserve the shape of the original astronomical object.

An alternative method is the isophotal segmentation, first introduced
by \citet{papa02}, and implemented for IFU data in
\citet{papa13}, and Gomes et al. (submitted).  The algorithm segmentates
the FoV on the basis of a set of isophotes, based on the
surface brightness distribution. Then each isophotal area is divided
in subsequent bins by aggregating adjacent pixels (spaxels) along the
azimuthal angle in order to achieve a goal S/N. Therefore, the area of
the final spatial bins grows with  galactocentric
distance (as the surface brightness decreases). The main problem with
this approach is that the resulting segmentation/binning depends
strongly on some arbitrary parameters, like the number and range in
surface brightness of original isophotes, and the original pixel
(spaxel) selected to start the aggregation in each isophote,
irrespectively of the goal S/N.

The most broadly used binning scheme in IFS data is the 
Voronoi binning procedure \citep{capp03}. This algorithm was developed to
satisfy three requirements, in addition to the main goal of all these
algorithms indicated above: (i) the bins should properly tessellate
the FoV (i.e., there should not be holes or overlapping areas), (ii)
the bin shape has to be as compact or round as possible, and (iii) the
scatter of the S/N after the binning should be as small as possible.
Under this basic assumption the authors developed an algorithm in
which, on the basis of a set of points within the FoV (called generators),
a tessellation based on the Voronoi algorithm is generated. This
guarantees that all pixels (spaxels) in a certain spatial bin are the nearest
ones to the point that has generated the considered bin. The
generators are selected on the basis of a ranking order S/N of the
pixels and a distance criterion \citep{capp03}. 

By construction this algorithm guarantees a very homogeneous
distribution of the S/N, which has made it very popular within the
community. However, it does not preserve the original shape of the
astronomical object, in particular for galaxies with sharp
structures. Even more, since the aggregation is based mostly on a S/N
criteria, it may include spaxels corresponding to areas of the galaxy
with very different physical properties (like spiral arms and
inter-arm regions). This issue was never a concern when the algorithm
was created since it was developed under the umbrella of the SAURON
project \citep{bacon01}, whose main (initial) goal was to explore the
central regions of a sample of early type galaxies, and mostly focused
on the study of their kinematical properties. It is expected that the
light distribution of an early type galaxy follows a smooth shape,
and the kinematics does not present abrupt changes. Therefore,
imposing further criteria to force the spatial bins to follow the
shape of the light (like the isophotal), was not needed. For similar
reasons it was broadly adopted in the analysis of the Atlas3D data
\citep{cappellari10ar}, and subsequently used in hundreds of studies.

An additional issue regarding the Voronoi binning algorithm is that it
assumes that the S/N follows the light distribution. In general, this
is the case for dataset acquired with IFUs that cover the complete
FoV, like the case of the lensarray systems of SAURON
\citep{bacon01}. In those cases, when the noise budget is dominated by
the intrinsic Poissonian noise due to light coming from the
astronomical target, the S/N is a function of the surface brightness. 
In the case of SAURON and Altas3D data this was the
case for most of the targets, since the FoV of the instrument rarely
covers more than $\sim$1.5 effective radius. Thus the noise produced
by the sky subtraction and other electronic effects of the detectors
are negligible.

\begin{figure*}[!t]
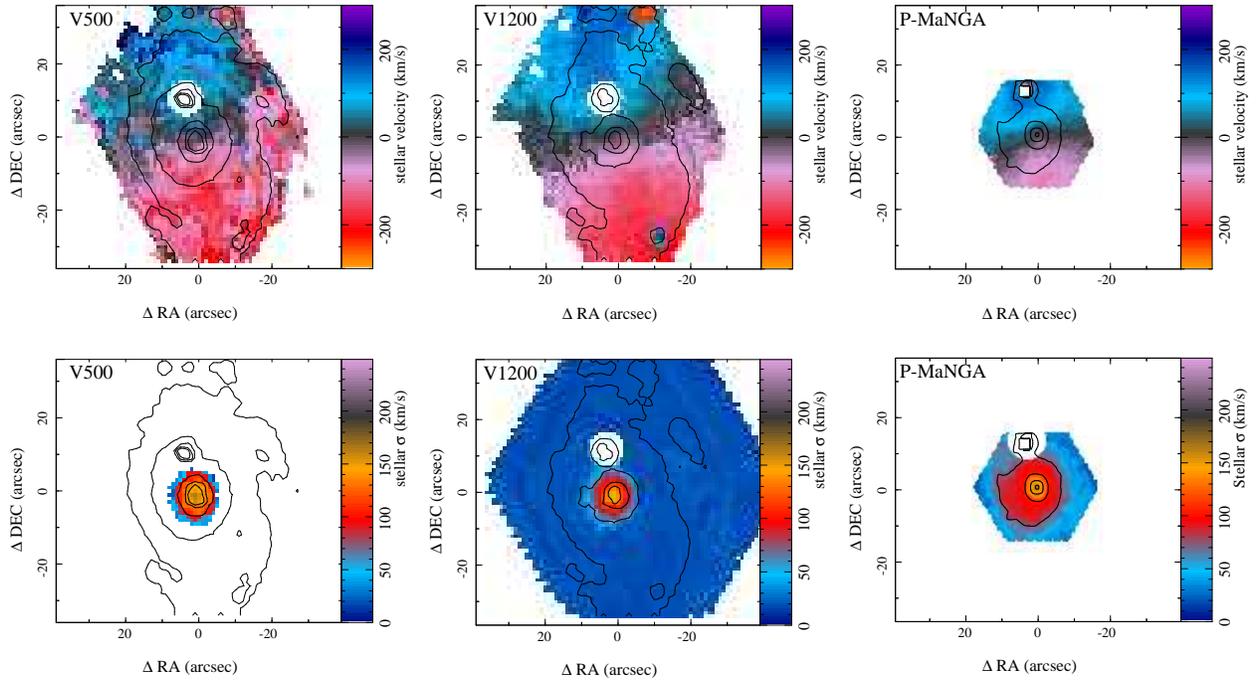

  \includegraphics[angle=270,width=0.33\linewidth]{figs/vel_V500.NGC2916.ps}\includegraphics[angle=270,width=0.33\linewidth]{figs/vel_V1200.NGC2916.ps}\includegraphics[angle=270,width=0.33\linewidth]{figs/vel_MaNGA.NGC2916.ps}
  \includegraphics[angle=270,width=0.33\linewidth]{figs/disp_V500.NGC2916.ps}\includegraphics[angle=270,width=0.33\linewidth]{figs/disp_V1200.NGC2916.ps}\includegraphics[angle=270,width=0.33\linewidth]{figs/disp_MaNGA.NGC2916.ps}
  \caption{Stellar velocity ({\it top panels}) and velocity dispersion maps ({\it bottom panels}) derived using the three  datasets for NGC2916: {\it left} CALIFA V500 setup; {\it central} CALIFA V1200 setup; {\it right}: P-MaNGA dataset. For the velocity dispersion the values below the instrumental velocity dispersion have been masked.}\label{fig:ssp_kin}
\end{figure*}

However, most of the current ongoing IFU surveys adopt a different
IFU technology (fiber bundle with an incomplete coverage of the FoV),
and the targets are sampled up to 2.5 effective radii and beyond
\citep[e.g.][]{walcher14}. As a consequence, these basic assumptions
do not hold. First, due to the use of fiber bundles current IFU
surveys adopt a dithering scheme in order to cover the complete FoV.
In most of the cases this approach creates intrinsic inhomogeneous
distribution of the S/N, even for exposures of totally flat
targets. In the case of the three pointing dithering pattern the
spaxels can be covered by one, two, or even three fibers. Therefore,
there could be a factor $\sqrt{3}$ in the S/N ratio of adjacent
spaxels of the same intensity. The Voronoi binning, that takes into
account only the S/N as the basic metric will aggregate spaxels from
different physical regions to compensate for that inhomogeneity.

The published version of the Voronoi binning does not take into
account the covariance between adjacent spaxels that is inherent
to the image reconstruction schemes required to obtain a datacube from
a dithering observation using a fiber bundle. It is known that when
co-adding $N$ adjacent spectra, the noise does decrease following a
$\sqrt(N)$ law. On the contrary the decrease in the error is
shallower, due to the covariance between adjacent spaxels. This was
nicely described in \citet{husemann13}, where a
functional form was proposed for the correction of the noise propagation when
taking into account the covariance. The Voronoi binning algorithm
could be corrected for this effect in a simple way.

In {\sc Pipe3D} we depart from the widely used Voronoi binning scheme
and we propose a different algorithm, based on both a continuity
criteria in the surface brightness and a goal in the signal-to-noise
ratio (Continuum plus S/N binning, CS-binning hereafter). Like the
Voronoi binning, CS-binning requires as input a signal-map, a noise-map, 
and a S/N goal. In addition it requires the fraction of flux that a
given spaxel differs from an adjacent one in order to be aggregated. In
principle, the algorithm looks initially for all the spaxels/pixel for
which the S/N is already above the minimum S/N required. Those ones are
selected as spatial bins with a single pixel. Then, for the remaining
pixels the algorithm looks for the one with the higher
intensity. This will be the seed of the next spatial bin. It derives
the S/N at this location and estimates the maximum number of adjacent
pixels required to increase that S/N to the target S/N, assuming
that adjacent pixels have similar S/N levels. It assumes
Poissonian statistics plus the effect of the covariance, and solves
$N$ (the number of adjacent pixels to co-add) from the equation:
\begin{equation}\label{eq:sn}
S/N_{goal} = S/N_{input} \sqrt{N}~ covar(N), 
\end{equation}
\noindent { where $S/N_{input}$ is the estimated signal-to-noise if the
noise distribution was Poissonian (i.e., no covariance between
adjacent spaxels), $N$ is the number of adjacent spaxels included in 
a particular spatial bin, and $covar(N)$ is the correction introduced
by the correlation of the noise between adjacent spaxels. This last
parameter is derived statistically in an empirical way as
described in \citet{husemann13} and more recently in \citet{rgb15},
by creating spatial bins of arbitary size coadding $N$ adjacent spaxels
computing $S/N_{input}$ and measuring the real $S/N$ from the coadded
spectra. Then a functional form for the dependence of $covar(N)$ with 
the number of coadded spaxels is derived as shown in Fig. 11  of \citet{rgb15}.} 

Then, it uses $N$ to estimate the radius of the circular aperture required
to be integrated to enclose this number of spaxels/pixels: 

\begin{equation}\label{eq:rmax}
 R_{max} = \sqrt{N/\pi} 
\end{equation}

Finally, it aggregates all the adjacent pixels within a maximum
distance of $R_{max}$ and for which the flux intensity is within the
predefined fraction to the initial {\it seed}.  In general this
creates spatial bins that are not round, since they tend to follow the
shape of the isophotes across the FoV. Due to the second criterion, in
general the $ S/N_{goal}$ is not reached for most of the
spatial bins. This segmentation/binning scheme is somehow a mix
between the isophotal and the Voronoi binning schemes.

Figure \ref{fig:bin} shows a comparison between the adopted CS-binning
and the Vorononi binning schemes for the V500 setup data extracted
from the CALIFA dataset of the NGC 2916. { The signal and noise
  maps adopted for both procedures were created by deriving the median
  and standard deviation of the flux intensity in each spaxel for the
  spectral pixels within the wavelength range between 5590-5680
  \AA. In the case of the Voronoi binning it is used only the $S/N$
  map. For the CS-binning the signal map is used for the continuity
  criterion. For both procedures the results depend a lot on the
  adopted wavelength regime to perform the spatial binning.} The
Voronoi binning was modified to take into account the spatial
co-variance between the data. It shows the distribution of spatial
bins when a S/N goal of 40 is selected for the Voronoi binning, and
S/N goal of 50 and a fractional flux variation of 20\% between
adjacent pixels is accepted for the CS-binning. The values were
selected to reach a S/N$>$30 in most of the FoV and to have a similar
number of spatial bins when using both algorithms to allow for a fair
comparison (391 in the case of the Voronoi and 439 in the case of the
CS-binning).

As expected, both algorithms create similar single pixel spatial bins for
those pixels already fulfilling the S/N criterium. Then, for pixels
below the S/N goal the Voronoi binning, less restrictive, creates
larger spatial bins, in particular in the outer regions of the galaxy.
We include in the Figure the spatial distribution of S/N after applying the
spatial binning, for both algorithms. By construction the distribution
is very homogeneous in the case of the Voronoi binning ($\langle
S/N\rangle=38.5\pm 4.7$), and present a clear structure with a larger
dispersion in the case of the CS-binning ($\langle S/N\rangle=30.7\pm
16.3$). The bottom-left panel of Figure \ref{fig:bin} shows the radial 
distribution of S/N for the original dataset
and for the two binning schemes. Up to $\sim$10$\arcsec$ the three
distributions are very similar (the regions where no binning is
needed). At larger galactocentric distances the distribution for the
Voronoi binning becomes almost flat, as expected from the results
presented by \citet{capp03}. In contrast, the CS-binning provides 
a S/N $\sim$40, between 10$\arcsec$ and 30$\arcsec$, covering a
wide range of S/N values (between $\sim$30 and $\sim$60). The average S/N
in this regime is very similar (but with twice the scatter) to the one provided by the Voronoi binning.

Beyond this distance, the CS-binning gives little
improvement in S/N with respect to the original data. However, at
those galactocentric distances the original data have S/N$<$3 in
most cases, and we regard those areas useless 
for the analysis of the underlying stellar population. If we try to
reach a S/N above $\sim$30 by co-adding individual spaxels with a S/N
below 3 the area required to be covered by the spatial bin would be so large
that the spectra would lose the coherence in their basic properties, as
indicated above. Therefore, although it may be mathematically correct
the interpretation of the physical properties derived will be always a problem.

\begin{figure}[!t]
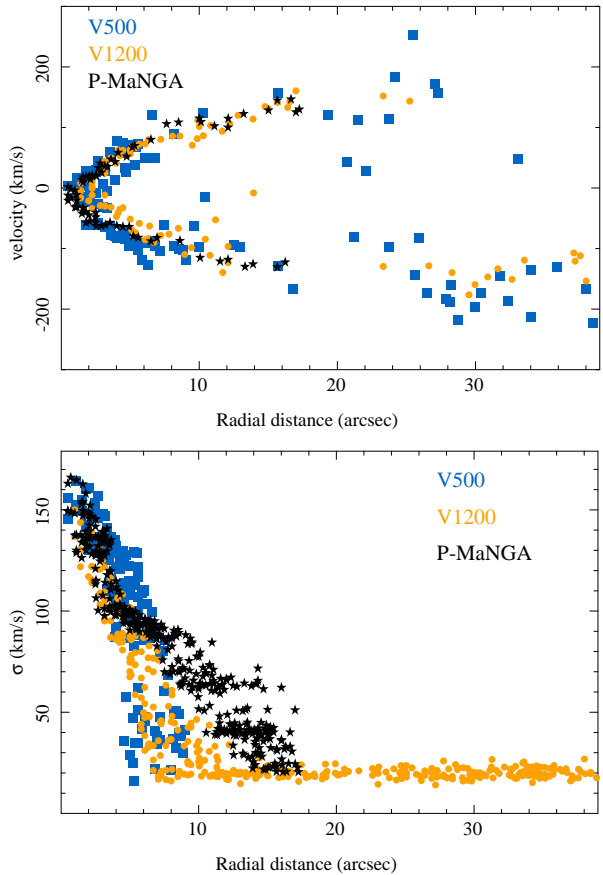

\includegraphics[angle=270,width=1\linewidth]{figs/rad_vel.ps}
\includegraphics[angle=270,width=1\linewidth]{figs/rad_disp.ps}
  \caption{Stellar velocity along a pseudo-slit located at the center of the galaxy and tilted 60$\degree$ (top-panel), and radial distribution of the velocity dispersion (bottom panel) extracted from the kinematic maps of the galaxy NGC2916 shown in Figure \ref{fig:ssp_kin} for the three  datasets: CALIFA V500 setup (blue squares), CALIFA V1200 setup (orange  circles), and P-MaNGA (black  stars).}\label{fig:rad_ssp_kin}
\end{figure}

This example does not demonstrate the superiority of any of these
methods, as this was never the intention. If the goal is to normalize
the S/N across the FoV of the data, definitely, Voronoi binning is (so
far) the best algorithm. However, for increasnig the S/N
preserving the shape of the original target, the CS-binning presents
significant advantages. For the current implementation of {\sc Pipe3D}
we adopted a S/N goal of 50 and a more restrictive { upper limit to the range of
relative fluxes between adjacent spaxels to be coadded, setting it to a value of 10\%}, 
prioritizing to keep as much as possible the
original shape of the data rather than the final S/N of the
spectra. By increasing the fractional flux variation one can achieve
S/N closer to the goal, with the corresponding lose of spatial
information. { If no limit is imposed to the fractional flux variation,} the
binning provided by both methods are very similar.

The procedure provides  a S/N map before and after  binning,
and a segmentation map in which each pixel corresponding to the same
spatial bin is labeled with the running index that identifies the spatial
bin. All those maps are stored as FITS format files.

\subsection{Analysis of the stellar population}
\label{ssp}

As described above, the original cube is spatially binned using the
CS-binning algorithm. The spectra corresponding to the spaxels within
each spatial bin are averaged and stored as a single spectrum, together with
the average spatial coordinates. { Thus, for each bin we obtain a
spectrum that corresponds to the mean of each individual spectra of all
the spaxels within that spatial bin, masking spectral pixels with bad values.} At the end of this process, 
a row stacked spectra (RSS)  is created and a position table for each
binned cube, following the order of the spatial bin indices (from the
brightest to the faintest areas in the cube, by construction). In
addition it provides an intensity map at the wavelength range
corresponding to the $V$-band before and after performing the binning.
The ratio between both maps is the relative contribution of each pixel
to the average intensity within the spatial bin where it is aggregated. This
ratio will be used later in the dezonification process, that will be explained below \citep{cid-fernandes13}.

\begin{figure*}[!t]
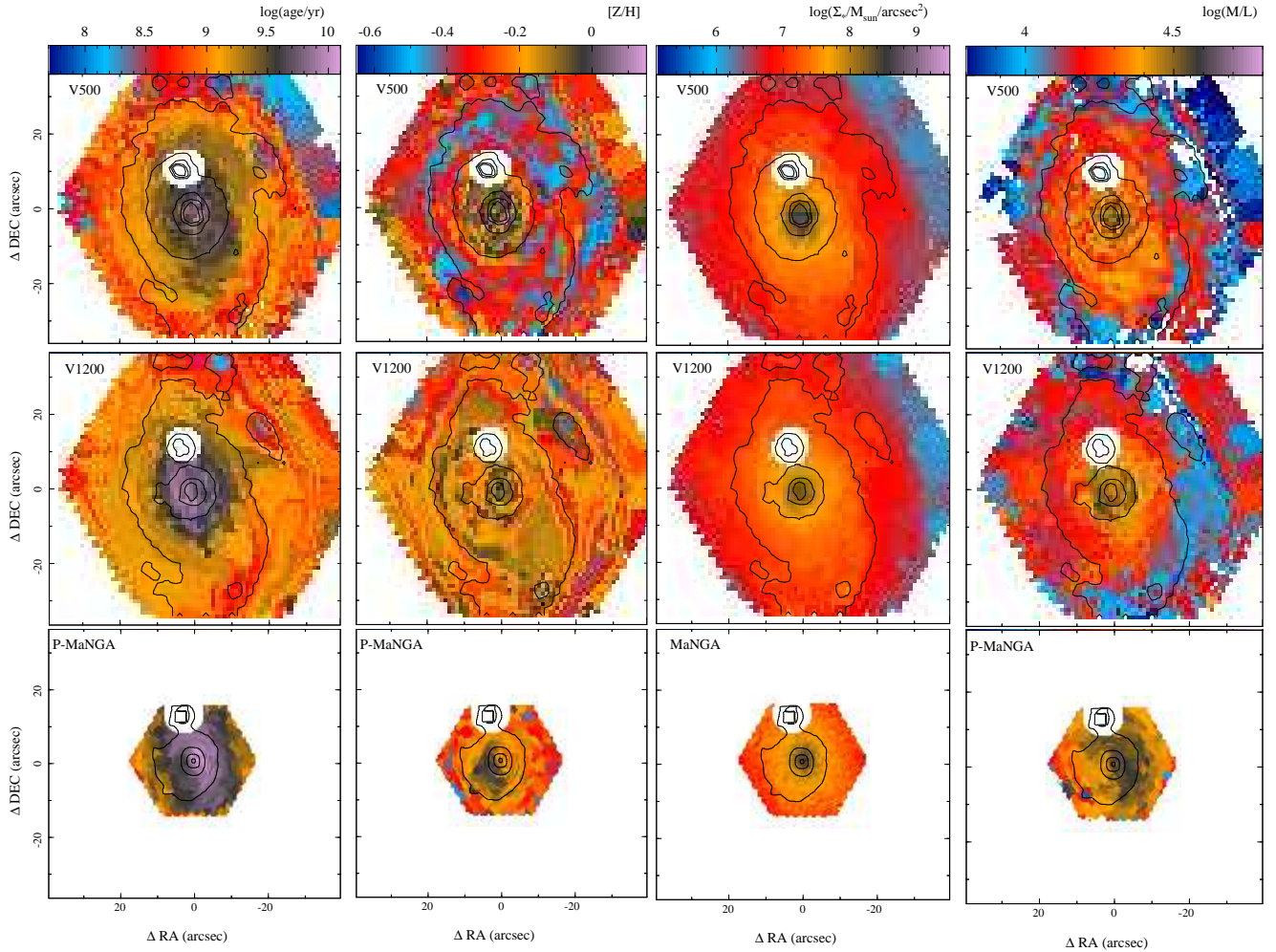

\includegraphics[angle=270,width=0.29\linewidth,clip,trim=0 0 92 0]{figs/Age_LM_V500.NGC2916.ps}\includegraphics[angle=270,width=0.247\linewidth,clip,trim=0 71 92 0]{figs/ZH_LM_V500.NGC2916.ps}\includegraphics[angle=270,width=0.255\linewidth,clip,trim=0 71 92 0]{figs/Mass_V500.ps}\includegraphics[angle=270,width=0.255\linewidth,clip,trim=0 71 92 0]{figs/ML_V500.ps}
\includegraphics[angle=270,width=0.29\linewidth,clip,trim=115 0 92 0]{figs/Age_LM_V1200.NGC2916.ps}\includegraphics[angle=270,width=0.247\linewidth,clip,trim=115 71 92 0]{figs/ZH_LM_V1200.NGC2916.ps}\includegraphics[angle=270,width=0.255\linewidth,clip,trim=115 71 92 0]{figs/Mass_V1200.ps}\includegraphics[angle=270,width=0.255\linewidth,clip,trim=115 71 92 0]{figs/ML_V1200.ps}
\includegraphics[angle=270,width=0.29\linewidth,clip,trim=117 0 0 0]{figs/Age_LM_MaNGA.NGC2916.ps}\includegraphics[angle=270,width=0.247\linewidth,clip,trim=117 71 0 0]{figs/ZH_LM_MaNGA.NGC2916.ps}\includegraphics[angle=270,width=0.255\linewidth,clip,trim=117 71 0 0]{figs/Mass_MaNGA.ps}\includegraphics[angle=270,width=0.255\linewidth,clip,trim=115 71 0 0]{figs/ML_MaNGA.ps}  
  \caption{From left to right, distribution of the luminosity weighted age and metallicity, mass surface density, and mass-to-light ratio of the stellar population in NGC 2916 across the FoV for the three datasets: {\it top panels:} CALIFA V500 setup; {\it central panels:} CALIFA V1200 setup; {\it bottom-right}: P-MaNGA dataset. Contours correspond to the same intensity level of the broad band images presented in Fig. \ref{fig:V_map}. The white masked region north-east of the center of the galaxy corresponds to a foreground star, visible in Fig. \ref{fig:V_map}.}\label{fig:age}
\end{figure*}

Each spectrum within the RSS file is analyzed following the same
procedures applied to the central spectrum, as described in Section
\ref{cen_ssp}.  The goals of this analysis are the following: (i) to obtain
the best representation of the underlying stellar population to
subtract it from the original data and provide a spectrum of the
emission lines (emission line pure spectrum); (ii) to characterize the main
properties of the underlying stellar population, as described in
PaperI {, Sec. 2.3}. 

Following the procedures discussed, the stellar continuum is first
fitted with a simple template of SSPs in order to derive the systemic
velocity, velocity dispersion, and dust attenuation ({\tt miles12}).
Then the main properties of the strong emission lines are derived by
fitting the residual spectrum (after the underlying stellar
population is subtracted) with a set of Gaussian functions. This first model of the
emission lines are subtracted from the original spectrum to remove the
effects of the strongest emission lines. Finally, this spectrum is
fitted with the {\tt gsd156} template library, defined in
Sec. \ref{cen_ssp}, to derive the main properties of the stellar
populations (age, metallicity, star-formation history, etc).  As
described in PaperI,  { Sec. 2.2}, the procedure may be iterated until it fulfills
a certain convergence criterion (i.e., that the $\chi^2$ decreases
less than a certain percent). In this particular implementation we
iterated just 2 times, to speed up the process, and due to the limited
improvement in terms of the $\chi^2$ between sucessive iterations.

The main differences with respect to the procedure described in
Sec. \ref{cen_ssp} and PaperI{ , Sec. 2 ,} are:

\begin{itemize}

\item The velocity dispersion ($\sigma$) for the first spectrum, that corresponds
  to the peak intensity of the galaxy and therefore the central
  region, is explored within a wide range of values up
  to 400 km/s (in addition to the instrumental dispersion that is
  first applied to convolve the SSP template). Then, for successive spectra,
  corresponding to spatial bins of lower flux intensity, the exploration of
  the velocity dispersion is restricted to a range between 0.5 and 1.5 the value
  of the previous iteration, i.e., 0.5$\sigma_i<\sigma_{i+1}<$1.5$\sigma_i$, where
$i$ is the index of the spatial bin.

  This procedure ensures that the velocity dispersion is kept within reasonable
  values for areas of lower S/N (lower intensity, i.e., in the outer
  part of the galaxies). It is known that at lower S/N all fitting procedures
  tend to increase the velocity dispersion to fit the average distribution
  of values that is dominated by the noise.

\item In the case of MaNGA data, the procedure is repeated twice,
  using a different template in the first { step}, due to the
  widest wavelength range covered by MaNGA. { First, we adopted a }
  template adopted extracted from the {\tt MIUSCAT} SSP library
  \citep{miuscat}. This library is an extension of {\tt MILES},
  covering the wavelength range 3465-9469 \AA, with a similar spectral
  resolution and spectrophotometric quality. We adopted a grid of {\tt
    MIUSCAT} SSPs including four stellar ages (0.06, 0.20, 2.00, and
  17.78 Gyr), and three metallicities (0.0004, 0.02, and 0.0331),
  subsolar, solar, or supersolar. For this particular library we
  include ages slightly younger than the ones included in {\tt
    miles12}, since we have seen that they tend to reproduce slightly
  better the blue end of the MaNGA spectra { (not covered by CALIFA
    and SAMI)}. The results of this first analysis are used only to
  characterize the underlying stellar population in the wider possible
  wavelength range, and provide the best emission line spectrum { (i.e., the orange 
spectrum in Fig. \ref{fig:fit_cen_spec}), that would provide with a GAS-pure cube over
almost the complete wavelength range covered by MaNGA. They are also used to derive the non-linear parameters of 
the stellar populations (velocity, velocity dispersion and dust attenuation) }

  In the second { step} the same parameters are used, wavelength
  ranges, stellar templates, and initial guess values for the three
  surveys (CALIFA, MaNGA, and SAMI), in order to homogenize the results
  as much as possible. { However, to speed-up the processes, in the
 case of MaNGA we do not repeat the derivation of the non-linear parameters
 of the stellar populations, using the result from the first step described before.}

\item For MUSE data \citep[e.g.][]{censusHII}  the same
  stellar templates and guess parameters were adopted as for MaNGA,
  but restricting the wavelength range to that MUSE.
  Since for low-z objects MUSE does not cover the 4000\AA\ break, we
  are still not sure about the accuracy of the parameters derived for
  the stellar populations, that should be compared with ad hoc simulations,
  similar to the ones shown in PaperI {, Sec. 3.2}.

\end{itemize}

The analysis of the stellar populations performed using {\sc FIT3D}
on the RSS file provides three different
dataproducts, two {\tt csv} files, and a FITS format cube: 

\begin{enumerate}

\item The first of the two {\tt csv} files, named {\tt
  auto\_ssp.CS.OBJECT.rss.out}, is an {\tt ascii} table. Each row
  comprises the main properties of the stellar population derived by
  the fitting procedure for each individual spectrum within the
  RSS file (and therefore each spatial bin within the binned
  cube). The parameters distributed in each column include the reduced
  $\chi^2$ of the fit, the luminosity and mass weighted log-ages
  and log-metallicities of the stellar populations, as defined in
  PaperI {, Sec. 2.3}. In addition it contains the dust attenuation, the systemic
  velocity, and velocity dispersion, with their corresponding
  errors. It also includes the average intensity and standard
  deviation of the residuals from the fitting procedure, and the
  average mass-to-light ratio within the spatial bin.

\item The second {\tt csv} file, named {\tt
  coeffs\_auto\_ssp.CS.OBJECT.rss.out}, is a table with one row for
  each SSP in the library and for each 
  spectrum in the RSS file (i.e., number of spatial bins).
  The columns include a running index corresponding to the 
  SSP,  age, metallicity, and mass-to-light ratio of this population,
  along with the fraction of light that it contributes to the
  original spectrum at the normalization wavelength,  with its
  estimated error. { This information is used to derive the
  luminosity and mass weighted parameters included in the first file.}

\item Finally a FITS format cube, named {\tt
  output.auto\_ssp.CS.OBJECT.rss.out.fits.gz} stores the original
  spectra, the best model spectra, emission line pure spectra, the residuals
  from the fit of the emission lines (as indicated below), and the spectra
  after subtracting the best model for the emission lines. In this FITS format cube
  each slice along the Z-axis comprises the results from the fitting procedure
  for each spectrum in the RSS file.

\end{enumerate}

\begin{figure*}[!t]
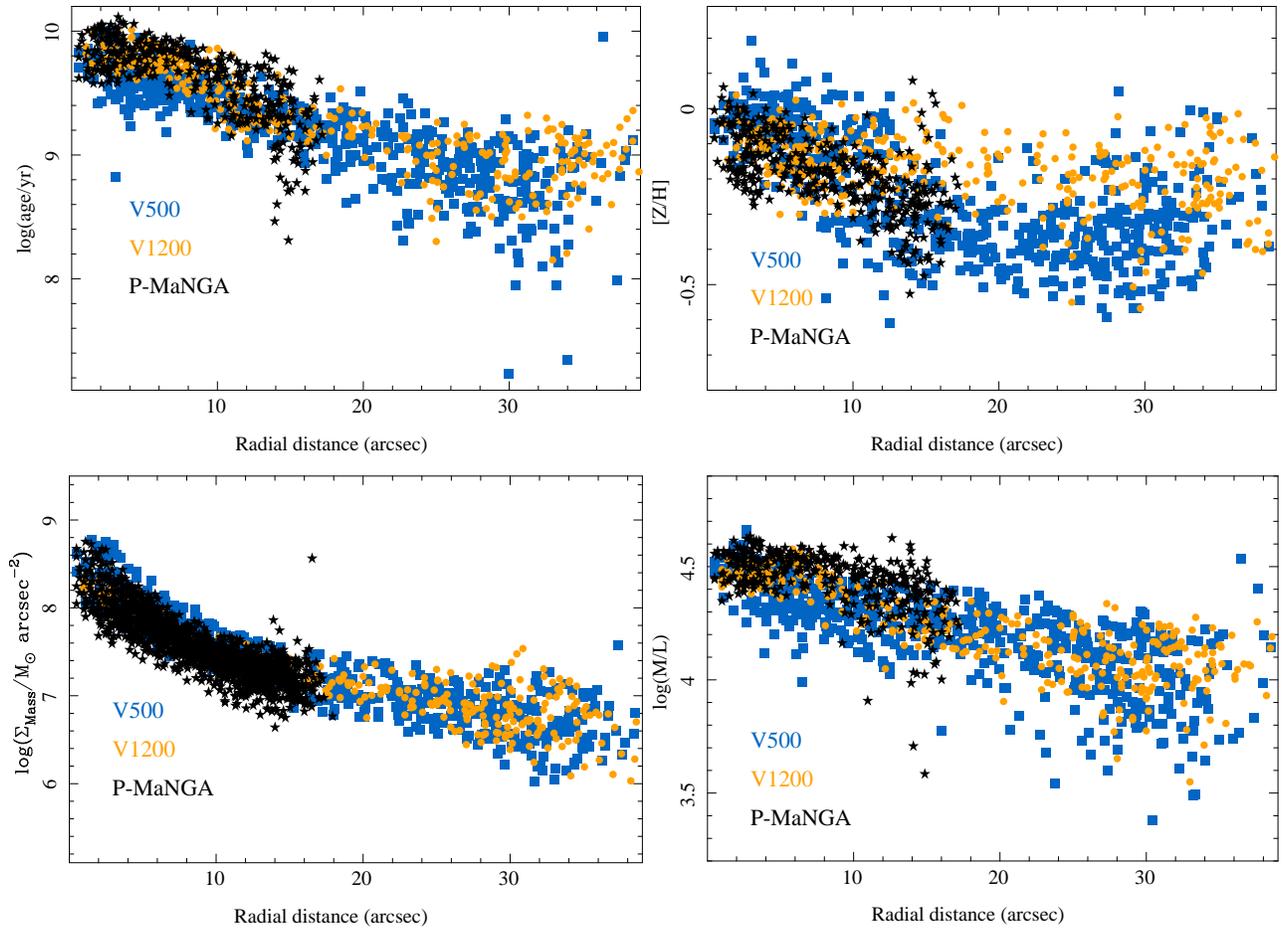

\includegraphics[angle=270,width=0.5\linewidth]{figs/rad_Age.ps}\includegraphics[angle=270,width=0.5\linewidth]{figs/rad_ZH.ps}
\includegraphics[angle=270,width=0.5\linewidth]{figs/rad_Mass.ps}\includegraphics[angle=270,width=0.5\linewidth]{figs/rad_ML.ps}
  \caption{Radial distribution of the luminosity weighted age (top-left panel), metallicity (top-right panel), mass surface density (bottom-left panel), and mass-to-light ratio (bottom-right panel) of NGC 2916 also shown in Figure \ref{fig:age} for the three datasets: CALIFA V500 setup (blue squares), CALIFA V1200 setup (orange circles), and P-MaNGA (black stars).}\label{fig:rad_age}
\end{figure*}

The derived dataproducts included in the two  {\tt csv} files
are rearranged into a set of maps (one for each 
dataproduct), following the original spatial shape of the datacubes,
by associating each value to the location in the 2D space
defined by the segmentation file, as described in Sec. \ref{binning}.
This format is convenient to store and share the data, to 
compare  different dataproducts, and for plotting
purposes. The maps of these dataproducts are stored in separate
FITS format files, named {\tt map.CS.OBJ\_PARAM\_ssp.fits.gz}, where {\tt
  OBJ} is the object name (as it appears in the name of the datacube)
and {\tt PARAM} is a label indicating each  of the derived
dataproducts.  For example, the FITS file
{\tt map.CS.NGC2916\_age\_ssp.fits.gz}  stores the luminosity
weighted age derived for the CALIFA V500 datacube of NGC 2916.
All the files generated by {\sc Pipe3D} for the V500 datacube
of NGC 2916 described in this section can be found in the 
FTP\footnote{\url{ftp://ftp.caha.es/CALIFA/dataproducts/DR2/Pipe3D_NGC2916}}.
In Section \ref{pack} we provide the correspondence of each FITS file 
with the measured parameter, for the distributed dataproducts.

\subsubsection{Stellar Kinematics}
\label{ssp_kin}

Figures \ref{fig:ssp_kin} and \ref{fig:rad_ssp_kin} illustrate the
results of the kinematics analysis for the stellar populations.  The
figures show the estimated  velocities agree within $\pm$30
km/s across the entire FoV of the three datasets. 
For the velocity dispersion the three datasets agree just
within the central $\sim$8$\arcsec$, as
expected from the simulations presented in PaperI { (e.g., Table 1 of that paper)}.  The agreement
between the V500 and V1200 datasets of the CALIFA survey reaches
$\sim$10$\arcsec$. Beyond that galactocentric distance, the velocity
dispersion derived for the V1200 data collapse to the minimum selected
value of 20 km/s fixed in the presented version of {\sc Pipe3D}, at
the limit of what is feasible at the resolution of the data. 
This indicates that we should re-analyze all the datasets again
allowing the exploration of lower velocity dispersion values. For the V500
dataset we have applied an overall quadratic offset of 120 km/s to
match the velocity dispersion; this indicates that offset 
between the SSP resolution and the instrumental resolution should be revised
{ and that in the current analysis we have a miss-match of the $\sim$30\%
in the assumed instrumental resolution for the V500 data. That offset
does not affect the derivation of the properties of the stellar populations,
since the final velocity profiles are well constrained, being affected only
the derivation of the velocity dispersions.} 
The offset was derived
from the comparison of the peak velocity dispersion, at the center of
the galaxies, obtained by the pipeline for the $\sim$500 objects in
common between the two CALIFA setups. After this correction the
velocity dispersion for the V500 dataset presents a cut at
$\sim$10$\arcsec$, a location at
which the values derived are dominated by the instrumental resolution.
For the P-MaNGA dataset the velocity dispersion measured beyond
9$\arcsec$ presents a large dispersion, with an offset respect to the
values derived using both the V1200 and V500 CALIFA datasets.
We still do not know the origin of this discrepancy {,  although most
probably it comes from the fact that the P-MaNGA data were taking
on an experimental phase of this project, using different fibers
and packing that may alter the nominal spectral resolution.}

\subsubsection{Composition of the stellar population}
\label{ssp_pop}

Figures \ref{fig:age} and \ref{fig:rad_age} illustrate the results
from the analysis of the properties of the stellar populations.  Both
figures show the 2D and radial distributions of the luminosity
weighted log-age and log-metallicity, the surface mass density, and
the Mass-to-Light ratio for the three different datasets.  The
log-ages agree within a range of $\pm$0.2 dex, for the central regions
($<$15$\arcsec$) of the three datasets, in agreement with the
expectations from the simulations presented in PaperI { (e.g.,
  Table 1 of that paper)}.  At larger radii, we are not able to
compare with the P-MaNGA (due to the smaller FoV of this dataset),
however, the analysis performed over the V1200 data seem to derive a
slightly higher log-ages ($\sim$0.2 dex) than the one derived from the
V500 one. { In average it is found an inaccuracy/offset between the
  derived log-ages using the CALIFA V500 and both the V1200 and
  P-MaNGA dataset of $\sim$0.1 dex, with a dispersion of $\sigma\sim$0.1
  dex. Between the V1200 and the P-MaNGA data there is a very good
agreement, with an offset of 0.03 dex and a dispersion of $\sigma=$ 0.06 dex. }

For the stellar metallicity we found an agreement within a range of
$\pm$0.1 dex for the three datasets in the shared FoV, although in the
inner regions the values derived for the V1200 are slighly lower. This
is consistent with the higher values derived for the ages and the
well-known age-metallicity degeneracy. For larger radii the derivation
based on V1200 data seem to present a slightly larger log-metallicity
($\sim$0.1 dex). { Indeed, the agreement between the results
  derived using the CALIFA V500 dataset and the P-MaNGA ones are
  remarkable good, with an offset of -0.02 dex and a dispersion of
  0.06 dex.} Taking into account the limited wavelength range of the
CALIFA V1200 data compared to the other two datasets (e.g.,
Fig. \ref{fig:cen_spec}), { a range that does not cover those
  features more sensitive to the variation of metallicities, like the
  \ion{Fe}{} and \ion{Mg}{} absorption features between 5100-5400\AA,
} this result is somehow expected.  This range does not cover the
stronger spectral features sensitive to the analyzed parameters, and
the wavelength range is too short to be sensitive to the dust
attenuation.

The surface mass density shows very good agreement, within the
range of the dispersion of each individual dataset, for the regions
covered by the three datasets. { The agreement is better between
the two CALIFA datasets than between them and the P-MaNGA data, with 
an offset of -0.01 dex and a dispersion of 0.12 dex, in the first case, compared with 
an offset of $\sim -$0.1 dex and a dispersion of 0.1 dex, in the second case.
In both cases the dispersion is consistent with the limit in the accuracy
of the mass estimation found by different authors using this methodology \citep[e.g]{rosa14}. The offset is most probably due to the different spectrophotometric calibration adopted in each survey \citep{rgb15,renbin16}, explaining while the two CALIFA datasets present a better agreement.}

Finally, the Mass-to-Light ratio presents a similar distribution for
the three datasets at the different galactocentric distances, although
there seems to be a systematic offset between the three
estimations. { Like in the case of the stellar mass density, the offset is larger
  for the CALIFA datasets with respect to the P-MaNGA ones
  (0.07-0.10$\pm$0.04-0.06 dex), than among the former two
  (0.03$\pm$0.05 dex).} 

All these Figures
confirm the consistency of the results for the stellar population
analysis obtained by {\sc PIPE3D} based on different datasets { with
differences within the range of the expected based on simulations (PaperI, Table 1), and previous results \citep[e.g.][]{cid-fernandes13,rosa14}}.

\subsubsection{Emission lines in the binned data}
\label{ssp_elines}

As  explained in PaperI  { (Sec. 2.4)} and briefly described in
Sec. \ref{ssp}, {\sc FIT3D} fits the emission lines in a
quasi-simultaneous way with the stellar populations, adopting an
iterative scheme. According to this, the residuals from the analysis of the
stellar population are fitted with a set of Gaussian functions to
characterize the properties of the emission lines, and the best model
of the emission lines is subtracted from the original spectra to perform
the analysis of the stellar population in a second iteration.

This iterative scheme was adopted for the analysis of the RSS file 
provided by the CS-binning. In the current implementation of {\sc
  Pipe3D} we included in the analysis loop the fitting to a set of
strong emission lines frequently observed in the optical range of
galaxies: [\ion{O}{ii}]$\lambda$3727, H$\delta$, H$\gamma$, H$\beta$,
[\ion{O}{iii}]$\lambda$4959, [\ion{O}{iii}]$\lambda$5007,
[\ion{N}{ii}]$\lambda$6548, [\ion{N}{ii}]$\lambda$6583, H$\alpha$,
[\ion{S}{ii}]$\lambda$6717, and [\ion{S}{ii}]$\lambda$6731. Each of
these emission lines were fitted with a single Gaussian profile for
the emission line pure spectrum at each spatial bin derived from 
the analysis of the stellar population.

\begin{figure}[!t]
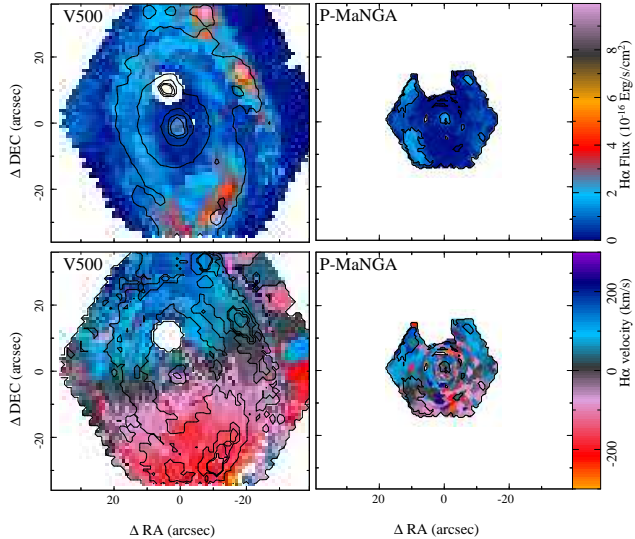

  \includegraphics[angle=270,clip,trim=9 0 100 125,width=0.51\linewidth]{figs/Ha_CS_V500.ps}\includegraphics[angle=270,clip,trim=9 70 100 0,width=0.56\linewidth]{figs/Ha_CS_MaNGA.ps}
  \includegraphics[angle=270,clip,trim=15 0 10 125,width=0.51\linewidth]{figs/gas_vel_CS_V500.NGC2916.ps}\includegraphics[angle=270,clip,trim=15 70 10 0,width=0.56\linewidth]{figs/gas_vel_CS_MaNGA.NGC2916.ps}
  \caption{H$\alpha$ intensity and velocity maps (top and bottom panels respectively) derived using the CS-binned RSS files derived from the CALIFA V500 (left panels) and the MaNGA (right panel) datasets of NGC 2916. In the left-hand panels contours correspond to the same intensity level of the broad band images presented in Fig. \ref{fig:V_map}. In the right-hand panels contours correspond to the H$\alpha$ intensity maps shown in the left-panels, starting at 0.05~\FunitsA\ with a constant step of 1~\FunitsA.}\label{fig:Ha_CS}
\end{figure}

The final product of this fitting procedure is an {\sc ascii} table
named {\tt elines\_auto\_ssp.CS.OBJ.rss.out} that comprises, for each
spectrum in the CS-file and for each emission line, a set of
columns including: (1) the nominal wavelength of the emission
line, (2) its integrated flux, (3) the $\sigma$ (dispersion in \AA) of the
Gaussian fitted, and (4) the systemic velocity with the
corresponding uncertainties estimated by {\sc FIT3D}. As in the case
of the stellar population, all dataproducts are
rearranged into a set of maps, following the original spatial shape
of the datacubes, by associating the given value to the location
in the 2D space, defined by the segmentation file described in
Sec. \ref{binning}. In a similar way to the analysis of the stellar
populations, the parameters derived for each emission line are stored
in separate FITS format files, named {\tt
  map.CS.OBJ\_PARAM\_WAVELENGTH.fits.gz}, where {\tt OBJ} is the
object name (as it appears in the name of the datacube), {\tt PARAM}
is a label that identifies each of the  dataproducts, and {\tt
  WAVELENGTH} is the nominal wavelength of the  emission
line.  For example,  the FITS file {\tt
  map.CS.NGC2916\_flux\_6562.fits.gz}  stores the flux density of
H$\alpha$ derived from the CS-binned RSS file extracted from the
CALIFA V500 datacube of NGC 2916. As indicated above, all the files
generated by {\sc Pipe3D} for the V500 datacube of NGC 2916 can be
found in the FTP indicated above.

\begin{figure}[!t]
\includegraphics[angle=270,clip,trim=9 0 100 123,width=0.51\linewidth]{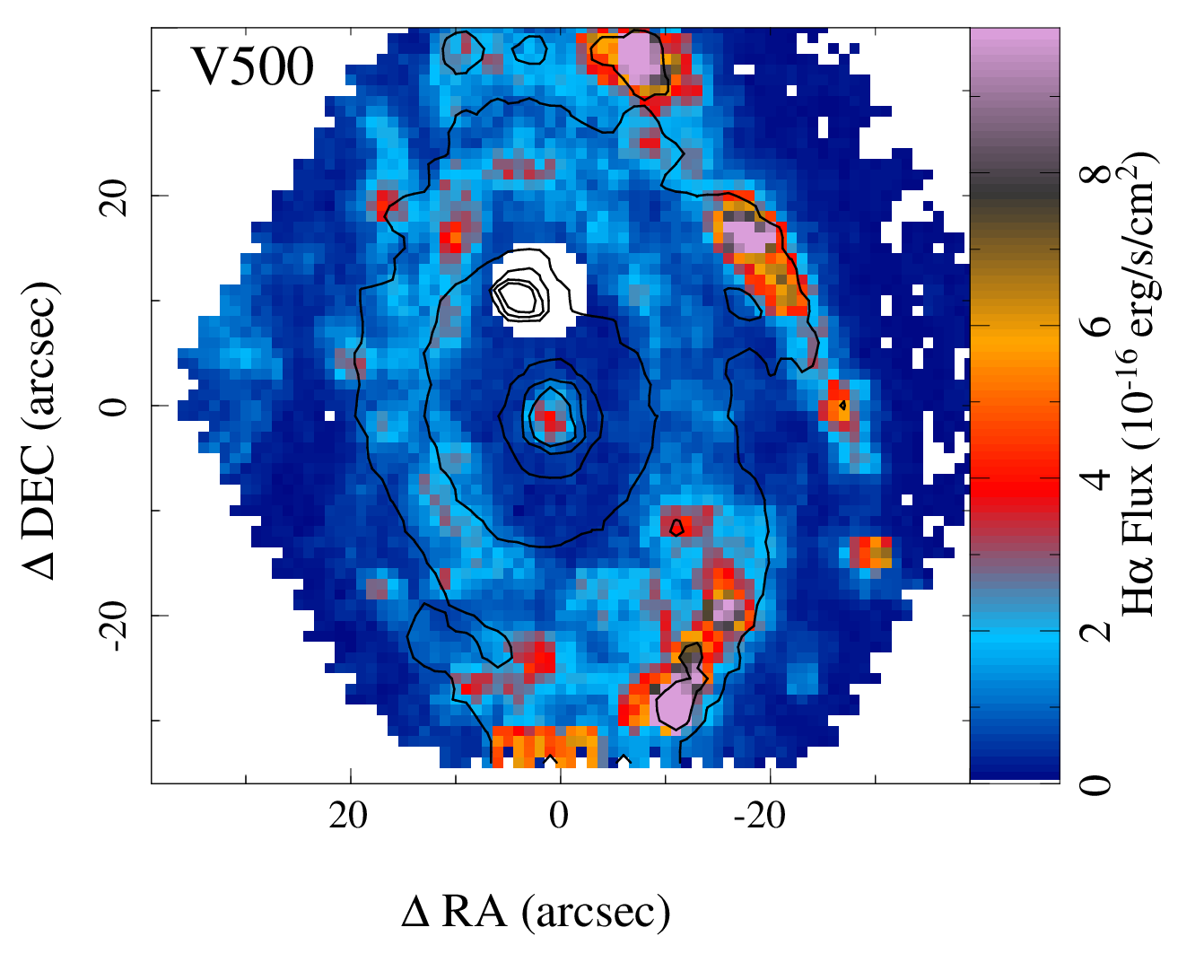}\includegraphics[angle=270,clip,trim=9 70 100 0,width=0.56\linewidth]{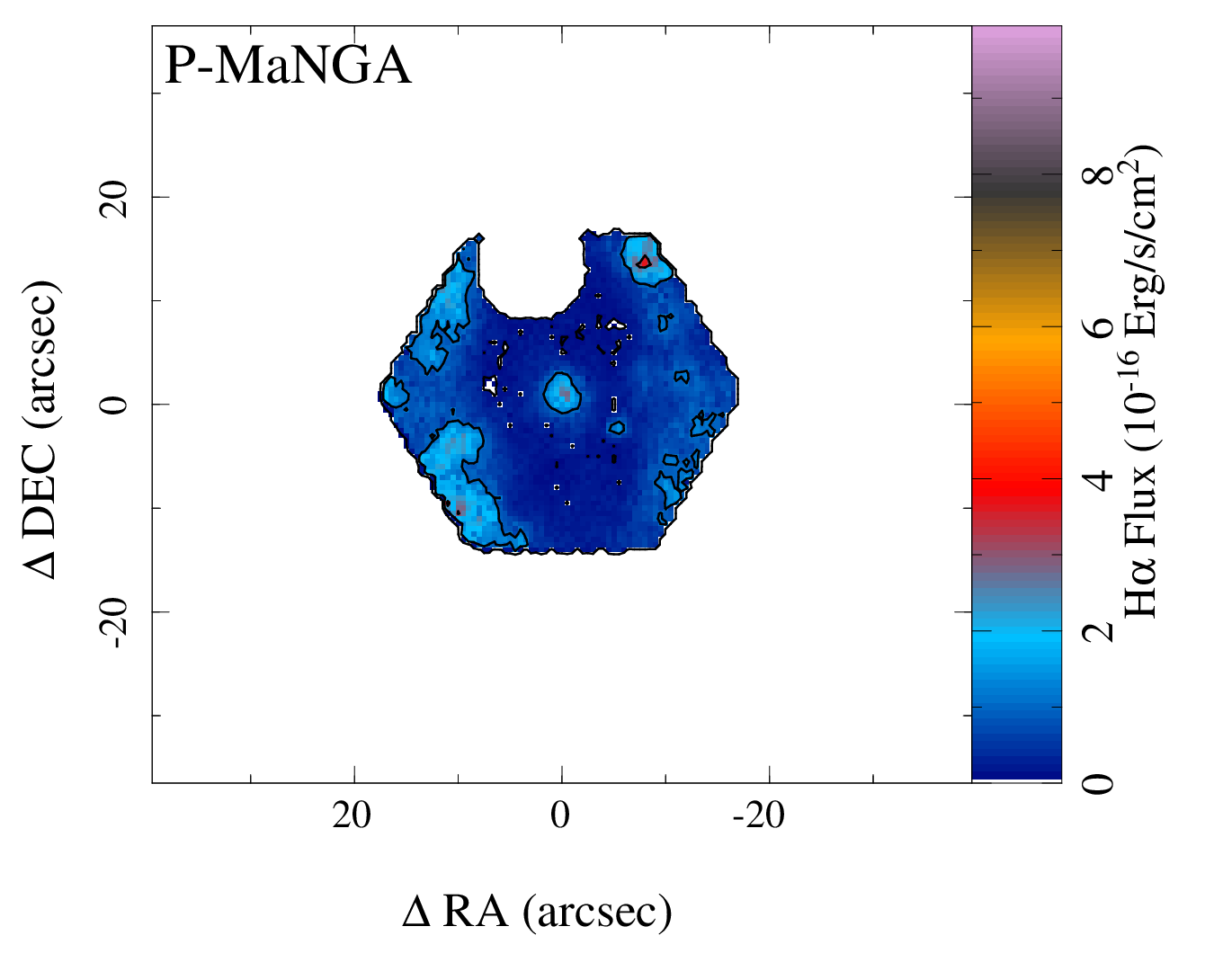}
  \includegraphics[angle=270,clip,trim=15 0 10 123,width=0.51\linewidth]{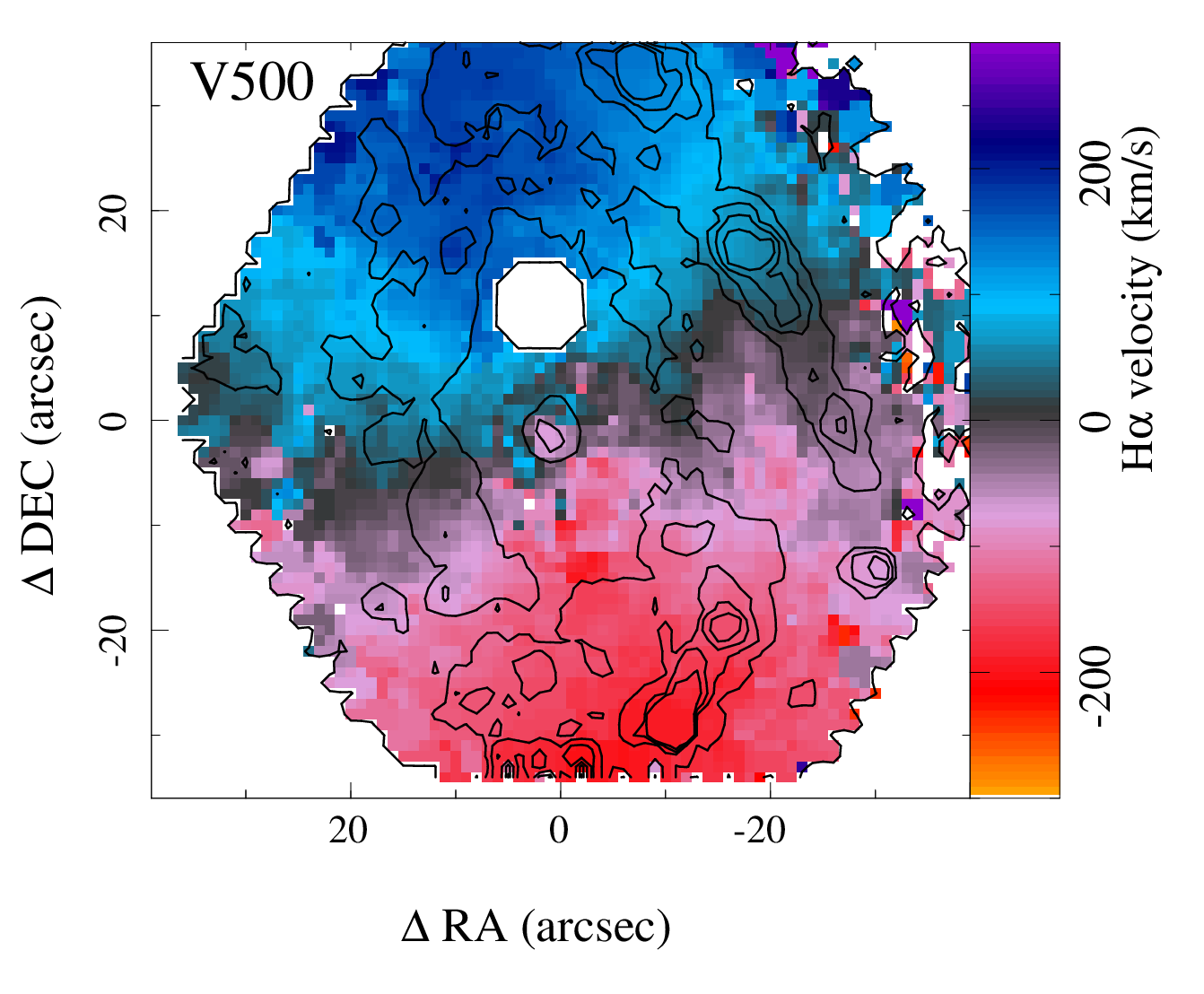}\includegraphics[angle=270,clip,trim=15 70 10 0,width=0.56\linewidth]{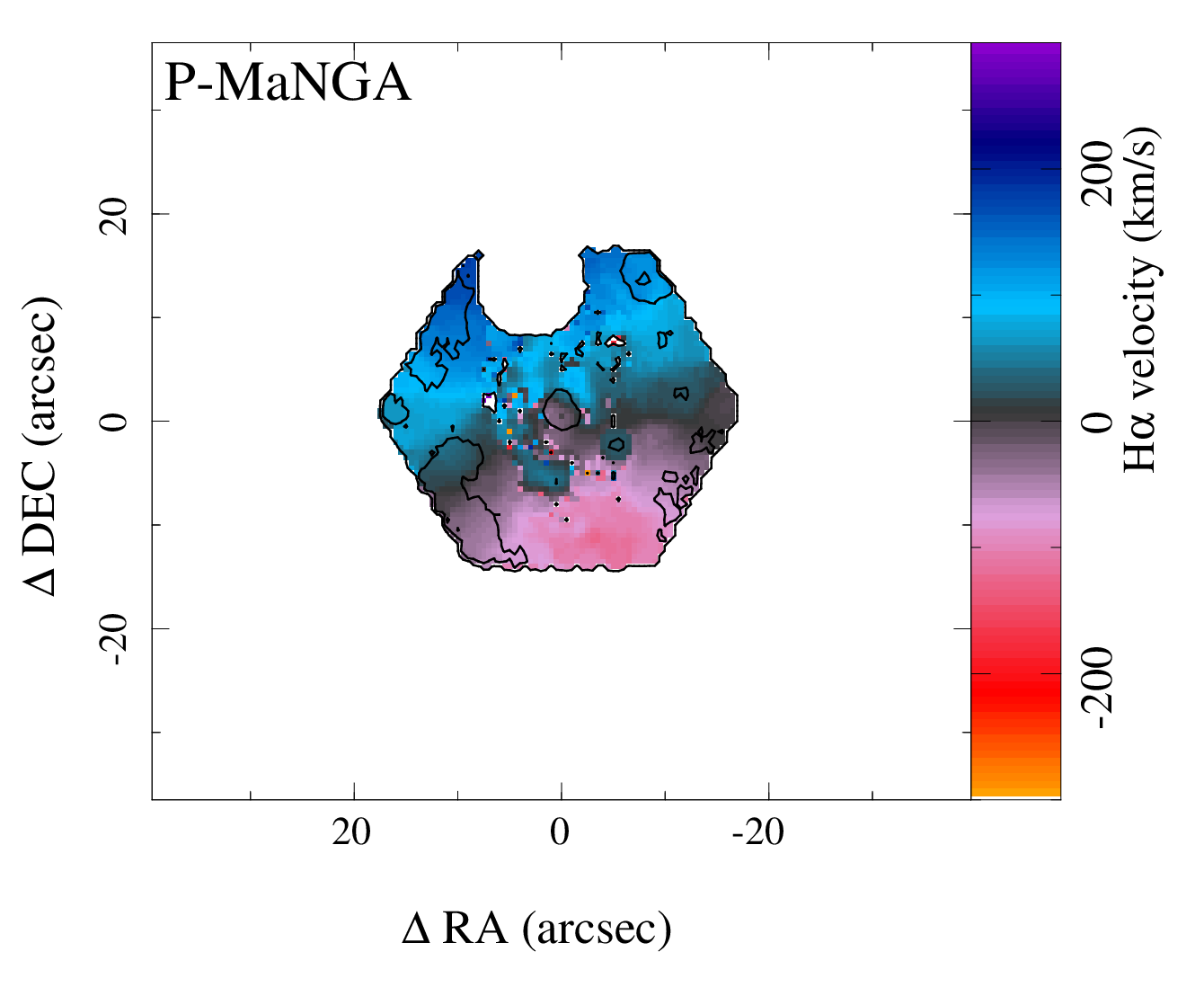}
  \caption{H$\alpha$ intensity and velocity maps (top and bottom panels respectively) derived using the emission line pure cubes derived from the CALIFA V500 (left panels) and the P-MaNGA (right panel) datasets of NGC 2916. In the left-hand panels contours correspond to the same intensity level of the broad band images presented in Fig. \ref{fig:V_map}. In the bottom panels the contours correspond to the H$\alpha$ intensity maps shown in the left-panels, starting at 0.05~\FunitsA\ with a constant step of 1~\FunitsA. { The parameteres presented in this figure were obtained after dezonification}.}\label{fig:Ha}
\end{figure}

Figure \ref{fig:Ha_CS} illustrates the result of this analysis, showing
the H$\alpha$ flux intensity and velocity maps for the CS-binned data,
after being rearranged into the original spatial shape of the datacubes. In
both panels it is possible to clearly identify  the original
CS segmentation.  This segmentation was created on the basis of the
flux intensity and S/N of the continuum, and in general, does not
reproduce the corresponding parameters for  the
emission lines. It does not only degrade unnecessarily the spatial
resolution of the emission line maps, but also it can blur the signature
of weak emission lines by co-adding in the same spatial bin emission lines
with different kinematics, and it may also affect significantly the estimated
equivalent width. This effect can be clearly observed in the
velocity maps of the areas displaying weak emission.

\subsubsection{Dezonification}
\label{ssp_des}

{ The {\tt dezonification} procedure was first presented by
  \citet{cid-fernandes13} in order to provide an accurate estimation
  of spatial distribution of the stellar properties. In {\sc Pipe3D} we use it
  to } decouple the analysis of the emission lines from the spatial
binning required to perform an accurate analysis of the stellar
continuum. This procedure takes into account the relative contribution
of each spaxel to the spatial bin in which it is aggregated, as
explained in Sec. \ref{ssp}. This is the so-called {\it dezonification
  map}.

{ 
The procedure is done performing the following steps:

\begin{itemize}

\item An empty datacube is created with the same spatial and spectral shape
  of the original cube. In this datacube it will be stored the result
  of the {\tt dezonification} procedure.

\item As indicated before (Sec. \ref{ssp} and \ref{ssp_elines}), for
  each cube it was perfomed a CS-binned, extracting a RSS-file that
  was fitted with a SSP stellar library (plus emission lines). This provides
  with a multi-SSP model for each spatial bin. 

\item For all the spaxels within the same spatial bin it is adopted
  the same multi-SSP model, that is stored in the empty datacube
  described before at the corresponding spatial coordinates of each
  spaxel.

\item Repeating this procedure for all the spatial bins, we end up
with a datacube where it is stored the SSP-model corresponding to each
spaxel. However, in this datacube the spectra corresponding to the
spaxels within the same spatial bin are all the same, keeping the 
spatial shape of the CS-segmentation.

\item This preliminar model datacube is multiplied by the
  dezonification map to match the flux
  intensity of each spectral model with that of the original cube,
  spaxel-by-spaxel. The dezonification map, explained in Sec. \ref{ssp},
  is the ratio between a broad-band intensity maps of the original and
  CS-segmented datacubes. Thus, it is the relative contribution of
  each spaxel to the intensity in corresponding spatial bin.

\item Then, in order to take into account the mismatch between adjacent spectra
corresponding to different spatial bins, the new cube is smoothed spatially
with a Gaussian Kernel having the size of the expected PSF of the datacubes
($\sim$2.5$\arcsec$-3$\arcsec$), preserving the flux intensity in each
spatial resolution element.

\item The product of this procedure is a cube comprising a
model of the underlying stellar population that presents a continuity
in the spectral shape and is adjusted to the flux intensity of the
original cube. This cube is stored in a FITS format file named {\tt SSP\_mod.OBJ.cube.fits.gz}.

\item Finally, this cube is subtracted from the original one providing 
with a set of spectra that comprise only the emission lines from the ionized gas
and the residuals from the analysis of the stellar population. A
low order polynomial is  fitted to the continuum of this
residual cube in order to remove inaccuracies in the
spectrophotometric calibration, or template mismatches
\citep[e.g.][]{husemann13,cid-fernandes13,rgb15}.

\end{itemize}

 The final product
of this analysis is the so called emission line pure cube, and it is stored
in a FITS format file named {\tt GAS.OBJ.cube.fits.gz}. 

}

\subsection{Analysis of the strong emission lines}
\label{strong}

An analysis of the emission lines using the emission line pure cube
is implemented in order to derive the properties of the ionized gas with the best
spatial resolution, and independently of the S/N required to analyze
the continuum. The strongest emission lines in the
wavelength range (from the list described in
Sec. \ref{ssp_elines}) are fitted with a single Gaussian function. 
This parametrization, implemented in the current version of {\sc Pipe3D} is
valid for most of the emission lines observed in a large fraction of the optical
extension of the galaxies. However this approach is too simplistic in some cases
(e.g., gas rich major mergers, overlapping foreground galaxies, or the core of AGNs). 
In future versions of the pipeline we will implement
multi-component analysis (already foreseen in {\sc FIT3D}). The only
limitation to implement this approach is the that the analysis will be more time consuming.

The emission lines are grouped in four group that are considered to be 
kinematically coupled (for simplicity). Each
group is fitted within a wavelength range, adjusted to the
observed frame on the basis of the galaxy redshift. The four
groups comprise the following emission lines and rest frame wavelength ranges:
(i) [\ion{O}{ii}]$\lambda$3727 (3700-3750); (ii) H$\beta$,
[\ion{O}{iii}]~$\lambda$4959, and [\ion{O}{iii}]~$\lambda$5007
(4800-5050); (iii) [\ion{N}{ii}]~$\lambda$6548, H$\alpha$, and
[\ion{N}{ii}]~$\lambda$6583 (6530-6630); and
[\ion{S}{ii}]$\lambda$6717 and [\ion{S}{ii}]$\lambda$6731 (6680-6770).
Before any of these lines is fitted with a single
Gaussian, a first guess of the kinematics is derived on the basis of
the expected H$\alpha$ wavelength at the galaxy redshift and
by performing a parabolic approximation to the centroid of the emission
line. This procedure is broadly used in the detection of peak
intensity fluxes, like in the case of the reduction of fiber fed
spectrographs, being fast and very reliable \citep[e.g.][]{sanchez06a}.
Then the emission lines are fitted using a narrow range of systemic
velocities centered in the initial guess, and limiting their width
to the nominal instrumental dispersion.

\begin{table}[!th]\label{simtab}
\begin{center}
\caption[Weak Emission Lines]{List of emission lines analyzed}
\begin{tabular}{ll|ll|ll}\hline\hline
$\lambda$ (\AA) & Id & $\lambda$ (\AA) & Id & $\lambda$(\AA) & Id$^1$ \\
\hline
3727.4   &  [\ion{O}{ii}]   & 4861.3  &  H$\beta$     &   7325.0  &   [\ion{O}{ii}] \\
3750.0  &  H12              & 4889.6  &  [\ion{Fe}{ii}]  &   7751.0  &   [\ion{Ar}{iii}] \\
3771.0  &  H11              & 4905.3  &  [\ion{Fe}{ii}]  &   9068.6  &   [\ion{S}{iii}] \\
3798.0  &  H10              & 4958.9  &  [\ion{O}{iii}]  &   9530.6  &   [\ion{S}{iii}] \\
3819.4  &  \ion{He}{i}     & 5006.8  &  [\ion{O}{iii}]  &   &  \\
3835.0  &  H9                & 5111.6   &  [\ion{Fe}{ii}] &   &  \\
3869.0  &  [\ion{Ne}{iii}] & 5158.8  &  [\ion{Fe}{ii}]  &   &  \\
3889.0  &  H8                & 5199.6  &  [NI]               &   &  \\
3967.0  &  [\ion{Ne}{iii}] & 5261.6  &  [\ion{Fe}{ii}]  &   &  \\
3970.1  &   He               & 5517.7  &  [\ion{Cl}{iii}]  &   &  \\
4026.3  &  \ion{He}{i}     & 5537.6  &  [\ion{Cl}{iii}]  &   &  \\
4069.2  &  [\ion{S}{ii}]    & 5554.9  &  \ion{O}{i}      &   &  \\
4076.7  &  [\ion{S}{ii}]    & 5577.3  &  [\ion{O}{i}]    &   &  \\
4101.7  &  H$\delta$     & 5754.5  &  [\ion{N}{ii}]    &   &  \\
4276.8  &  [\ion{Fe}{ii}]  & 5875.6  &  \ion{He}{i}    &   &  \\
4287.4  &  [\ion{Fe}{ii}]  & 6300.3  &  [\ion{O}{i}]    &   &  \\
4319.6  &  [\ion{Fe}{ii}]  & 6312.4  &  [\ion{S}{iii}]   &   &  \\
4340.5  &  H$\gamma$ & 6347.3  &  SiII               &   &  \\
4363.2  &  [\ion{O}{iii}]   & 6363.8  &  [\ion{O}{i}]    &   &  \\
4413.8  &  [\ion{Fe}{ii}]  & 6562.7  &  H$\alpha$    &   &  \\
4416.3  &  [\ion{Fe}{ii}]  & 6583.4  &  [\ion{N}{ii}]    &   &  \\
4471.0  &  \ion{He}{i}    & 6548.1  &  [\ion{N}{ii}]    &   &  \\
4657.9  &  [\ion{Fe}{iii}] & 6678.0  &  \ion{He}{i}     &   &  \\
4686.0  &  \ion{He}{ii}   & 6716.4  &  [\ion{S}{ii}]    &   &  \\
4713.0  &  \ion{He}{i}    & 6730.7  &  [\ion{S}{ii}]    &   &  \\
4922.2  &  \ion{He}{i}    & 7136.0  &  [\ion{Ar}{iii}]  &   &  \\
\hline
\end{tabular}
\end{center}
(1) Only accessible for MaNGA.    
\end{table}

\begin{figure*}[!t]
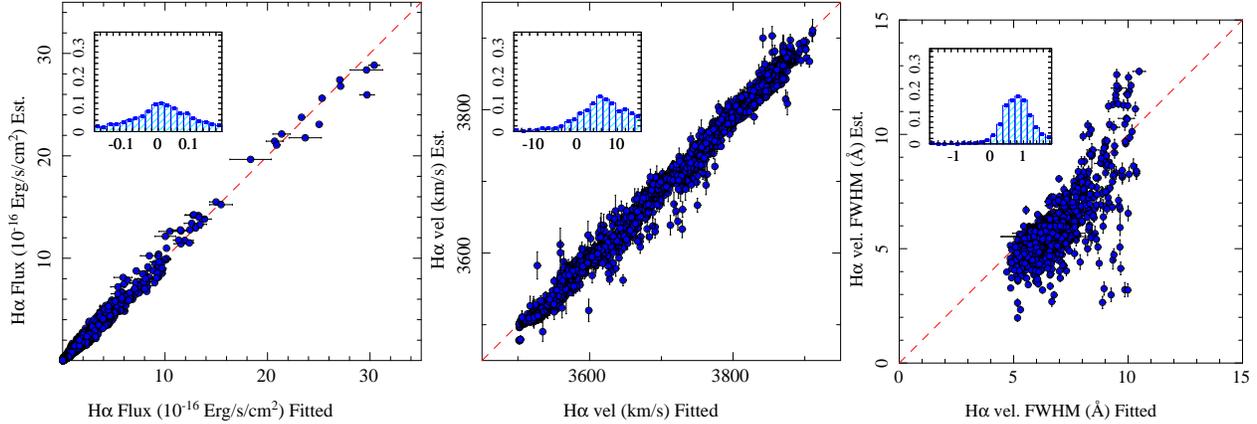

\includegraphics[angle=270,width=0.33\linewidth]{figs/comp_Ha_Fit_Est.ps}\includegraphics[angle=270,width=0.33\linewidth]{figs/comp_vel_Ha_Fit_Est.ps}\includegraphics[angle=270,width=0.33\linewidth]{figs/comp_disp_Ha_Fit_Est.ps}
  \caption{Comparison between the integrated flux intensity (left panel), velocity (central panel), and velocity dispersion (FWHM, right panel) for the H$\alpha$ emission line extracted from the emission line pure cubes  from the CALIFA V500 based on the Gaussian fits described in Sec. \ref{strong} (x-axis), versus the values derived using the algorithm described in Sec. \ref{weak} (y-axis). The error bars indicate the  errors estimated by each procedure. For the velocity and velocity dispersion we show only the $\sim$2700 spaxels for which the H$\alpha$ flux density is larger than 0.5 \fuDEN. In each panel the inset shows the normalized histogram of the difference between the two estimations.}\label{fig:Est}
\end{figure*}

The result of this analysis is a set of maps with the spatial shape
of the emission line pure cube, that include the various
 parameters derived for each emission line as described in Sec. \ref{ssp_elines}. 
These maps are stored in a set of FITS format files, named 
{\tt map.W1\_W2.OBJ\_PARAM\_NN.fits.gz}, where {\tt OBJ}
is the galaxy name (as it appears in the name of the datacube), {\tt PARAM}
is a label indicating each of the derived dataproducts, and {\tt W1} and {\tt W2} are the
wavelength ranges of each emission line group, as described above, and {\tt NN} is an
index indicating the order of the emission line within each group.
For example, the FITS file {\tt map.6530\_6630.NGC2916\_flux\_00.fits.gz}
stores the flux density of H$\alpha$ derived from the emission line pure cube derived for CALIFA V500 data of NGC 2916. The emission line  fluxes are not corrected for extinction, 
that should be derived using the usual procedures, e.g. analyzing the Balmer line ratios,
as we will describe in Sec. \ref{test}. Like in previous cases, an example of these files is
stored in the FTP indicated above and described in Sec.\ref{pack}.

Figure \ref{fig:Ha} illustrates the result of this analysis, showing
the H$\alpha$ flux intensity and velocity maps derived from the
emission line pure cube. On the other hand, Fig. \ref{fig:Ha_CS} highlights
the differences in the  parameters derived
when the analysis of the emission lines is coupled or not with the
spatial binning required to analyze the stellar population. 
As anticipated, the emission lines are blurred in those areas where
the continuum intensity is lower, and therefore it requires larger spatial bins
to achieve a sufficient S/N. In some cases the gas kinematics is clearly
affected too. { The effect is stronger in the P-MaNGA data than in the CALIFA ones, due to the lower S/N of former ones. The final MaNGA observing strategy \citep{law15}, with a minimum goal in the S/N, guarantees that this will not be the case for the final dataset. However, for the P-MaNGA data, it was selected a fixed exposure time what affects more the continuum S/N at this spectral resolution.}

\subsection{Analysis of the weak emission lines}
\label{weak}

So far we have characterized the strongest and more frequently observed
(and studied) emission lines within the  wavelength range considered.
However, there are many more weak emission lines. Table 1 lists
the usual emission lines observed in ionized regions of our Galaxy in
the common wavelength range between the three  IFU surveys considered here.
This list was extracted from those detected in classical \HII
regions, like the Orion nebula \citep[][]{bald91,sanchez07c}. Those
emission lines that are only accessible for MaNGA due to its larger wavelength
coverage are indicated.

It is not practical to perform a Gaussian fit, like the one
described in the previous Section, for all the $\sim$50 emission lines
and for all the spectra in each datacube, since it is very time
consuming. Therefore, we have adopted a different scheme to extract
the main properties of these emission lines: flux intensity, velocity 
and velocity dispersion, and equivalent width.

This procedure is not a Gaussian fit, but rather a direct estimation of
these parameters. It requires as input the emission line pure and the
stellar population model cubes described in Sec. \ref{ssp} and
\ref{ssp_des}, together with an error cube provided by the data
reduction. In addition, it requires a list of the emission lines to be
analyzed, with their corresponding identification and nominal wavelength
(like in Table 1), and an estimation of the gas velocity (in km/s) and
velocity dispersion, including the instrumental dispersion, in \AA. 
The output of the H$\alpha$ emission analysis described in Sec. \ref{strong} 
is adopted for these latter entries.

After reading the required input, the algorithm performs the following
steps: (i) For each emission line in the list, and for
each spectra in the emission line pure cube, it estimates the expected
observed central wavelength of the emission line taking into account
the initial guessed velocity ($\lambda_{obs}$).  Then a
wavelength range is selected within $\pm$FWHM of the emission
line, derived from the initial guessed dispersion ($\sigma_{in}$):
[$\lambda_{obs}-$2.354$\sigma_{in}$, $\lambda_{obs}+$2.354$\sigma_{in}$];
(ii) Within this wavelength range a set of 50 MC
realizations of the spectra are performed, by co-adding to the original flux
 the error noise multiplied by a random number between
$\pm$0.5; (iii) For each MC realization ($mc$), and for each
spectral pixel ($i$) at a wavelength $\lambda_i$, the
extended flux intensity is estimated if the emission line was well characterized by
a Gaussian function centered at $\lambda_{obs}$ with a dispersion 
$\sigma_{in}$, using the formula:

\begin{equation}\label{eq1}
 F^{mc}_{0,i} = I^{mc}_{i} \sigma_{in} \sqrt{2\pi} ~ {\rm exp}\left( 0.5\frac{(\lambda_i-\lambda_{obs})^2}{\sigma_{in}^2} \right)  
\end{equation}

\noindent where $F^{mc}_{0,i}$ is the integrated flux intensity of the
emission line estimated from the measured flux density ($I^{mc}_{i}$) at
the spectral point $i$, and for the MC realization ($mc$); (iv) Then,
for each MC loop, an average of the integrated
flux intensities is derived for all the spectral points in the
wavelength range considered. The estimation is more
accurate for those spectral points near the peak intensity of the
emission line (due to the higher S/N). Therefore  a
weighted average is performed, with the weights following a Gaussian distribution
centered in the observed wavelength of the emission line and with a
dispersion $\sigma_{in}$; (v) Once  the integrated flux
intensity is derived, the procedure is repeated solving Equation \ref{eq1} for
the velocity ($v$), given in km/s:

\begin{equation}\label{eq:lambda}
 \lambda^{mc}_{obs,i}=\lambda_{i}-\sqrt{2\sigma_{in} ln\left(\frac{F_{0}}{I^{mc}_{i} \sigma_{in} \sqrt{2\pi}}\right)}
\end{equation}

\noindent where:

\begin{equation}\label{eq:lambda_mc}
\lambda^{mc}_{obs,i} = \lambda_{rest} \left( 1 + \frac{v^{mc}_{i}}{c}\right)
\end{equation}

\noindent where $\lambda_{rest}$ is the rest frame nominal wavelength of the
emission line, $c$ is the speed of light, and
$v^{mc}_{i}$ is the velocity estimated for each MC realization at each
spectral point $i$. As in the previous case, a weighted average is derived
as the best estimation of $\lambda_{obs}$ and $v$; 

(vi) Finally the velocity dispersion ($\sigma$)  is
estimated on the basis of the second order moment of the
distribution, for each MC realization.  This approach is adopted 
due to the complexity of solving Equation
\ref{eq1} for this parameter:

\begin{equation}\label{eq:sigma}
\sigma^2_{mc,i} =  \frac{\Sigma I^{mc}_{i}(\lambda_{i}-\lambda_{obs})^2}{\Sigma I^{mc}_{i}} 
\end{equation}

\noindent and then transformed to FWHM by the scaling factor ($FWHM = 2.354 \sigma$).
The dispersion also includes the
instrumental resolution, that should be subtracted quadratically in any
further analysis; (vii) In addition, the EW of the
corresponding emission line is derived  by dividing the intensity
by the flux density of the underlying continuum, derived as the
average within two 30\AA\ wide spectral windows  centered at
$\pm$60\AA\ from $\lambda_{obs}$, of the spectrum extracted from the
stellar population model (i.e., once subtracted the emission lines).
The band width is large enough to smooth any significant contribution
by most of the stellar absorption features, although some effect is
impossible to avoid;
(vi) The average and the standard deviation of the
four parameters obtained for each MC realization and for each spaxel
are derived and stored in a set of 2D arrays with the same spatial shape
as the original cube; (viii) The final dataproducts for each emission
line comprise eight 2D arrays, four for the parameters derived and
four more for the errors. The complete set of 2D arrays for
all the emission lines analyzed  are stored in a datacube named {\tt
  flux\_elines.OBJ.cube.fits.gz}, in which each 2D slice corresponds
to a particular dataproduct (or its error) for each of the
emission lines  analyzed. The header comprises a set of keywords named {\tt
  NAMEXX} that store the correspondence of the slice {\tt XX} to 
a particular emission line and dataproduct. We note once more
that emission line fluxes are not corrected for internal extinction in the galaxy.

\begin{table*}[!th]\label{st_index}
\begin{center}
\caption[Stellar Indices]{List of stellar indices analyzed}
\begin{tabular}{llllll}\hline\hline
Index & Index $\lambda$ range (\AA) & blue $\lambda$ range (\AA) & red $\lambda$ range (\AA) \\
\hline
H$\delta$     & 4083.500-4122.250 &4041.600-4079.750 &4128.500-4161.000\\
H$\delta$mod  & 4083.500-4122.250 &4079.000-4083.000 &4128.500-4161.000\\
H$\gamma$     & 4319.750-4363.50  &4283.500-4319.75  &4367.250-4419.750\\
H$\beta$     & 4847.875-4876.625 &4827.875-4847.875 &4876.625-4891.625\\
Mg$b$    & 5160.125-5192.625 &5142.625-5161.375 &5191.375-5206.375\\ 
Fe5270 & 5245.650-5285.650 &5233.150-5248.150 &5285.650-5318.150\\
Fe5335 & 5312.125-5352.125 &5304.625-5315.875 &5353.375-5363.375\\
\hline
D4000  & 4050.000-4250.000 &3750.000-3950.000 & \\
\hline
\end{tabular}
\end{center}
\end{table*}

Figure \ref{fig:Est} shows the comparison between the parameters
derived using this algorithm and the values derived using the Gaussian
fitting procedure for H$\alpha$ when analyzing the emission line pure cube for
the CALIFA V500 dataset. The integrated flux intensity is the
parameter that presents the smaller differences between the two procedures
($\Delta F $=0.02$\pm$0.25\,\Funits). For the velocity the agreement is
within the expectated errors ($\Delta vel$=9.6$\pm$11.7 km/s). The largest relative
differences are found in the velocity dispersion, although they are
within the expectations from the estimated errors ($\Delta
\sigma$=0.8$\pm$0.66 \AA, that corresponds to $\sigma_{vel}$=37$\pm$31
km/s). { No correction is applied based on these differences, since a priori
we do not know which of the two results is more accurate. More simulations are required on this regards. }

In general, when the emission lines are well deblended this
algorithm produces reliable results. Indeed, based on extensive 
simulations as described in PaperI  { (Sec. 3.3)}, the accuracy of
the  parameters recovered is very similar to the one estimated for the
Gaussian fits. However, it seems that there is a non negligible systematic offset
between the  kinematic parameters derived from both methods; this should be
clarified based on simulations. Like in the case of the method described in
Sec. \ref{strong} (assuming a single Gaussian function per emission
line), this procedure is not valid to analyze heavily blended
emission lines, like in the case of multi-component kinematics and/or
broad emission lines due to outflows or AGNs.

The major advantage of this procedure is speed. Using a single
core i7 processor it takes about 1 hour to analyze a single emission
line using the Gaussian fitting algorithm described in
Sec. \ref{strong} for a CALIFA-like datacube (or a MaNGA datacube for
the bundles with the largest FoVs). However, for the direct
estimation procedure it takes $\sim$3 minutes to analyze the $\sim$50
emission lines listed in Table 1. Its disadvantage 
is that it requires a prior estimation of the properties of the gas
kinematics. For this reason, in {\sc Pipe3D} we first perform a
Gaussian fit for a set of strong emission lines and we adopted the
new algorithm for a much wider set of weaker (in general) emission lines.
We are exploring alternative solutions to speed up the process even more.

\subsubsection{Stellar Indices}
\label{indices}

A classical technique to characterize the properties of the stellar
population in galaxies is to measure certain line strength indices,
such as the Lick/IDS index system
\citep[e.g.][]{Burstein:1984p3764,Faber:1985p3766,Burstein:1986p3765,Gorgas:1993p3767,Worthey:1994p3434}.
When comparing with the expected values derived using stellar population synthesis
models, indices can be used to infer stellar population parameters
such as age, metallicity, and $\alpha$ enhancement
\citep[e.g.][]{trager+00,gallazzi+05}. They provide robust,
model-independent, information, complementary to that provided by
fitting the full spectrum with multi-SSP templates, as described in Sec. \ref{ssp}.
In general, the method employs a combination of  
indices mostly orthogonal in the physical parameter space (i.e. age and
metallicity), like D4000 or H$\delta$ (sensitive to the age), and Mg$b$ or [MgFe]' 
(sensitive to the  metallicity),
where [MgFe]' is a combined stellar index, given by the formula:

\begin{equation}\label{eq:Mgb}
[{\rm MgFe}]'~=~ \sqrt{{\rm Mg}b ~ (0.72 {\rm Fe}_{5270}~+~0.28 {\rm
    Fe}_{5335})}
\end{equation}

\begin{figure*}[!t]
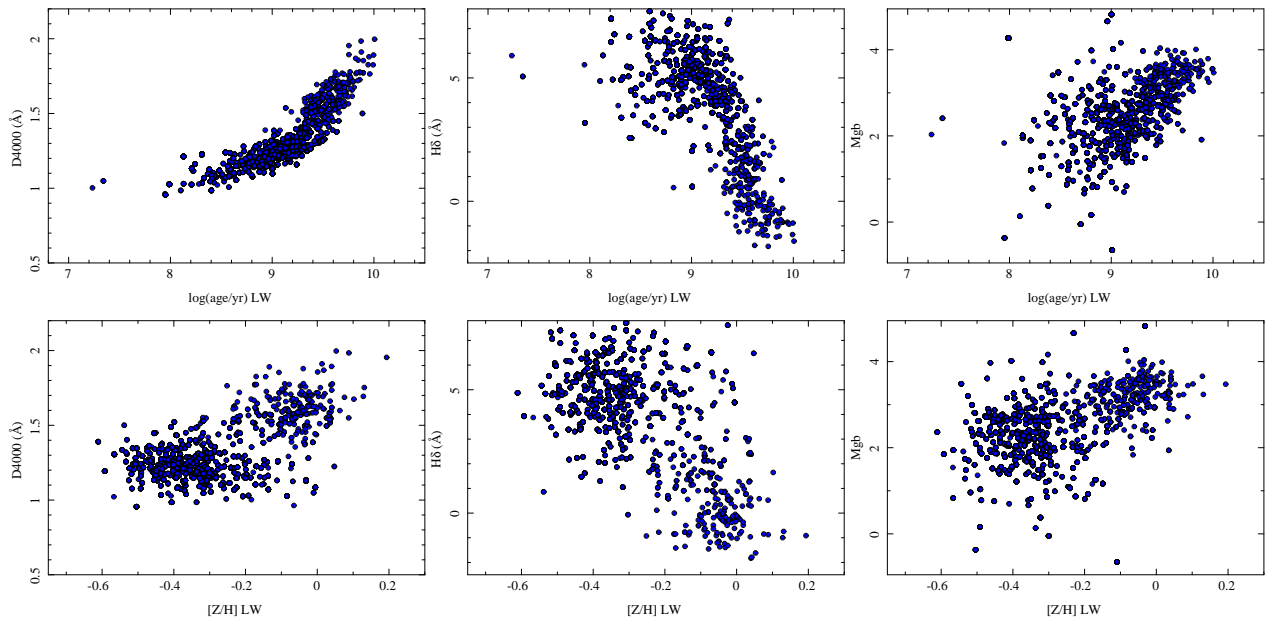

\includegraphics[angle=270,width=0.33\linewidth]{figs/ind_D4000_age.ps}\includegraphics[angle=270,width=0.33\linewidth]{figs/ind_Hd_age.ps}\includegraphics[angle=270,width=0.33\linewidth]{figs/ind_Mgb_age.ps}
\includegraphics[angle=270,width=0.33\linewidth]{figs/ind_D4000_ZH.ps}\includegraphics[angle=270,width=0.33\linewidth]{figs/ind_Hd_ZH.ps}\includegraphics[angle=270,width=0.33\linewidth]{figs/ind_Mgb_ZH.ps}
  \caption{Comparison of a set of stellar indices (D4000, H$\delta$
    and Mg$b$) and the luminosity weighted ages and metallicities
    derived for the spatially binned spectra extracted from the
    CALIFA V500 datacube of NGC 2916.}\label{fig:ind}
\end{figure*}

As briefly described in PaperI, { Sec. 4}, {\sc FIT3D} includes a
script to derive the equivalent widths of a predefined set of stellar
indices. The algorithm follows the prescriptions implemented in {\sc
  indexf} \citep{Cardiel:2003p3435}, slightly modified to take into
account the format of the dataset analyzed. It requires as input the
output from the previous analysis, described in Sec. \ref{ssp} and
Sec. \ref{ssp_elines}.  In particular, it uses the files {\tt
  output.auto\_ssp.CS.OBJECT.rss.out.fits.gz} and {\tt
  auto\_ssp.CS.OBJECT.rss.out}, and a number of MC simulations.  The
algorithm uses the spectrum after subtracting the strong emission lines,
for each spatial-bin, and it takes the residuals from the analysis of the
stellar population as a hint of the noise pattern to perform a set of MC
realizations of the data. Then it estimates the equivalent width for
each of the stellar indices using the formula included in
{\sc indexf} for each of the MC realizations. The bandwidths adopted
to derive the equivalent width are redshifted to the observed
wavelength range using the velocity estimated for each spectrum from the
analysis of the stellar population, included in the input files. 
Finally, the average and the
standard deviation for the different values estimated for each MC
simulation are derived. These parameters are stored in an ASCII file named {\tt
  indices.CS.OBJ.rss.out}, that later is transformed to a datacube
named {\tt indices.CS.OBJ.cube.fits.gz } by associating the 
value to the location in the 2D space defined by the segmentation file
described in Sec. \ref{binning}.  This final datacube comprises a set
of slices, each one including the value derived for the stellar index
(mean value) and its estimated error (standard deviation).

\begin{table*}[!th]\label{felines}
\begin{center}
\caption[Description example flux elines]{Description of the {\tt
    flux\_elines.OBJ.cube.fits.gz} dataproduct.}
\begin{tabular}{lll}\hline\hline
Keyword/Slice & Value  & Description\\
\hline
NAME0 & flux [OII]3727 & Integrated flux of [\ion{O}{ii}]$\lambda$3727 in \Funits \\
NAME51 & vel [OII]3727 & Velocity of [\ion{O}{ii}]$\lambda$3727 in km/s \\
NAME102 & disp [OII]3727 &  Velocity dispersion FWHM of [\ion{O}{ii}]$\lambda$3727 in \AA \\
NAME153 & EW [OII]3727  & Equivalent width of [\ion{O}{ii}]$\lambda$3727 in \AA \\
NAME204 & e\_flux [OII]3727 & Estimated error of the integrated flux of [\ion{O}{ii}]$\lambda$3727 in \Funits \\
NAME255 & e\_vel [OII]3727 & Estimated error of the velocity of [\ion{O}{ii}]$\lambda$3727 in km/s \\
NAME306 & e\_disp [OII]3727 & Estimated error of the velocity dispersion of [\ion{O}{ii}]$\lambda$3727 in \AA \\
NAME357 & e\_EW [OII]3727 & Estimated error of the Equivalent width of [\ion{O}{ii}]$\lambda$3727 in \AA \\
 ...  & ... & ... \\
NAME50 & flux [SII]6731  & Integrated flux of [\ion{S}{ii}]$\lambda$6731 in \Funits \\
NAME101 & vel [SII]6731  & Velocity of [\ion{S}{ii}]$\lambda$6731 in km/s \\
NAME152 & disp [SII]6731 &  Velocity dispersion FWHM of [\ion{S}{ii}]$\lambda$6731 in \AA \\
NAME203 & EW [SII]6731   & Equivalent width of [\ion{S}{ii}]$\lambda$6731 in \AA \\
NAME254 & e\_flux [SII]6731 & Estimated error of the integrated flux of [\ion{S}{ii}]$\lambda$6731 in \Funits \\
NAME305 & e\_vel [SII]6731  & Estimated error of the velocity dispersion of [\ion{S}{ii}]$\lambda$6731 in \AA \\
NAME356 & e\_disp [SII]6731 & Estimated error of the velocity dispersion of [\ion{S}{ii}]$\lambda$6731 in \AA \\
NAME407 & e\_EW [SII]6731 & Estimated error of the equivalent width of [\ion{S}{ii}]$\lambda$6731 in \AA \\
\hline
\end{tabular}
\end{center}
\end{table*}

\begin{table*}[!th]\label{SFH_cube}
\begin{center}
\caption[Description example SFH cube]{Description of the {\tt OBJ.SFH.cube.fits.gz} dataproduct.}
\begin{tabular}{ll}\hline\hline
Keyword/Slice & Value/Description\\
\hline
DESC\_0 & Luminosity Fraction for age-met 0.0010-0.0037 SSP \\
DESC\_1 & Luminosity Fraction for age-met 0.0010-0.0076 SSP \\
... & ... \\
DESC\_154 & Luminosity Fraction for age-met 7.9433-0.0190 SSP \\
DESC\_155 & Luminosity Fraction for age-met 7.9433-0.0315 SSP \\
\hline
DESC\_156 & Luminosity Fraction for age 0.0010 SSP \\
DESC\_157 & Luminosity Fraction for age 0.0030 SSP \\
... & ... \\
DESC\_193 & Luminosity Fraction for age 12.5893 SSP \\
DESC\_194 & Luminosity Fraction for age 14.1254 SSP \\
\hline
DESC\_195 & Luminosity Fraction for met 0.0037 SSP \\
DESC\_196 & Luminosity Fraction for met 0.0076 SSP \\
DESC\_197 & Luminosity Fraction for met 0.0190 SSP \\
DESC\_198 & Luminosity Fraction for met 0.0315 SSP \\
\hline
\end{tabular}
\end{center}
\end{table*}

\begin{table*}[!th]\label{SFH_cube}
\begin{center}
\caption[Description example SSP cube]{Description of the {\tt OBJ.SSP.cube.fits.gz} dataproduct.}
\begin{tabular}{ll}\hline\hline
Keyword/Slice & Value/Description\\
\hline
DESC\_0 &  $V$-band map reconstructed from the original cube.  \\
DESC\_1 &  CS segmentation map. \\
DESC\_2 &  Dezonification map. \\
DESC\_3 &  Average intensity flux within the wavelength range, in \funits. \\
DESC\_4 &  Standard deviation of the  flux within the wavelength range, in \funits. \\
DESC\_5 &  Luminosity weighted age of the stellar population in log(age/yr). \\
DESC\_6 &  Mass weighted age of the stellar population in log(age/yr). \\
DESC\_7 &  Error of the age of the stellar population in $\Delta$age/age. \\
DESC\_8 &  Luminosity weighted metallicity of the stellar population in log(Z/Z$_\odot$). \\
DESC\_9 &  Mass weighted metallicity of the stellar population in log(Z/Z$_\odot$). \\
DESC\_10 &  Error metallicity of the stellar population in $\Delta$Z/Z. \\
DESC\_11 &  Dust attenuation of the stellar population (A$_{V,stars}$) in mag. \\
DESC\_12 &  Error of the average dust attenuation of the stellar population in mag. \\
DESC\_13 &  Velocity of the stellar population in km/s. \\
DESC\_14 &  Error in the velocity of the stellar population in km/s. \\
DESC\_15 &  Velocity dispersion of the stellar population in km/s. \\
DESC\_16 &  Error in velocity dispersion of the stellar population in km/s. \\
DESC\_17 &  Average mass-to-light ratio of the stellar population in Solar Units. \\
DESC\_18 &  Stellar mass density in $M_\odot/arcsec^2$, not dust corrected. \\
DESC\_19 &  Stellar mass density in $M_\odot/arcsec^2$, dust corrected using the A$_{v,stars}$. \\
\hline
\end{tabular}
\end{center}
\end{table*}

\begin{table*}[!th]\label{ELINES_cube}
\begin{center}
\caption[Description example ELINES cube]{Description of the {\tt OBJ.ELINES.cube.fits.gz} dataproduct.}
\begin{tabular}{ll}\hline\hline
Keyword/Slice & Value/Description\\
\hline
DESC\_0 &  H$\alpha$ velocity map, km/s.\\
DESC\_1 &  H$\alpha$ emission line velocity dispersion, FWHM in \AA .\\
DESC\_2 &  [\ion{O}{ii}]$\lambda$3727 emission line flux in \Funits . \\
DESC\_3 &  [\ion{O}{III}]$\lambda$5007 emission line flux in \Funits . \\
DESC\_4 &  [\ion{O}{III}]$\lambda$4959 emission line flux in \Funits . \\
DESC\_5 &  H$\beta$ emission line flux in \Funits . \\
DESC\_6 &  H$\alpha$ emission line flux in \Funits . \\
DESC\_7 &  [\ion{N}{II}]$\lambda$6583 emission line flux in \Funits . \\
DESC\_8 &  [\ion{N}{II}]$\lambda$6548 emission line flux in \Funits . \\
DESC\_9 &  [\ion{S}{II}]$\lambda$6731 emission line flux in \Funits . \\
DESC\_10 &  [\ion{S}{II}]$\lambda$6717 emission line flux in \Funits . \\
DESC\_11 &  [\ion{O}{ii}]$\lambda$3727 emission line flux error in \Funits . \\
DESC\_12 &  [\ion{O}{III}]$\lambda$5007 emission line flux error in \Funits . \\
DESC\_13 &  [\ion{O}{III}]$\lambda$4959 emission line flux error in \Funits . \\
DESC\_14 &  H$\beta$ emission line flux error in \Funits . \\
DESC\_15 &  H$\alpha$ emission line flux error in \Funits . \\
DESC\_16 &  [\ion{N}{II}]$\lambda$6583 emission line flux error in \Funits . \\
DESC\_17 &  [\ion{N}{II}]$\lambda$6548 emission line flux error in \Funits . \\
DESC\_18 &  [\ion{S}{II}]$\lambda$6731 emission line flux error in \Funits . \\
DESC\_19 &  [\ion{S}{II}]$\lambda$6717 emission line flux error in \Funits . \\
\hline
\end{tabular}
\end{center}
\end{table*}

Table 2 shows the list of stellar indices included in the analysis,
along with the adopted bandwidths for each index, and the
blue and red wavelength ranges from which  the
continuum is estimated. By construction, these bandwidths were selected to be
compatible with {\sc rmodel} \citep{Cardiel:2003p3435}, an algorithm
that allows to estimate the age and metallicity of the stellar
population based on a comparison with the expected values for a pair
of indices with the corresponding ones for a stellar template, as
described in PaperI { , Sec. 4}.  This procedure was already used in the
analysis of previous IFU data, as described in \citet{sanchez11} and
\citet{sanchez12a}.

The dependence of the different stellar indices with the age and
metallicity of single stellar populations has been broadly explored in
detail using SSP models or more complex SFHs
\citep[e.g.][]{poggianti97}. To illustrate how the  stellar indices derived compare
with the luminosity weighted ages and metallicities derived from the
procedure described in Sec. \ref{ssp} we show a few of them in Figure
\ref{fig:ind}. As expected, there is a clear trend between the stellar
indices and the corresponding physical parameter that they are sensitive
to. It is also known that those indices are sensitive to other physical parameters.
It is beyond the scope of this article to study this effect in detail, since 
our aim is to show the different dataproducts delivered by {\tt Pipe3D}. 
These results will be analyzed in future studies.

\subsection{Packing of the dataproducts}
\label{pack}

{\sc Pipe3D} produces a large number of intermediate dataproducts that
are usually stored either as FITS files, corresponding to the maps
comprising the values and errors of each of the estimated parameters,
or as ASCII files, as tables with the different parameters listed
either for each spaxel or for each spatial bin. Since the main goal of
this pipeline is to produce dataproducts that are easily distributed,
shared and compared between different researchers and for different
surveys, we have foreseen a simple solution: (1) All dataproducts are
stored in 2D maps following the spatial shape of the original
datacubes, and keeping the original WCS; (2) then, those
corresponding to a similar kind of analysis are packed together and
stored in datacubes with the same spatial shape of the dataproducts
(and the original cube), that corresponds to a set of maps arbitrarily ordered
along the 3rd dimension, with a header keyword ({\tt DEC\_XX})
indicating which parameter is stored at each slice ({\tt XX}) of the
datacube; (3) the dataproducts corresponding to spatial bins are
resampled to the original spatial shape without any interpolation,
just associating the same value to all the spaxels corresponding to
the same spatial bin. The segmentation and dezonification files are
stored in order to allow the user to dezonify a particular
integrated property and identify in an easy way which spaxels are
binned together; (4) A total of three dataproduct cubes, in addition
to those already described corresponding to the stellar indices
(Sec. \ref{indices}), and the weak emission lines ones
(Sec. \ref{weak}), are delivered comprising: (i) the dataproducts
derived from the analysis of the stellar population ({\tt
  OBJ.SSP.cube.fits.gz}); (ii) the dataproducts describing the stellar
decomposition or star formation history ({\tt OBJ.SFH.cube.fits.gz});
(iii) the dataproducts describing the properties of the strong
emission lines ({\tt OBJ.ELINES.cube.fits.gz}). 
Tables 3, 4, 5, and 6 list the descriptions of
the different dataproducts stored in each slice for each of the 
datacubes, as indicated in the header keywords.

\section{A practical implementation of Pipe3D}
\label{test}

Along this article we have described in detail the different
procedures implemented in {\sc Pipe3D} to characterize the physical
properties of the stellar populations and ionized gas of galaxies
observed by the major currently on-going IFU surveys.  In this section
we present a practical implementation showing the results of applying
{\sc Pipe3D} to a particular dataset: the 200 cubes that comprise
the CALIFA DR2 for the V500 setup \citep{rgb15}\footnote{\url{http://califa.caha.es/DR2/}}.

\begin{figure*}[!t]
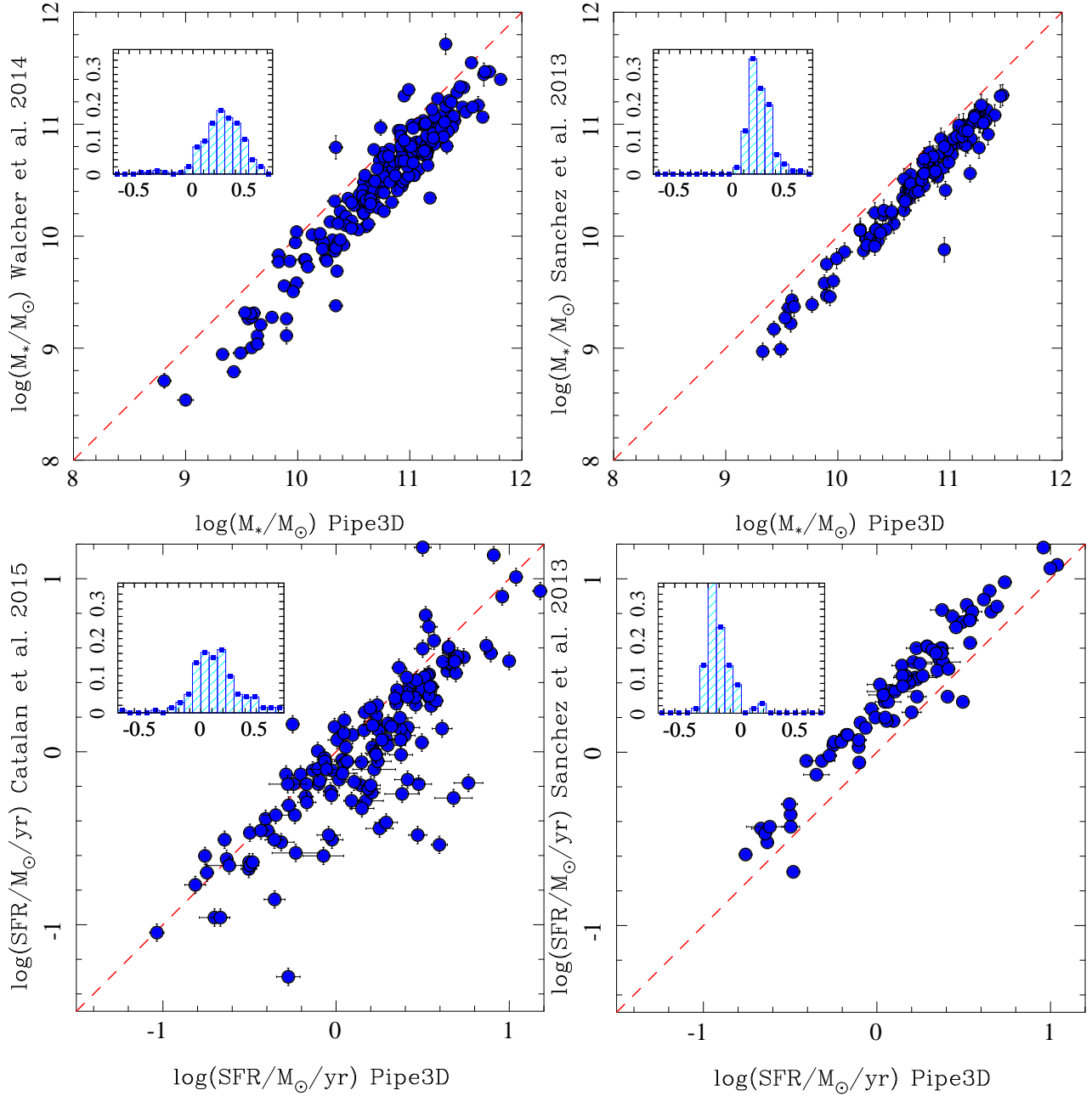

  \includegraphics[angle=270,width=0.5\linewidth]{figs/comp_Mass_DR2_walcher.ps}\includegraphics[angle=270,width=0.5\linewidth]{figs/comp_Mass_DR2_sanchez.ps}
  \includegraphics[angle=270,width=0.5\linewidth]{figs/comp_SFR_DR2_catalan.ps}\includegraphics[angle=270,width=0.5\linewidth]{figs/comp_SFR_DR2_sanchez.ps}
  \caption{Comparison of the integrated properties derived using {\sc Pipe3D} listed in Table 7 for the galaxies comprising the CALIFA DR2 sample with previous published results using different procedures.  In each panel the inset shows the normalized histogram of the difference between the two estimations of the considered parameter.}\label{fig:comp_DR2}
\end{figure*}

The list of individual objects, including their nominal names and the
corresponding CALIFA-IDs, together with the coordinates are listed in
Appendix \ref{app:char}, Table \ref{DR2_obj}.1.  In addition, it
includes some of the main global properties derived by {\tt Pipe3D}:
(i) the redshift, (ii) the stellar mass integrated
within the FoV of the datacubes; (ii) the corresponding star formation
rate\footnote{The H$\alpha$ flux for each spaxel in the FoV,
is  corrected for dust attenuation, derived from the Balmer ratio
(H$\alpha$/H$\beta$) spaxel by spaxel, assuming a canonical line
ratio of 2.86 and the Milky-Way extinction law \citep{cardelli89}, and
a Milky-Way specific dust attenuation of $R_V=$3.1. Then it
is transformed to absolute luminosity using the standard cosmology (H$_0$=71 km/s/Mpc, $\Omega_M$=0.27, $\Omega_\lambda$=0.73),
and transformed to SFR adopting the \citet{kennicutt98} empirical
relation}. We should note that we have co-added all the emission
within the FoV irrespectively of the nature of the
ionization. Therefore, the SFR listed here should be considered as a
linear transformation of the H$\alpha$ luminosity in a general
sense. The errors derived by {\tt Pipe3D} have been propagated and
included in the table.

\begin{figure*}[!t]
  \includegraphics[angle=270,width=1\linewidth]{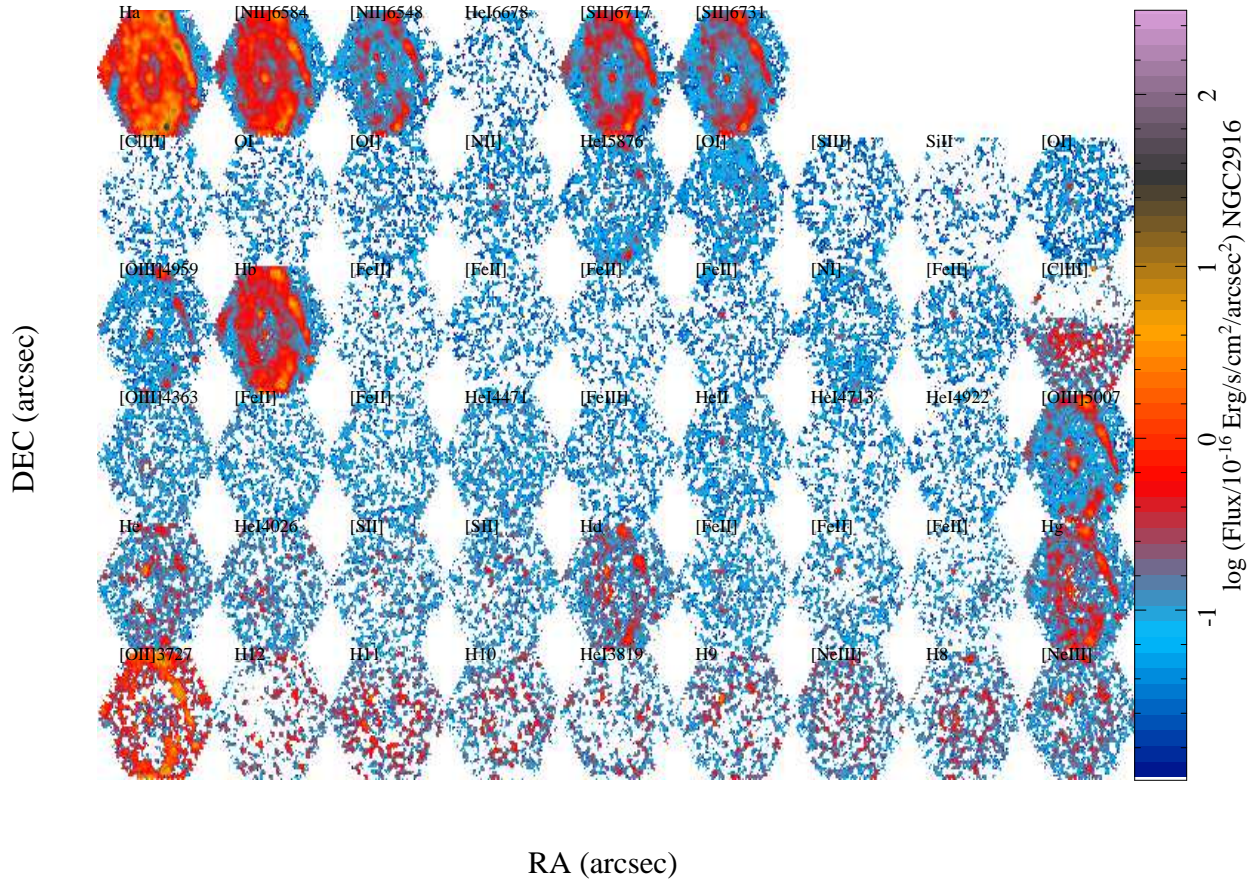}
  \caption{Flux intensity maps for all the emission lines listed in Table 1 and analyzed following the procedure described in Sec. \ref{weak}, in units of \fuden, derived for NGC 2916.}\label{fig:Ha_cubes}
\end{figure*}

Figure \ref{fig:comp_DR2} shows a comparison of these parameters with
those reported in previous studies
\citep{sanchez13,walcher14,catalan15}, for the galaxies in
common. \citet{walcher14} derived the stellar mass density for the
full CALIFA mother sample using the full available photometric dataset
for these galaxies, from the UV to the NIR, and performing a SED
fitting using {\sc Paradise} (PI: Walcher). By construction, all
galaxies listed in Table 7 were analyzed by \citet{walcher14}. We
found an offset of 0.29 dex, with a dispersion of $\pm$0.18 dex around
this offset. \citet{sanchez13} derived the stellar mass using the SDSS
photometry and the \citet{bell01} M/L relations with the color for the
galaxies currently available for the on-going CALIFA survey. For the
110 galaxies in common, we found a similar offset with a lower
dispersion around it ($\Delta log(M)= 0.24\pm0.13$ dex). The offset
between the two derived masses is expected since the former ones are
derived using the \citet{Chabrier:2003p3777} IMF, which produces
masses of the order of 0.55 lower than the adopted in the current
implementation of Pipe3D \citet{Salpeter:1955p3438}, as indicated
before. Once removed this offset the agreement between the different
estimations of the masses is remarkable good taking into account the
overall photometric accuracy of the CALIFA datacubes
\citep[$\sim$5\%,][]{rgb15}. \citet{catalan15} derived the star
formation rate for a subset of the CALIFA galaxies, 147 of them
included in the DR2. The SFR was estimated using the H$\alpha$ flux
\citep{kennicutt98} derived over an integrated elliptical aperture,
performing a global dust attenuation correction using the
H$\alpha$/H$\beta$ ratio, and an aperture correction for those
galaxies larger than the CALIFA FoV. Although there is a clear
correlation between the SFRs derived here and those reported by
\citet{catalan15}, there is a non-negligible dispersion around the
one-to-one relation and a trend towards lower values compared with
those derived by {\sc Pipe3D} is present ($\Delta$log(SFR)
=0.12$\pm$0.26 dex). In that regard, we should note that since
\citet{catalan15} carried out a comparison with other SFR tracers that
could only be measured as global properties, the ionized gas
extinction values and the extinction corrected SFR were derived using
aperture corrected integrated spectra. In our case, the SFR were
computed spaxel-by-spaxel, which commonly require slightly different
(local) SFR recipes \citep{calz01}, and were not corrected for the
fraction of SFR coming from outside of the PPAK FoV. These differences
could by themselves explain the dispersion and offset found for a
fraction of the targets in our sample. Finally, \citet{sanchez13}
derived the SFR by co-adding the H$\alpha$ flux of the \ion{H}{ii}
regions within the FoV of the galaxies, performing a global dust
attenuation correction using the average dust extinction derived for
those regions in each galaxy.  Although there is a clear offset
between the SFR derived, the dispersion around this offset is rather
low ($\Delta log(SFR)=-$0.23$\pm$0.11 dex). Two reasons account for
this difference: (i) first, in \citet{sanchez13} it was assumed
a relation between the SFR and the H$\alpha$ luminosity of:

\begin{equation}\label{eq:SFR}
SFR{\rm (M_\cdot yr^{-1})} = 8.9 \times 10^{\rm -42}~L_{\rm H_\alpha}(erg~s^{-1}) 
\end{equation}

\noindent while in here we assumed the more standard one of:

\begin{equation}\label{eq:SFR_Ha}
SFR{\rm (M_\cdot yr^{-1})} = 7.9 \times 10^{\rm -42}~L_{\rm H_\alpha}(erg~s^{-1}) 
\end{equation}

\noindent and (ii) in \citet{sanchez13} it was used the Hyperleda
distance modulus \citep[][, \url{http://leda.univ-lyon1.fr}]{patu03}, instead
of the ones derived by the estimated reshift. In general, we consider that the properties derived using
{\sc Pipe3D} present a good agreement with the previous reported
values when taking into account the different nature of the analysis.

\begin{figure*}[!t]
  \includegraphics[angle=270,width=1\linewidth]{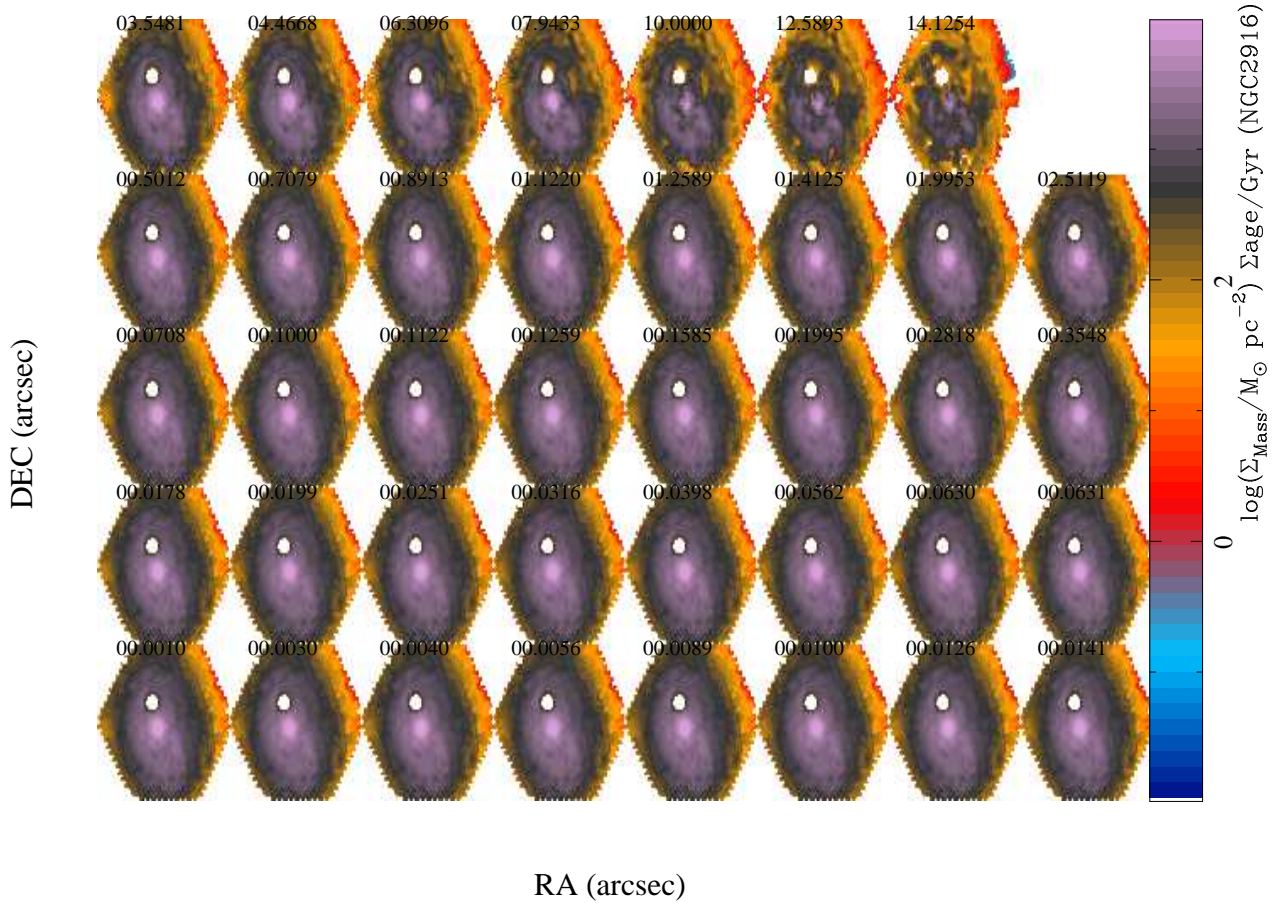}
  \caption{Cumulative stellar mass density at different lookback times derived from the analysis of the stellar population of NGC 2916 }\label{fig:SMA}
\end{figure*}

The dataproduct cubes from which we have derived these integrated
properties, as described in Sec \ref{pack}, are freely accessible at the
following FTP address
\url{ftp://ftp.caha.es/CALIFA/dataproducts/DR2/Pipe3D}, including a
table summarizing the properties of these
galaxies\footnote{\url{ftp://ftp.caha.es/CALIFA/dataproducts/DR2/Pipe3D/table_DR2_Pipe3D.csv}}.
The H$\alpha$ intensity maps for all the galaxies  analyzed with {\sc Pipe3D}
have been included in  Appendix \ref{app:Ha_maps}, illustrating the content of the 
dataproduct cubes delivered. We may notice that, up to our detection limit, all
the galaxies present ionized gas, with a distribution that reflects
the nature of the ionization. In spiral galaxies the gas presents the
typical clumpy distribution associated with the \ion{H}{ii} regions,
that follows the spiral arms. In earlier type galaxies the ionization
is dominated by a low-intensity/diffuse component, most probably
associated with pAGB stars \citep[e.g.][]{papa13}, and/or an AGN.

However, the  dataproducts distributed comprise similar information
for several ion species, not only H$\alpha$. Fig. \ref{fig:Ha_cubes}
shows the intensity maps of all the  emission lines analyzed  for NGC 2916 listed in
Table 1. For most of the weak emission lines the maps 
just show a noisy pattern, as expected since not all the emission lines are bright enough everywhere.

Finally, along this article we have illustrated the average quantities
derived from the analysis of the stellar population: luminosity
weighted age and metallicity, and stellar dust
attenuation. However, as indicated in PaperI {\sc Fit3D} provides
 the full star formation and chemical enrichment histories { (e.g., Fig.7 \& 9 of that article)}, and
when implemented in {\sc Pipe3D} it also provides  the corresponding
spatially resolved versions of both dataproducts.  As an example of this
analysis we present in Figure \ref{fig:SMA} the cumulative stellar
mass density ($\Sigma_{*,t}$) along  cosmological time, i.e., the stellar mass
density as a function of lookback time, derived from the stellar population
analysis of the CALIFA V500 setup data of NGC 2916. This
$\Sigma_{*,t}$ distribution has been analyzed to determine whether
galaxies growth their stellar mass inside-out or out-side, and the
difference in mass assembly depending on galaxy properties
\citep[e.g.][ Ibarra-Mede et al., submitted]{eperez13}. { We should note
here that this procedure is unable to determine where those stars were actually formed. Thus, it is unsensitive to radial movements. However, by describing the spatial distribution of stellar mass at different epochs it shows how the mass is assambled, irrespective of the origin of those stars (in situ starformation or migration).}

\subsection{The properties at the effective radius}
\label{Re}

Different previous studies have reported that some properties of both
stellar populations and ionized gas at the effective radius are
representative of the average properties of galaxies
\citep[e.g.][]{mous10,sanchez13,rosa15a}. As an example of the
practical use of the  dataset analyzed we will explore this result here,
by comparing five of the main dataproducts derived by {\sc Pipe3D}:
The luminosity weighted stellar log age and metallicity, the dust
attenuation of the stellar component, the dust attenuation derived for
the ionized gas, and the gas phase oxygen abundance.

\begin{figure}[!t]
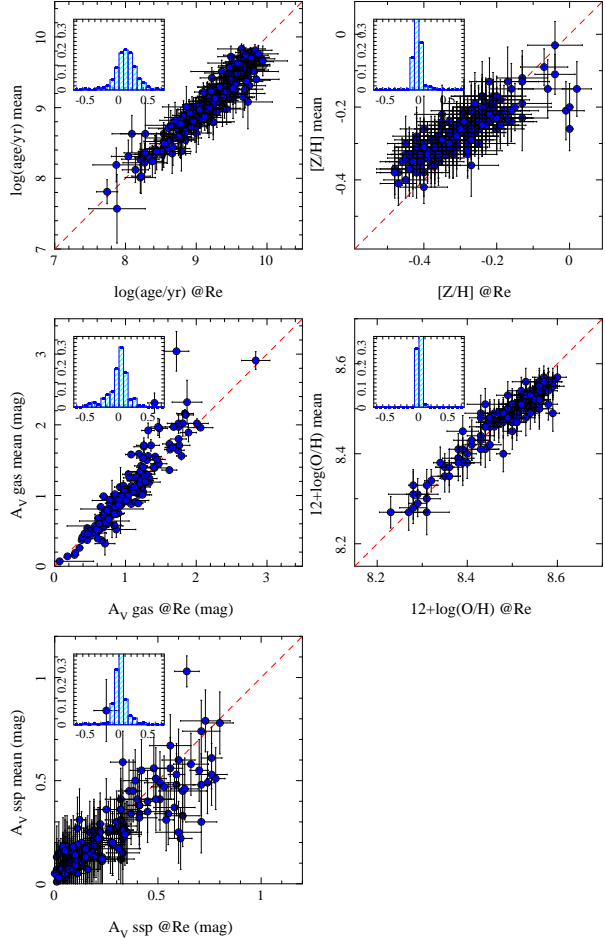

  \includegraphics[angle=270,width=0.5\linewidth]{figs/age_DR2_mean_Re.ps}\includegraphics[angle=270,width=0.5\linewidth]{figs/ZH_DR2_mean_Re.ps}
  \includegraphics[angle=270,width=0.5\linewidth]{figs/Av_gas_DR2_mean_Re.ps}\includegraphics[angle=270,width=0.5\linewidth]{figs/OH_DR2_mean_Re.ps}
  \includegraphics[angle=270,width=0.5\linewidth]{figs/Av_ssp_DR2_mean_Re.ps}
  \caption{Comparison of the average value and the value at the effective radius for four properties of the galaxies: {\it top-left:} Luminosity weighted log age, {\it top-right:} Luminosity weighted metalliciy, {\it central-left:} Ionized gas dust attenuation, and {\it central-right:} gas phase oxygen abundance. {\it bottom panel:} Dust attenuation of the stellar population. For this later parameter the errors have been truncated to a maximum value of 0.15 mag, for the sake of clarity.  In each panel the inset shows the normalized histogram of the difference between the two estimations of the  parameter.}\label{fig:Re}
\end{figure}

The average values have been derived as the mean values across the
optical extension of the galaxies, without  weighing, or
rejection process for the stellar population as derived directly from
{\sc FIT3D} (PaperI  {, Sec. 2.3}) and shown in Fig. \ref{fig:age}. The gas phase dust
attenuation was derived across the optical extension of the galaxy
based on the spaxel-by-spaxel H$\alpha$ to H$\beta$ line ratio and
using the prescriptions described in Sec. \ref{test}.  For the
gas phase oxygen abundance we adopted the O3N2 line ratio and the
calibrator proposed by \citet{marino13}. This line ratio involves the
use of flux intensities of the following emission lines:
[\ion{O}{iii}], [\ion{N}{ii}], H$\alpha$, and H$\beta$. For both, the
dust attenuation and the oxygen abundance, a 3$\sigma$ selection
criterion is applied to all the emission lines involved in the
derivation of the parameters, spaxel-by-spaxel. Once the map
of the different properties is derived, the mean value was derived without any
further rejection criteria.  Finally, we exclude those average values
derived for galaxies that do not have at least 50 spaxels (an area of
$\sim$7 arcsec$^2$) fulfilling the accuracy criteria indicated above
to derive the considered parameter. The individual values derived
by this analysis are listed in Appendix \ref{app:char}, Table \ref{DR2_mean}.2 .

\begin{table}\label{mean_Re}
\begin{center}
\caption[Average and effective parameters.]{Average and effective parameters.}
\begin{tabular}{ccccr}\hline\hline
Par. & $r$ & $\alpha$ & $\chi^2/\nu$ & $\Delta$par\\
 (1)   & (2) & (3) & (4)     & (5)          \\
\hline
log(age)     & 0.98 & 0.88$\pm$0.07 & 0.87 & 0.11$\pm$0.16\\
$[$Z/H$]$        & 0.84 & 0.53$\pm$0.07 & 1.36 & $-$0.03$\pm$0.06\\
A$_{\rm V,gas}$ & 0.93 & 1.15$\pm$0.17 & 7.10 & 0.00$\pm$0.21\\
12+log(O/H)  & 0.95 & 0.81$\pm$0.01 & 1.01 & 0.01$\pm$0.03\\
A$_{\rm V,ssp}$ & 0.83 & 0.74$\pm$0.08 & 0.41 & 0.01$\pm$0.11\\
\hline                    
 \end{tabular}
 \end{center}
(1) Compared parameter; (2) correlation coefficient; (3) slope of the
correlation coefficient; (4) reduced $\chi^2$ with respect to the one-to-one relation; (5) offset with respect
to the one-to-one relation.
\end{table}

The values at the effective radius have been derived as the average
within an annular ring of 0.75-1.25 effective
radius from the center of the galaxy, after deprojecting the 2D distribution of the
 parameter. For the effective radius we
adopted the value described in \citet{walcher14}.  For the
deprojection we adopted the procedure described in \citet{sanchez14},
using the average semi-minor to semi-major axis ratio and ellipticity
derived from the isophotal analysis described in the Appendix of that
article. An intrinsic ellipticity of $q_o$=0.13 has been taken into
account in the deprojection procedure. Finally, we exclude those
average values at the effective radius derived for galaxies that do
not have at least 10 spaxels in the  annular ring fulfilling
the accuracy criteria indicated above to evaluate the 
parameter.  The individual values derived by this analysis are listed
in Appendix \ref{app:char}, Table \ref{DR2_Re}.3 .

Figure \ref{fig:Re} shows the comparison between the average values
derived across the optical extension of the galaxies and the values at
the effective radius as described before. For each of the 
parameters we perform a least square linear regression, deriving both
the correlation coefficient and the slope of the 
correlation.  In addition, we derive the reduced $\chi^2/\nu$ with
respect to the one-to-one relation, and the mean and standard deviation
of the difference between the two parameters. 

The results of this analysis are included in Table 7.  As
expected from the inspection of Fig. \ref{fig:Re}, all the 
parameters analyzed show a clear and strong correlation, with large
correlation coefficients. In almost all the cases the
correlation has a slope near to the one-to-one relation, except the
stellar abundance ([Z/H]), which seems to have a lower slope.  When
compared with the one-to-one relation, three parameters (log(age),
[Z/H], and 12+log(O/H)), show a $\chi^2/\nu$ near to one,
indicating that their correlation is completely compatible with the unity
relation. For the gas phase dust attenuation, the $\chi^2/\nu$ is too
large, taking into account that the two derivations of this parameter
present a strong correlation with a slope very close to one. We
consider that this is due to a clear underestimation of the error of
A$_{\rm V}$, that requires a re-evaluation, rather than an indication
of a clear deviation from the one-to-one relation.  Indeed, including
a systematic error of 0.15 dex in both parameters, the $\chi^2/\nu$
reaches a value near to one. We find the opposite effect for the 
stellar dust attenuation. It is clear from Fig. \ref{fig:Re} that the
estimated errors are much too large. This is maybe due
to the inclusion of the standard deviation with respect to the mean
values in the error budget, which indicates that the individual
derivation of the dust attenuation is not very stable at each spatial
bin. A global reduction of the estimated errors in these parameters by
a $\sim$50\% makes the $\chi^2/\nu$ reach a value near to one.  In
general, the difference between each pair of explored parameters is
almost compatible with zero, or show very small offsets
compared with the overall dispersion and the individual estimated errors.

This study demonstrates that at least for the parameters analyzed  the
values derived at the effective radius are very representative of the
average values across the entire optical extension of galaxies.  In
other words, it seems that the effective radius is indeed effective,
as already noticed by \citet{sanchez13} and \citet{rosa15a}.  This
result is particularly important for those IFU surveys that do not
cover the entire optical extension for all the sampled
galaxies but whose FoV reach at least one effective radius.

\section{Summary and Conclusions}
\label{sum}

Along this article we have described the analysis sequence of {\sc
  Pipe3D}, a pipeline based on {\sc FIT3D}, developed to extract the
main properties of the ionized gas and stellar populations from IFS
data. In particular we present the different steps and describe the
intermediate and final dataproducts for a set of datacubes extracted
from the MaNGA and CALIFA surveys. This includes: (i) cube
pre-processing; (ii) analysis of the central spectrum; (iii) spatial
binning  to achieve the required S/N level per spectrum,
comparing with the most frequently used binning procedure in the
field; (iv) detailed analysis of the stellar population based on 
multi-SSP fitting in each spatial bin, including a description of the
derived dataproducts; (v) analysis of the emission lines in the
spatially binned spectra; (vi) the required dezonification to recover a
spaxel-wise model of the underlying stellar population and a emission line pure
datacube; (vii) the spaxel-wise analysis of the strong and weak
emission lines, including a comparison between the derived parameters;
(viii) the analysis of the stellar indices, showing a
comparison/correspondence with the results derived from the multi-SSP
fitting procedure; and finally (ix) a description of the format of the
derived dataproducts provided by each individual step of the analysis.
In summary, we present here the current status of the {\sc Pipe3D}
pipeline, currently implemented to analyze the data from the
three major on-going IFU surveys: CALIFA, MaNGA, and SAMI. 
{\sc Pipe3D} produces reliable estimations (as demonstrated in PaperI {, Sec. 3}) 
of the parameters analyzed that are coherent within the differences between the 
instrumental setups for the three surveys. The use of a single
data format for the dataproducts derived  will help to make comparisons
between the three surveys in a coherent and simple way. {\sc Pipe3D}
has already demonstrated its scientific validity since it was already used
to analyze: (i) the effects of galaxy interaction in the enhancement
of the star formation rate and the onset of galactic outflows, using
CALIFA data \citep{jkbb15b}; (ii) the exploration of the local version
of the star forming main sequence in galaxies using CALIFA data
(Cano-D\'\i az et al. in press); and (iii) the exploration of the evidence
of the inside-out scenario in the mass assembly history of galaxies,
using MaNGA data (Ibarra Mede et al. in preparation).

 As a practical implementation we present the dataproducts provided by
 {\sc Pipe3D} for the publicly accessible datacubes of the 200
 galaxies that comprise the V500 setup of the CALIFA DR2. We
 illustrate the content of the dataproducts delivered  showing the
 H$\alpha$ intensity maps for all of them, as well as the intensity
 maps of all the  emission lines analyzed for NGC 2916, together with
 the stellar mass assembly history of this galaxy. For all these
 galaxies we list a set of parameters that characterize their
 properties, including the integrated stellar mass and star formation
 within the aperture of the IFU, the mean luminosity weighted log age
 and metallicity, the dust attenuation for the stellar populations,
 as well as the oxygen abundance and dust attenuation of the ionized
 gas, both derived across the full optical extension of the galaxies
 and at the effective radius. Finally, we demonstrate that, as
 suggested by previous studies, the values at the effective radius are
 indeed characteristic of the average values of the
 parameters across the entire optical extension of the galaxies.

\appendices

\section{H$\alpha$ intensity maps}
\label{app:Ha_maps}

Figure \ref{fig:Ha_maps} shows the individual H$\alpha$ intensity maps
for all the  galaxies analyzed, derived using {\sc Pipe3D} as described
in Section \ref{weak}. As indicated above, similar intensity maps and
their corresponding errors and kinematic properties have been derived
for all the emission lines listed in Table 1, and included in the {\tt
  flux\_elines.OBJ.cube.fits.gz} datacube (where {\tt OBJ} is the name
of the corresponding object).

\begin{figure*}[!t]
\includegraphics[width=1\linewidth]{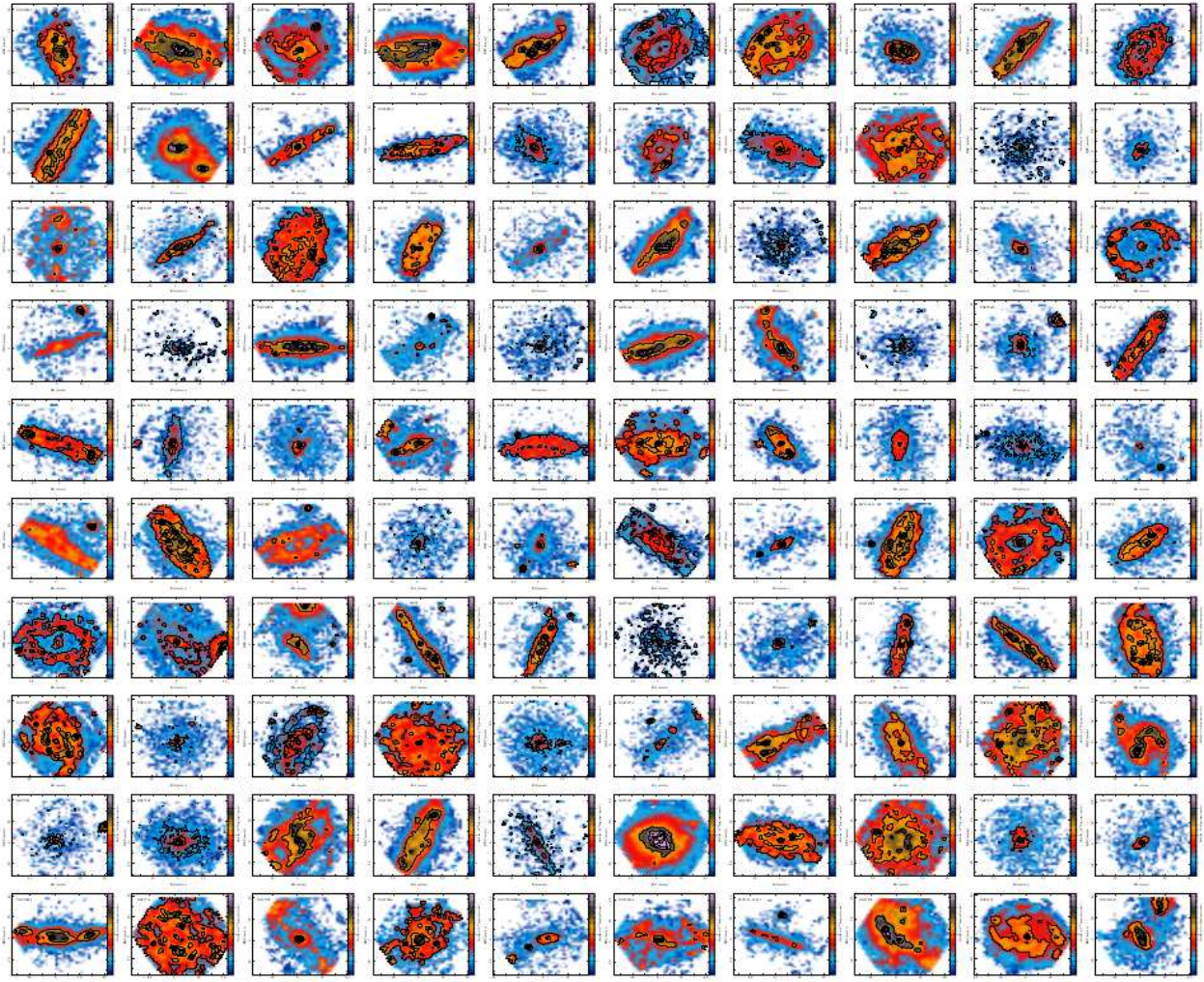}
  \caption{H$\alpha$ intensity maps of the galaxies listed in Table 7, in units of \fuden. The color scale is the same for all the galaxies. The contours are different for each galaxy. The first contour represents the median intensity level, and each successive contour is off by 1/5th of the median value.}\label{fig:Ha_maps}
\end{figure*}

\addtocounter{figure}{-1}\clearpage

\begin{figure*}[!t]
\includegraphics[width=1\linewidth]{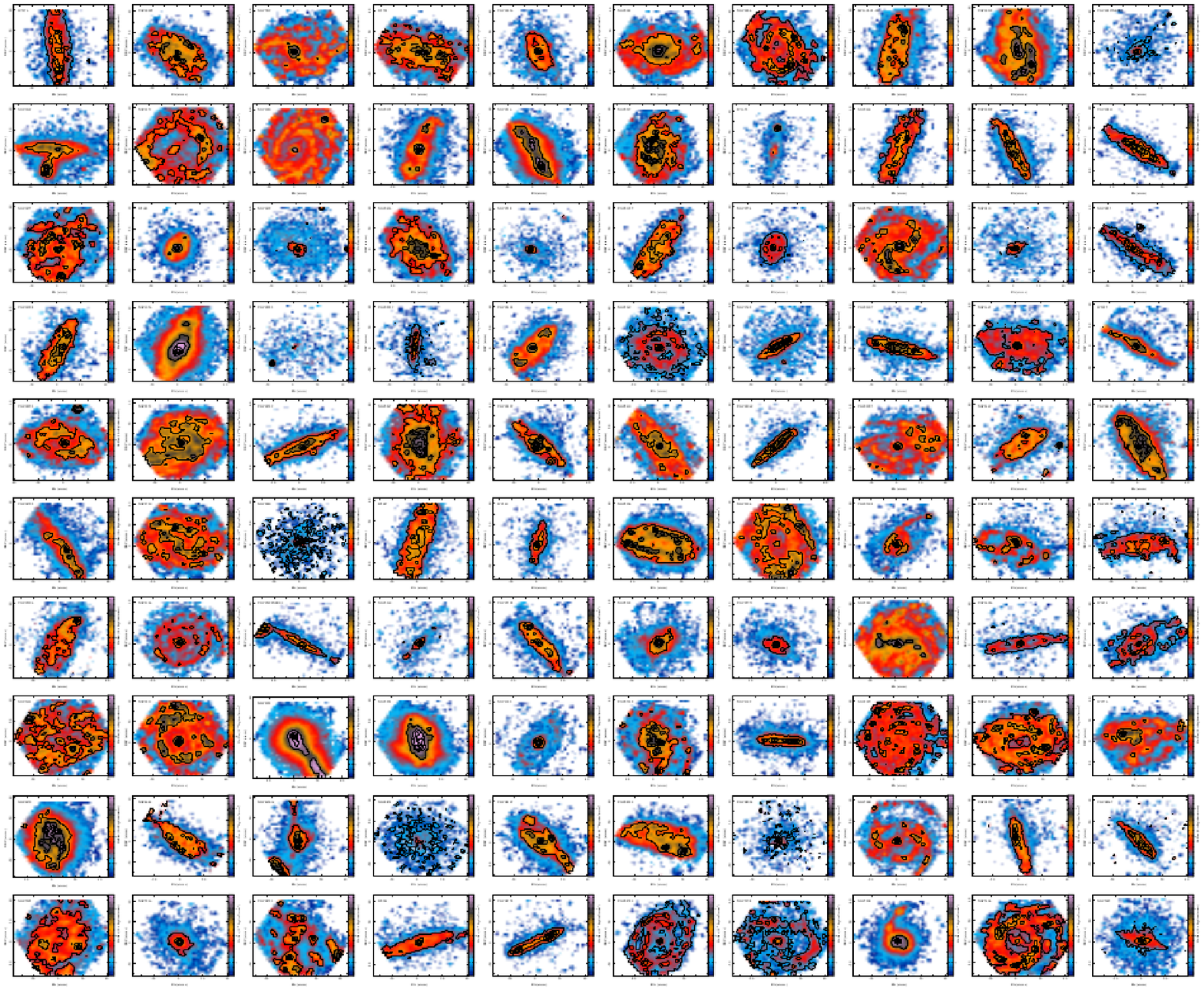}
  \caption{Continued}\label{fig:Ha_maps}
\end{figure*}

\clearpage

\section{Characteristic properties of the galaxies.}
\label{app:char}

Table \ref{DR2_obj}.1 shows the list of individual galaxies
distributed in the CALIFA DR2,  including their nominal
names and the corresponding CALIFA-ID, together with the coordinates.
In addition it has been included a few of the main global properties
derived by {\tt Pipe3D} as described in Section \ref{test}.

Table \ref{DR2_mean}.2 and \ref{DR2_Re}.3 list the parameters explored in
Sec. \ref{Re} for all the galaxies included in the CALIFA DR2. Each
table includes the average values and the  errors, derived by
considering both the individual errors and the standard deviation of
the distribution. Table \ref{DR2_mean}.2 lists the average parameters
across the optical extension of the galaxies, while Table \ref{DR2_Re}.3
lists the corresponding parameter derived at the effective radius. In
most  cases the individual errors dominate the error budget,
being the dust attenuation of the stellar component the exception. In
this case the fluctuations along the average values are much larger
than the typical individual estimation of the error.

An electronic version of these tables can be found in the FTP:
({\url{ftp://ftp.caha.es/CALIFA/dataproducts/DR2/Pipe3D/}}).

\begin{table*}\label{DR2_obj}
\begin{center}
\caption[Integrated properties of the CALIFA DR2 galaxies.]{Integrated properties of the CALIFA DR2 galaxies.}
\begin{tabular}{lcccclr}\hline\hline
NAME & CAL-ID & RA & DEC & $z$ & log(M/M$\odot$) & log(SFR/M$\odot$/yr)\\
\hline
IC5376 & 001 & 00:01:19.77 & +34:31:32.52 & 0.0166 & 10.65 $\pm$ 0.10 & 0.006 $\pm$ 0.105\\
UGC00005 & 002 & 00:03:05.63 & -01:54:49.67 & 0.0240 & 11.16 $\pm$ 0.08 & 0.867 $\pm$ 0.066\\
NGC7819 & 003 & 00:04:24.50 & +31:28:19.20 & 0.0165 & 10.61 $\pm$ 0.08 & 0.412 $\pm$ 0.067\\
IC1528 & 005 & 00:05:05.37 & -07:05:36.23 & 0.0125 & 10.54 $\pm$ 0.11 & 0.197 $\pm$ 0.064\\
UGC00036 & 007 & 00:05:13.87 & +06:46:19.20 & 0.0208 & 11.06 $\pm$ 0.09 & 0.186 $\pm$ 0.134\\
NGC0001 & 008 & 00:07:15.86 & +27:42:29.16 & 0.0149 & 10.82 $\pm$ 0.10 & 0.649 $\pm$ 0.063\\
NGC0036 & 010 & 00:11:22.29 & +06:23:21.84 & 0.0197 & 11.22 $\pm$ 0.10 & 0.535 $\pm$ 0.096\\
\hline                    
 \end{tabular}
 \end{center}
\end{table*}

\addtocounter{table}{-1}\clearpage

\begin{table*}\label{DR2_obj}
\begin{center}
\caption[Integrated properties of the CALIFA DR2 galaxies.]{Integrated properties of the CALIFA DR2 galaxies. ({\it continued})}
\begin{tabular}{lcccclr}\hline\hline
NAME & CAL-ID & RA & DEC & $z$ & log(M/M$\odot$) & log(SFR/M$\odot$/yr)\\
\hline
UGC03253 & 146 & 05:19:41.87 & +84:03:09.35 & 0.0136 & 10.63 $\pm$ 0.10 & 0.231 $\pm$ 0.106\\
NGC2253 & 147 & 06:43:41.83 & +65:12:23.04 & 0.0118 & 10.80 $\pm$ 0.10 & 0.495 $\pm$ 0.061\\
UGC03539 & 148 & 06:48:53.99 & +66:15:41.76 & 0.0107 & 9.83 $\pm$ 0.12 & -0.169 $\pm$ 0.089\\
NGC2347 & 149 & 07:16:04.07 & +64:42:40.67 & 0.0146 & 11.04 $\pm$ 0.10 & 0.542 $\pm$ 0.066\\
UGC03899 & 150 & 07:32:37.75 & +35:36:52.19 & 0.0128 & 9.59 $\pm$ 0.13 & -0.238 $\pm$ 0.072\\
NGC2410 & 151 & 07:35:02.25 & +32:49:19.55 & 0.0153 & 11.03 $\pm$ 0.09 & 0.549 $\pm$ 0.112\\
UGC03969 & 153 & 07:41:14.35 & +27:36:50.76 & 0.0269 & 10.73 $\pm$ 0.10 & 0.551 $\pm$ 0.084\\
\hline                    
 \end{tabular}
 \end{center}
\end{table*}

\addtocounter{table}{-1} \clearpage

\begin{table*}\label{DR2_obj}
\begin{center}
\caption[Integrated properties of the CALIFA DR2 galaxies.]{Integrated properties of the CALIFA DR2 galaxies. ({\it continued})}
\begin{tabular}{lcccclr}\hline\hline
NAME & CAL-ID & RA & DEC & $z$ & log(M/M$\odot$) & log(SFR/M$\odot$/yr)\\
\hline
NGC4470 & 548 & 12:29:37.77 & +07:49:27.12 & 0.0076 & 10.23 $\pm$ 0.09 & -0.009 $\pm$ 0.053\\
NGC4644 & 569 & 12:42:42.67 & +55:08:43.79 & 0.0164 & 10.68 $\pm$ 0.11 & 0.089 $\pm$ 0.089\\
NGC4676A & 577 & 12:46:10.10 & +30:43:54.83 & 0.0222 & 10.85 $\pm$ 0.10 & 0.524 $\pm$ 0.087\\
NGC4874 & 592 & 12:59:35.71 & +27:57:33.47 & 0.0242 & 11.66 $\pm$ 0.08 & -0.757 $\pm$ 0.601\\
UGC08107 & 593 & 12:59:39.76 & +53:20:28.32 & 0.0273 & 11.20 $\pm$ 0.09 & 0.893 $\pm$ 0.078\\
UGC08231 & 606 & 13:08:37.55 & +54:04:27.83 & 0.0080 & 9.64 $\pm$ 0.12 & -0.395 $\pm$ 0.060\\
UGC08234 & 607 & 13:08:46.51 & +62:16:18.12 & 0.0270 & 11.39 $\pm$ 0.10 & -0.955 $\pm$ 0.768\\
\hline                    
 \end{tabular}
 \end{center}
\end{table*}

\addtocounter{table}{-1} \clearpage

\begin{table*}\label{DR2_obj}
\begin{center}
\caption[Integrated properties of the CALIFA DR2 galaxies.]{Integrated properties of the CALIFA DR2 galaxies. ({\it continued})}
\begin{tabular}{lcccclr}\hline\hline
NAME & CAL-ID & RA & DEC & $z$ & log(M/M$\odot$) & log(SFR/M$\odot$/yr)\\
\hline
NGC6032 & 820 & 16:03:01.12 & +20:57:21.23 & 0.0143 & 10.65 $\pm$ 0.13 & -0.056 $\pm$ 0.230\\
UGC10205 & 822 & 16:06:40.17 & +30:05:56.75 & 0.0216 & 11.08 $\pm$ 0.10 & 0.381 $\pm$ 0.198\\
NGC6063 & 823 & 16:07:12.98 & +07:58:44.40 & 0.0094 & 10.36 $\pm$ 0.11 & -0.274 $\pm$ 0.083\\
IC1199 & 824 & 16:10:34.34 & +10:02:25. & 0.0156 & 10.77 $\pm$ 0.10 & 0.162 $\pm$ 0.070\\
NGC6081 & 826 & 16:12:56.85 & +09:52:01.56 & 0.0166 & 11.03 $\pm$ 0.08 & -0.326 $\pm$ 0.282\\
UGC10331 & 828 & 16:17:21.12 & +59:19:12.35 & 0.0150 & 10.26 $\pm$ 0.10 & 0.537 $\pm$ 0.054\\
NGC6125 & 829 & 16:19:11.54 & +57:59:02.75 & 0.0157 & 11.35 $\pm$ 0.09 & -1.269 $\pm$ 0.577\\
\hline                    
 \end{tabular}
 \end{center}
\end{table*}

\addtocounter{table}{-1} \clearpage

\begin{table*}\label{DR2_obj}
\begin{center}
\caption[Integrated properties of the CALIFA DR2 galaxies.]{Integrated properties of the CALIFA DR2 galaxies. ({\it continued})}
\begin{tabular}{lcccclr}\hline\hline
NAME & CAL-ID & RA & DEC & $z$ & log(M/M$\odot$) & log(SFR/M$\odot$/yr)\\
\hline
UGC11649 & 872 & 20:55:27.62 & -01:13:30.72 & 0.0124 & 10.75 $\pm$ 0.09 & -0.349 $\pm$ 0.148\\
NGC7025 & 874 & 21:07:47.32 & +16:20:09.23 & 0.0162 & 11.50 $\pm$ 0.09 & 0.065 $\pm$ 0.293\\
UGC11717 & 877 & 21:18:35.42 & +19:43:07.31 & 0.0209 & 11.16 $\pm$ 0.09 & 0.764 $\pm$ 0.166\\
MCG-01-54-016 & 878 & 21:25:59.97 & -03:48:32.40 & 0.0095 & 9.43 $\pm$ 0.12 & -0.434 $\pm$ 0.093\\
UGC11792 & 880 & 21:42:12.69 & +05:36:55.44 & 0.0158 & 10.34 $\pm$ 0.12 & 0.264 $\pm$ 0.115\\
NGC7194 & 881 & 22:03:30.93 & +12:38:12.48 & 0.0268 & 11.56 $\pm$ 0.09 & -0.462 $\pm$ 0.636\\
UGC11958 & 883 & 22:14:46.87 & +13:50:27.23 & 0.0259 & 11.32 $\pm$ 0.08 & -0.395 $\pm$ 0.409\\
\hline                    
 \end{tabular}
 \end{center}
\end{table*}
\clearpage


\begin{table*}\label{DR2_mean}
\begin{center}
\caption[Average properties of the CALIFA DR2 galaxies.]{Average properties of the CALIFA DR2 galaxies.}
\begin{tabular}{lrrrrr}\hline\hline
NAME & log(age/yr) & [Z/H] & A$_{\rm V}^{ssp}$ & 12+log(O/H) & A$_{\rm V}^{gas}$\\
\hline
IC5376 & 8.99 $\pm$ 0.51 & -0.35 $\pm$ 0.16 & 0.56 $\pm$ 1.02 & 8.50 $\pm$ 0.05 & 1.24 $\pm$ 0.03\\
UGC00005 & 8.97 $\pm$ 0.42 & -0.31 $\pm$ 0.13 & 0.26 $\pm$ 0.75 & 8.54 $\pm$ 0.06 & 1.30 $\pm$ 0.04\\
NGC7819 & 8.68 $\pm$ 0.45 & -0.29 $\pm$ 0.11 & 0.10 $\pm$ 0.53 & 8.47 $\pm$ 0.07 & 0.90 $\pm$ 0.08\\
IC1528 & 8.75 $\pm$ 0.54 & -0.34 $\pm$ 0.14 & 0.22 $\pm$ 0.74 & 8.47 $\pm$ 0.08 & 0.77 $\pm$ 0.03\\
UGC00036 & 9.42 $\pm$ 0.26 & -0.17 $\pm$ 0.12 & 0.10 $\pm$ 0.61 & 8.52 $\pm$ 0.06 & 1.33 $\pm$ 0.09\\
NGC0001 & 8.89 $\pm$ 0.49 & -0.35 $\pm$ 0.13 & 0.51 $\pm$ 0.89 & 8.55 $\pm$ 0.06 & 1.59 $\pm$ 0.06\\
NGC0036 & 9.30 $\pm$ 0.42 & -0.34 $\pm$ 0.14 & 0.21 $\pm$ 0.69 & 8.53 $\pm$ 0.06 & 1.31 $\pm$ 0.03\\
\hline
\end{tabular}
\end{center}
\end{table*}

\addtocounter{table}{-1} \clearpage

\begin{table*}\label{DR2_mean_2}
\begin{center}
\caption[Average properties of the CALIFA DR2 galaxies.]{Average properties of the CALIFA DR2 galaxies. ({\it continued})}
\begin{tabular}{lrrrrr}\hline\hline
NAME & log(age/yr) & [Z/H] & A$_{\rm V}^{ssp}$ & 12+log(O/H) & A$_{\rm V}^{gas}$\\
\hline
UGC03253 & 8.29 $\pm$ 0.30 & -0.26 $\pm$ 0.10 & 1.19 $\pm$ 1.68 & 8.50 $\pm$ 0.07 & 1.12 $\pm$ 0.05\\
NGC2253 & 8.80 $\pm$ 0.37 & -0.36 $\pm$ 0.12 & 0.30 $\pm$ 0.67 & 8.55 $\pm$ 0.05 & 1.13 $\pm$ 0.04\\
UGC03539 & 8.63 $\pm$ 0.52 & -0.28 $\pm$ 0.14 & 0.74 $\pm$ 1.35 & 8.45 $\pm$ 0.07 & 1.30 $\pm$ 0.06\\
NGC2347 & 8.64 $\pm$ 0.54 & -0.33 $\pm$ 0.14 & 0.25 $\pm$ 0.61 & 8.50 $\pm$ 0.08 & 0.68 $\pm$ 0.02\\
UGC03899 & 8.01 $\pm$ 0.43 & -0.26 $\pm$ 0.13 & 0.25 $\pm$ 0.57 & 8.27 $\pm$ 0.08 & 0.55 $\pm$ 0.04\\
NGC2410 & 9.20 $\pm$ 0.29 & -0.27 $\pm$ 0.13 & 0.34 $\pm$ 0.93 & 8.49 $\pm$ 0.05 & 1.56 $\pm$ 0.10\\
UGC03969 & 9.09 $\pm$ 0.37 & -0.22 $\pm$ 0.16 & 0.58 $\pm$ 1.63 & 8.50 $\pm$ 0.05 & 1.64 $\pm$ 0.08\\
\hline
\end{tabular}
\end{center}
\end{table*}

\addtocounter{table}{-1} \clearpage

\begin{table*}\label{DR2_mean_3}
\begin{center}
\caption[Average properties of the CALIFA DR2 galaxies.]{Average properties of the CALIFA DR2 galaxies. ({\it continued})}
\begin{tabular}{lrrrrr}\hline\hline
NAME & log(age/yr) & [Z/H] & A$_{\rm V}^{ssp}$ & 12+log(O/H) & A$_{\rm V}^{gas}$\\
\hline
NGC4470 & 9.01 $\pm$ 0.29 & -0.38 $\pm$ 0.14 & 0.10 $\pm$ 0.72 & 8.46 $\pm$ 0.05 & 0.48 $\pm$ 0.01\\
NGC4644 & 8.98 $\pm$ 0.43 & -0.13 $\pm$ 0.13 & 0.16 $\pm$ 0.74 & 8.55 $\pm$ 0.06 & 1.14 $\pm$ 0.14\\
NGC4676A & 9.30 $\pm$ 0.28 & -0.20 $\pm$ 0.13 & 0.26 $\pm$ 1.40 & 8.50 $\pm$ 0.05 & 1.42 $\pm$ 0.10\\
NGC4874 & 9.83 $\pm$ 0.14 & -0.21 $\pm$ 0.17 & 0.00 $\pm$ 0.15 & $---$ & $---$\\
UGC08107 & 9.16 $\pm$ 0.35 & -0.26 $\pm$ 0.15 & 0.34 $\pm$ 1.27 & 8.47 $\pm$ 0.05 & 1.51 $\pm$ 0.08\\
UGC08231 & 8.12 $\pm$ 0.36 & -0.26 $\pm$ 0.12 & 0.10 $\pm$ 0.39 & 8.27 $\pm$ 0.07 & 0.14 $\pm$ 0.01\\
UGC08234 & 9.10 $\pm$ 0.42 & -0.17 $\pm$ 0.13 & 0.15 $\pm$ 0.61 & $---$ & $---$\\
\hline
\end{tabular}
\end{center}
\end{table*}

\addtocounter{table}{-1} \clearpage

\begin{table*}\label{DR2_mean_4}
\begin{center}
\caption[Average properties of the CALIFA DR2 galaxies.]{Average properties of the CALIFA DR2 galaxies. ({\it continued})}
\begin{tabular}{lrrrrr}\hline\hline
NAME & log(age/yr) & [Z/H] & A$_{\rm V}^{ssp}$ & 12+log(O/H) & A$_{\rm V}^{gas}$\\
\hline
NGC6032 & 8.94 $\pm$ 0.44 & -0.20 $\pm$ 0.13 & 0.15 $\pm$ 0.72 & 8.49 $\pm$ 0.08 & 1.07 $\pm$ 0.16\\
UGC10205 & 9.14 $\pm$ 0.42 & -0.21 $\pm$ 0.14 & 0.36 $\pm$ 1.18 & 8.48 $\pm$ 0.05 & 2.02 $\pm$ 0.13\\
NGC6063 & 9.04 $\pm$ 0.38 & -0.38 $\pm$ 0.14 & 0.10 $\pm$ 0.58 & 8.48 $\pm$ 0.07 & 0.62 $\pm$ 0.01\\
IC1199 & 9.03 $\pm$ 0.64 & -0.28 $\pm$ 0.16 & 0.25 $\pm$ 0.67 & 8.56 $\pm$ 0.06 & 0.99 $\pm$ 0.05\\
NGC6081 & 9.53 $\pm$ 0.29 & -0.22 $\pm$ 0.12 & 0.11 $\pm$ 0.66 & $---$ & 1.59 $\pm$ 0.30\\
UGC10331 & 8.73 $\pm$ 0.42 & -0.38 $\pm$ 0.14 & 0.45 $\pm$ 0.88 & 8.39 $\pm$ 0.05 & 0.93 $\pm$ 0.06\\
NGC6125 & 9.78 $\pm$ 0.21 & -0.24 $\pm$ 0.19 & 0.05 $\pm$ 0.54 & $---$ & $---$\\
\hline
\end{tabular}
\end{center}
\end{table*}

\addtocounter{table}{-1} \clearpage

\begin{table*}\label{DR2_mean_5}
\begin{center}
\caption[Average properties of the CALIFA DR2 galaxies.]{Average properties of the CALIFA DR2 galaxies. ({\it continued})}
\begin{tabular}{lrrrrr}\hline\hline
NAME & log(age/yr) & [Z/H] & A$_{\rm V}^{ssp}$ & 12+log(O/H) & A$_{\rm V}^{gas}$\\
\hline
UGC11649 & 9.22 $\pm$ 0.39 & -0.22 $\pm$ 0.12 & 0.05 $\pm$ 0.45 & 8.52 $\pm$ 0.07 & 0.61 $\pm$ 0.02\\
NGC7025 & 9.72 $\pm$ 0.21 & -0.27 $\pm$ 0.19 & 0.14 $\pm$ 1.28 & 8.51 $\pm$ 0.07 & 1.21 $\pm$ 0.10\\
UGC11717 & 8.63 $\pm$ 0.87 & -0.25 $\pm$ 0.16 & 0.36 $\pm$ 0.28 & 8.56 $\pm$ 0.05 & 1.89 $\pm$ 0.26\\
MCG-01-54-016 & 8.36 $\pm$ 0.55 & -0.26 $\pm$ 0.15 & 0.25 $\pm$ 0.66 & 8.31 $\pm$ 0.11 & 0.50 $\pm$ 0.03\\
UGC11792 & 8.38 $\pm$ 0.53 & -0.35 $\pm$ 0.15 & 1.20 $\pm$ 1.77 & 8.49 $\pm$ 0.06 & 1.97 $\pm$ 0.08\\
NGC7194 & 9.67 $\pm$ 0.23 & -0.13 $\pm$ 0.14 & 0.03 $\pm$ 0.39 & $---$ & 1.40 $\pm$ 0.88\\
UGC11958 & 9.63 $\pm$ 0.41 & -0.24 $\pm$ 0.16 & -0.01 $\pm$ 1.13 & $---$ & 0.57 $\pm$ 0.33\\
\hline
\end{tabular}
\end{center}
\end{table*}

\clearpage


\begin{table*}\label{DR2_Re}
\begin{center}
\caption[Properties of the CALIFA DR2 galaxies at the effective radius.]{Properties of the CALIFA DR2 galaxies at the effective radius.}
\begin{tabular}{lrrrrr}\hline\hline
NAME & log(age/yr) & [Z/H] & A$_{\rm V}^{ssp}$ & 12+log(O/H) & A$_{\rm V}^{gas}$\\
\hline
IC5376 & 9.37 $\pm$ 0.50 & -0.44 $\pm$ 0.13 & 0.48 $\pm$ 0.11 & 8.53 $\pm$ 0.04 & 1.34 $\pm$ 0.21\\
UGC00005 & 9.18 $\pm$ 0.28 & -0.37 $\pm$ 0.11 & 0.29 $\pm$ 0.09 & 8.57 $\pm$ 0.04 & 1.28 $\pm$ 0.18\\
NGC7819 & 8.81 $\pm$ 0.44 & -0.28 $\pm$ 0.14 & 0.06 $\pm$ 0.06 & 8.47 $\pm$ 0.07 & 0.82 $\pm$ 0.17\\
IC1528 & 9.03 $\pm$ 0.31 & -0.41 $\pm$ 0.14 & 0.19 $\pm$ 0.10 & 8.51 $\pm$ 0.04 & 0.77 $\pm$ 0.14\\
UGC00036 & 9.56 $\pm$ 0.44 & -0.24 $\pm$ 0.15 & 0.10 $\pm$ 0.06 & 8.53 $\pm$ 0.05 & 1.22 $\pm$ 0.25\\
NGC0001 & 9.18 $\pm$ 0.32 & -0.46 $\pm$ 0.12 & 0.49 $\pm$ 0.10 & 8.58 $\pm$ 0.05 & 1.19 $\pm$ 0.17\\
NGC0036 & 9.33 $\pm$ 0.44 & -0.37 $\pm$ 0.15 & 0.21 $\pm$ 0.09 & 8.53 $\pm$ 0.05 & 1.26 $\pm$ 0.22\\
\hline
\end{tabular}
\end{center}
\end{table*}

\addtocounter{table}{-1} \clearpage

\begin{table*}\label{DR2_Re}
\begin{center}
\caption[Properties of the CALIFA DR2 galaxies at the effective radius.]{Properties of the CALIFA DR2 galaxies at the effective radius. ({\it continued})}
\begin{tabular}{lrrrrr}\hline\hline
NAME & log(age/yr) & [Z/H] & A$_{\rm V}^{ssp}$ & 12+log(O/H) & A$_{\rm V}^{gas}$\\
\hline
UGC03253 & 8.34 $\pm$ 0.41 & -0.28 $\pm$ 0.13 & 1.22 $\pm$ 0.08 & 8.50 $\pm$ 0.07 & 1.11 $\pm$ 0.22\\
NGC2253 & 9.24 $\pm$ 0.59 & -0.38 $\pm$ 0.10 & 0.30 $\pm$ 0.06 & 8.59 $\pm$ 0.03 & 1.23 $\pm$ 0.12\\
UGC03539 & 8.09 $\pm$ 0.56 & -0.28 $\pm$ 0.16 & 0.71 $\pm$ 0.18 & 8.39 $\pm$ 0.07 & 1.30 $\pm$ 0.23\\
NGC2347 & 9.00 $\pm$ 0.42 & -0.45 $\pm$ 0.10 & 0.32 $\pm$ 0.08 & 8.56 $\pm$ 0.04 & 0.62 $\pm$ 0.11\\
UGC03899 & 8.21 $\pm$ 0.43 & -0.28 $\pm$ 0.17 & 0.35 $\pm$ 0.14 & 8.27 $\pm$ 0.07 & 0.69 $\pm$ 0.15\\
NGC2410 & 9.29 $\pm$ 0.46 & -0.29 $\pm$ 0.14 & 0.41 $\pm$ 0.08 & 8.51 $\pm$ 0.04 & 1.32 $\pm$ 0.21\\
UGC03969 & 9.10 $\pm$ 0.42 & -0.19 $\pm$ 0.17 & 0.66 $\pm$ 0.17 & 8.49 $\pm$ 0.05 & 1.74 $\pm$ 0.18\\
\hline
\end{tabular}
\end{center}
\end{table*}

\addtocounter{table}{-1} \clearpage

\begin{table*}\label{DR2_Re_2}
\begin{center}
\caption[Properties of the CALIFA DR2 galaxies at the effective radius.]{Properties of the CALIFA DR2 galaxies at the effective radius. ({\it continued})}
\begin{tabular}{lrrrrr}\hline\hline
NAME & log(age/yr) & [Z/H] & A$_{\rm V}^{ssp}$ & 12+log(O/H) & A$_{\rm V}^{gas}$\\
\hline
NGC4470 & 8.94 $\pm$ 0.26 & -0.47 $\pm$ 0.10 & 0.11 $\pm$ 0.07 & 8.45 $\pm$ 0.02 & 0.45 $\pm$ 0.07\\
NGC4644 & 9.43 $\pm$ 0.50 & -0.22 $\pm$ 0.11 & 0.12 $\pm$ 0.08 & 8.58 $\pm$ 0.04 & 0.97 $\pm$ 0.13\\
NGC4676A & 0. $\pm$ 0. & 0. $\pm$ 0. & 0. $\pm$ 0. & $---$ & $---$\\
NGC4874 & 9.64 $\pm$ 0.67 & -0.23 $\pm$ 0.19 & 1e-12. $\pm$ 3.19 & $---$ & $---$\\
UGC08107 & 9.18 $\pm$ 0.33 & -0.35 $\pm$ 0.15 & 0.37 $\pm$ 0.13 & 8.45 $\pm$ 0.05 & 1.45 $\pm$ 0.22\\
UGC08231 & 8.14 $\pm$ 0.37 & -0.32 $\pm$ 0.14 & 0.11 $\pm$ 0.08 & 8.23 $\pm$ 0.05 & 0.18 $\pm$ 0.04\\
UGC08234 & 9.37 $\pm$ 0.17 & -0.21 $\pm$ 0.15 & 0.02 $\pm$ 0.03 & $---$ & $---$\\
\hline
\end{tabular}
\end{center}
\end{table*}

\addtocounter{table}{-1} \clearpage

\begin{table*}\label{DR2_Re_3}
\begin{center}
\caption[Properties of the CALIFA DR2 galaxies at the effective radius.]{Properties of the CALIFA DR2 galaxies at the effective radius. ({\it continued})}
\begin{tabular}{lrrrrr}\hline\hline
NAME & log(age/yr) & [Z/H] & A$_{\rm V}^{ssp}$ & 12+log(O/H) & A$_{\rm V}^{gas}$\\
\hline
NGC6032 & 9.29 $\pm$ 0.31 & -0.15 $\pm$ 0.20 & 0.15 $\pm$ 0.11 & 8.51 $\pm$ 0.07 & 1.17 $\pm$ 0.17\\
UGC10205 & 9.43 $\pm$ 0.34 & -0.24 $\pm$ 0.18 & 0.25 $\pm$ 0.10 & 8.49 $\pm$ 0.04 & 2.01 $\pm$ 0.32\\
NGC6063 & 9.19 $\pm$ 0.25 & -0.42 $\pm$ 0.14 & 0.08 $\pm$ 0.08 & 8.49 $\pm$ 0.05 & 0.66 $\pm$ 0.13\\
IC1199 & 9.15 $\pm$ 0.51 & -0.28 $\pm$ 0.17 & 0.32 $\pm$ 0.15 & 8.53 $\pm$ 0.06 & 0.95 $\pm$ 0.15\\
NGC6081 & 9.72 $\pm$ 0.60 & -0.20 $\pm$ 0.14 & 0.09 $\pm$ 0.05 & $---$ & $---$\\
UGC10331 & 8.93 $\pm$ 0.24 & -0.48 $\pm$ 0.10 & 0.38 $\pm$ 0.08 & 8.38 $\pm$ 0.03 & 0.91 $\pm$ 0.08\\
NGC6125 & 9.67 $\pm$ 0.23 & -0.24 $\pm$ 0.18 & 0.11 $\pm$ 0.08 & $---$ & $---$\\
\hline
\end{tabular}
\end{center}
\end{table*}

\addtocounter{table}{-1} \clearpage

\begin{table*}\label{DR2_Re_4}
\begin{center}
\caption[Properties of the CALIFA DR2 galaxies at the effective radius.]{Properties of the CALIFA DR2 galaxies at the effective radius. ({\it continued})}
\begin{tabular}{lrrrrr}\hline\hline
NAME & log(age/yr) & [Z/H] & A$_{\rm V}^{ssp}$ & 12+log(O/H) & A$_{\rm V}^{gas}$\\
\hline
UGC11649 & 9.43 $\pm$ 0.20 & -0.22 $\pm$ 0.13 & 0.01 $\pm$ 0.02 & 8.55 $\pm$ 0.06 & 0.75 $\pm$ 0.17\\
NGC7025 & 9.73 $\pm$ 0.46 & -0.33 $\pm$ 0.16 & 0.17 $\pm$ 0.11 & 8.50 $\pm$ 0.05 & 1.21 $\pm$ 0.37\\
UGC11717 & 8.28 $\pm$ 0.69 & -0.23 $\pm$ 0.18 & 0.32 $\pm$ 0.17 & 8.57 $\pm$ 0.04 & 1.89 $\pm$ 0.23\\
MCG-01-54-016 & 8.28 $\pm$ 0.60 & -0.29 $\pm$ 0.17 & 0.60 $\pm$ 0.22 & 8.28 $\pm$ 0.11 & 0.62 $\pm$ 0.12\\
UGC11792 & 8.52 $\pm$ 0.48 & -0.43 $\pm$ 0.13 & 1.39 $\pm$ 0.08 & 8.47 $\pm$ 0.05 & 1.75 $\pm$ 0.26\\
NGC7194 & 9.78 $\pm$ 0.38 & -0.13 $\pm$ 0.15 & 0.00 $\pm$ 0.01 & $---$ & $---$\\
UGC11958 & 9.59 $\pm$ 0.41 & -0.24 $\pm$ 0.21 & 0.20 $\pm$ 0.12 & $---$ & 0.57 $\pm$ 0.79\\
\hline
\end{tabular}
\end{center}
\end{table*}

\begin{acknowledgements}

SFS thanks the director of CEFCA, M. Moles, for his sincere support to
this project.

SFS thanks the CONACYT-125180 and DGAPA-IA100815 projects for
providing him support in this study.

We thanks Dr. C.J. Walcher for all his comments and suggestions that has improved the article.

We thanks the anonymous referee for all the comments that has improved the article.

We acknowledge support from the Spanish Ministerio de Econom\'\i a y Competitividad, 
through projects AYA2010-15081 and AYA2010-10904E, and Junta de Andaluc\'\i a FQ1580.
EP acknowledges support from the IA-UNAM and from the Guillermo Haro program at INAOE.

Support for LG is provided by the Ministry of Economy, Development,
and Tourism's Millennium Science Initiative through grant IC120009,
awarded to The Millennium Institute of Astrophysics, MAS. LG
acknowledges support by CONICYT through FONDECYT grant 3140566.

This study  uses data provided by the Calar Alto Legacy
Integral Field Area (CALIFA) survey (http://califa.caha.es/).

CALIFA is the first legacy survey performed at Calar Alto. The
CALIFA collaboration would like to thank the IAA-CSIC and MPIA-MPG as
major partners of the observatory, and CAHA itself, for the unique
access to telescope time and support in manpower and infrastructures.
The CALIFA collaboration also thanks the CAHA staff for the dedication
to this project.

Based on observations collected at the Centro Astron\'omico Hispano
Alem\'an (CAHA) at Calar Alto, operated jointly by the
Max-Planck-Institut f\"ur Astronomie and the Instituto de Astrof\'\i sica de
Andaluc\'\i a  (CSIC).

Funding for SDSS-III has been provided by the Alfred P. Sloan
Foundation, the Participating Institutions, the National Science
Foundation, and the U.S. Department of Energy Office of Science. The
SDSS-III web site is http://www.sdss3.org/.

SDSS-III is managed by the Astrophysical Research Consortium for the
Participating Institutions of the SDSS-III Collaboration including the
University of Arizona, the Brazilian Participation Group, Brookhaven
National Laboratory, Carnegie Mellon University, University of
Florida, the French Participation Group, the German Participation
Group, Harvard University, the Instituto de Astrofisica de Canarias,
the Michigan State/Notre Dame/JINA Participation Group, Johns Hopkins
University, Lawrence Berkeley National Laboratory, Max Planck
Institute for Astrophysics, Max Planck Institute for Extraterrestrial
Physics, New Mexico State University, New York University, Ohio State
University, Pennsylvania State University, University of Portsmouth,
Princeton University, the Spanish Participation Group, University of
Tokyo, University of Utah, Vanderbilt University, University of
Virginia, University of Washington, and Yale University.

\end{acknowledgements}

\bibliography{CALIFAI}

\end{document}